\documentclass[11pt,british]{article}
\usepackage[T1]{fontenc}
\usepackage[latin9]{inputenc}
\usepackage{geometry}
\usepackage{color}
\usepackage{babel}
\usepackage{booktabs}
\usepackage{mathrsfs}
\usepackage{amsmath}
\usepackage{amssymb}
\usepackage{cancel}
\usepackage{graphicx}
\usepackage[unicode=true]
 {hyperref}

\makeatletter

\providecommand{\tabularnewline}{\\}

\newcommand{\lyxaddress}[1]{
\par {\raggedright #1
\vspace{1.4em}
\noindent\par}
}

\@ifundefined{date}{}{\date{}}
\usepackage{cite}
\usepackage{upgreek}

\makeatother

\begin{document}

\title{\textbf{Analog Programmable-Photonic Computation}}

\author{Andr\'es Macho-Ortiz\textsuperscript{1,$\ast$}, Daniel P\'erez-L\'opez\textsuperscript{1,2},
Jos\'e Aza\~{n}a\textsuperscript{3} and Jos\'e Capmany\textsuperscript{1,2,$\ast$} }
\maketitle

\lyxaddress{\begin{center}
\textsuperscript{1}{\small{}ITEAM Research Institute, Universitat
Polit\`ecnica de Val\`encia, Valencia 46022, Spain}\\
\textsuperscript{2}{\small{}iPronics, Programmable Photonics, S.L,
Camino de Vera s/n, Valencia 46022, Spain}\\
\textsuperscript{3}{\small{}Institut National de la Recherche Scientifique
\textendash{} \'Energie, Mat\'eriaux, et T\'el\'ecommunications
(INRS-EMT), Montr\'eal, QC, Canada}
\par\end{center}}

\noindent $\ast$corresponding author: \href{mailto:amachor@iteam.upv.es}{amachor@iteam.upv.es},
\href{mailto:jcapmany@iteam.upv.es}{jcapmany@iteam.upv.es}

\vspace{1cm}

\subsection*{Abstract}

Digital electronics is a technological cornerstone in our modern society
which has covered the increasing demand in computing power during
the last decades thanks to a periodic doubling of transistor density
and power efficiency in integrated circuits. Currently, such scaling
laws are reaching their fundamental limits, leading to the emergence
of a large gamut of applications that cannot be supported by digital
electronics, specifically, those that involve real-time \emph{analog}
multi-data processing, e.g., medical diagnostic imaging, robotic control
and remote sensing, among others. In this scenario, an analog computing
approach implemented in a reconfigurable non-electronic hardware such
as programmable integrated photonics (PIP) can be more efficient than
digital electronics to perform these emerging applications. However,
actual analog computing models such as quantum and neuromorphic computation
were not conceived to extract the benefits of PIP technology (and
integrated photonics\linebreak{}
in general). In this work, we present the foundations of a new computation
theory, termed\linebreak{}
\emph{Analog Programmable-Photonic Computation} (APC), explicitly
designed to unleash the full potential of PIP. Interestingly, APC
enables overcoming some of the basic theoretical and technological
limitations of existing computational models, can be implemented in
other technologies (e.g. in electronics, acoustics or using metamaterials)
and, consequently, exhibits the potential to spark a ground-breaking
impact on our information society. 
\noindent \begin{center}
\newpage{}
\par\end{center}

\subsection*{Introduction}

\noindent Over the last decades, digital electronic technology has
supported the increasing demand in signal processing and computing
power thanks to an exponential performance scaling in microelectronics.
In particular, this progress is embodied in Moore's and Dennard\textquoteright s
laws by which the density of transistors, power efficiency and clock
frequency in microprocessors has approximately doubled every 18-24
months \cite{key-P1,key-P2}. Nevertheless, as seen in Fig.\,1a,
these scaling laws are reaching their fundamental limits. As a result,
there is currently a wide range of emerging real-time signal processing
and computing applications (including medical diagnostic imaging,
robotic control, remote sensing, smart homes, and autonomous driving,
among others) which may not be efficiently dealt with using the dominant
digital electronic paradigm \cite{key-P1,key-P2,key-P3,key-P4,key-P5,key-P6}.

Despite the fact that the electronic industry has proposed to circumvent
the end of Moore's and Dennard\textquoteright s laws by introducing
multi-core technology, there is a limit in the number of cores that
can simultaneously be powered on with a fixed power budget and a constant
heat extraction rate (Amdahl\textquoteright s law) \cite{key-P2,key-P3}.
Moreover, as the bandwidth limitations of silicon electronics and
printed metallic tracks are reached, the time and power consumed in
data transport in an electrical circuit cannot be further reduced
\cite{key-P2,key-P4}. These physical bottlenecks $-$ in combination
with the fact that conventional computational models are conceived
as serialized and centralized processing architectures (von Neumann
machines) implementing the nonlinear Boolean algebra $-$ severely
limit the performance of digital electronic computers \cite{key-P1,key-P2,key-P7,key-P8,key-P9}.
In general, such schemes are inefficient to perform multi-linear analog
operations and computational architectures that are distributed, parallel
and adaptive (Fig.\,1b); for instance, those used to perform real-time
matrix operations, requiring low latency, high bandwidth, low energy
consumption and high reconfigurability (such as the applications mentioned
above) \cite{key-P5,key-P6}.

Although from the Church-Turing thesis can be inferred that any class
of computational problem (or computable function) can be solved by
a digital electronic computer \cite{key-P2,key-P10}, this does not
imply that digital computation (DC) \cite{key-P9} and electronic
technology always lead to the most suitable marriage between a mathematical
computing theory and a hardware platform. Other non-digital computing
approaches implemented in alternative system-on-chip technologies
can be mathematically more efficient than DC to solve the aforementioned
computational scenarios and may provide hardware advantages over electronics
in basic performance parameters (latency, bandwidth, parallelism,
power consumption, or reconfigurability) \cite{key-P1,key-P5,key-P6,key-P10,key-P11}.
Specifically, technologies that are inherently capable of performing
analog operations offering complementary hardware requirements to
those of electronics and being CMOS-compatible are a priority \cite{key-P4,key-P12,key-P13}. 

In this context, programmable integrated photonics (PIP) is the ideal
technology to explore an analog computation paradigm since it is a
hardware platform with complementary features to electronics (providing
lower latency, higher bandwidth, massive parallelism via wavelength-division
multiplexing (WDM), lower power consumption and higher reconfigurability)
and which may be integrated into existing microelectronic processors
by exploiting its CMOS compatibility via silicon photonic platforms
(Fig.\,1c) \cite{key-P12,key-P13,key-P14}. Furthermore, PIP benefits
from the scalable fabrication methods of integrated circuits and its
manufacturing could achieve economies of scale comparable with microelectronic
industry in the next decades \cite{key-P12}.

\newpage{}

So far, in the same vein as other optical computing platforms \cite{key-P4,key-P5,key-P6,key-P7,key-P11,key-P15,key-P16,key-P17,key-P18,key-P19},
PIP has essentially been explored as a hardware acceleration solution
for existing computational models such as DC \cite{key-P9}, quantum
computation (QC) \cite{key-P10} and neuromorphic computation (NC)
\cite{key-P20} implemented in electronic circuits. Here, PIP only
carries out analog signal processing tasks (i.e. wave transformations)
which involve a high degree of complexity in electronics (in particular,
multi-dimensional wave transformations via vector-by-matrix multiplications
\cite{key-P12,key-P13,key-P14}), but does not perform true computational
tasks (i.e. operations among units of information, e.g., among digital
bits). In fact, at present, there is no specific computation theory
available explicitly designed for PIP (and integrated photonics in
general) that allows us to exploit this technology to implement true
optical computing, in the same way as DC sparked a paradigm shift
in commercial electronics. Moreover, DC, QC and NC were not conceived
to extract the benefits of PIP since these models were originally
built without considering the complexity of their implementation in
integrated optics \cite{key-P4,key-P6,key-P11,key-P16,key-P21,key-P22}. 

Being PIP a hardware technology that naturally performs analog matrix
transformations on optical signals, and whose building block may be
designed by using a mathematical framework similar to QC \cite{key-P23},
one could ask whether a classical version of QC might be proposed
within the realm of classical wave-optics. Different works have been
reported revolving around this idea in order to \cite{key-P24,key-P25,key-P26,key-P27,key-P28,key-P29,key-P30}:
(\emph{i}) simulate a quantum computer with a classical computer and
(\emph{ii}) dig into the fundamental differences between quantum and
classical systems. However, to our knowledge, the QC formalism has
never been extrapolated to a classical scenario to construct an analog
computing landscape, based on deterministic physical laws, which allows
us to overcome some of the main theoretical and technological limitations
of QC \cite{key-P10,key-P25,key-P31,key-P32} (e.g., the need to operate
with extreme low temperatures, the practical difficulties to scale
the capabilities of a quantum computer to a large number of quantum
bits (qubits), the wave function collapse in data measurement, and
the impossibility of performing cloning, summation, feedback and non-unitary
operations) and being explicitly designed to harness the full potential
of PIP technology.

To this end, here we present the foundations of an entire new class
of computation theory, termed as Analog Programmable-Photonic Computation
(APC). To achieve this overall aim, we will follow the steps sketched
in Fig.\,1d. Firstly, we will propose a unit of information named
as analog bit (or anbit) and defined as a two-dimensional (2D) analog
function (similar to the qubit, but with essential differences, as
detailed below). Secondly, we will introduce the basic operations
(i.e. gates) among anbits formalising the underlying algebra. Thirdly,
we will design the circuit implementations of these gates using PIP
technology and, fourthly, we will specify a roadmap to further develop
the APC in future works. Finally, a qualitative comparison among the
main properties of APC, DC, QC and NC is discussed, assessing the
unique potential and versatility offered by this computing paradigm.

\subsection*{Results}

\subsubsection*{\emph{Unit of information: the analog bit}}

\noindent APC emerges around the concept of performing analog operations
on a new unit of information, the anbit, which must be easily implementable
using PIP technology. In this vein, considering that the building
block of PIP is usually an optical circuit carrying out $2\times2$
matrix transformations \cite{key-P12,key-P13,key-P14,key-P23}, we
define an anbit as a 2D vector function $\boldsymbol{\uppsi}\left(t\right)=\psi_{0}\left(t\right)\hat{\mathbf{e}}_{0}+\psi_{1}\left(t\right)\hat{\mathbf{e}}_{1}$,
where $\psi_{0,1}$ are scalar complex functions referred to as the
anbit amplitudes and $\hat{\mathbf{e}}_{0,1}$ are constant orthonormal
vectors. The user information is encoded in the moduli and phases
of $\psi_{0,1}$, which may be directly implementable in PIP using
two classical optical wave packets propagated in the fundamental modes
of two parallel uncoupled waveguides, a technique termed as space-encoding
modulation (see Fig.\,2a and Methods). In this work, for the sake
of simplicity, the temporal shape of the wave packets are assumed
rectangular (quasi-rectangular in practice, e.g., using super-Gaussian
pulses). In this fashion, the anbit amplitudes can be regarded as
time-invariant functions (complex numbers) within the time intervals
where $\psi_{0,1}$ are defined. Thus, in contrast to the discrete
1D nature of the bit used in DC (whose value may be 0 or 1), $\psi_{0,1}$
can actually take on a continuous range of complex values. Alternative
physical implementations of an anbit can be proposed by generating
diverse shapes of the wave packets in the space, mode, polarisation,
frequency and time domains, giving rise to different anbit modulation
formats (Supplementary Note 1). Moreover, the following noteworthy
features of an anbit should be highlighted:
\begin{itemize}
\item \emph{Hilbert space}. The single-anbit vector space $\mathscr{E}_{1}=\mathrm{span}\{\hat{\mathbf{e}}_{0},\hat{\mathbf{e}}_{1}\}$
in combination with the standard complex inner product $\bigl\langle\cdot|\cdot\bigr\rangle$
lead to a Hilbert space with a finite norm\linebreak{}
($0\leq\left\Vert \boldsymbol{\uppsi}\right\Vert <\infty$) whose
square provides information about the optical power ($\mathcal{P}$)
propagated by the waveguides depicted in Fig.\,2a: $\mathcal{P}\equiv\left\Vert \boldsymbol{\uppsi}\right\Vert ^{2}=\bigl\langle\boldsymbol{\uppsi}|\boldsymbol{\uppsi}\bigr\rangle=\left|\psi_{0}\right|^{2}+\left|\psi_{1}\right|^{2}$
(see Methods). 
\item \emph{Dimension}. Although, in general, we will work in a Hilbert
space with dimension $d=2$, we have the possibility of defining the
unit of information in a Hilbert space with $d\geq1$, leading to
different versions of APC termed as $d$-APC (the usual case with
$d=2$ will be referred to as APC for short). In Supplementary Note
4, we discuss how to construct the theory with $d\neq2$.
\item \emph{Anbit period}. $\boldsymbol{\uppsi}$ is defined in a time interval
$T_{\mathrm{ANBIT}}$, termed as the anbit period, encompassing from
the beginning of $\psi_{0}$ or $\psi_{1}$ to the end of $\psi_{0}$
and $\psi_{1}$ (Fig.\,2a). The anbit amplitudes are defined in different
time intervals $T_{0}$ and $T_{1}$ (with the possibility of setting
$T_{0}=T_{1}$ or $T_{0}\neq T_{1}$), and the time delay $\Delta T$
between $\psi_{0}=\left|\psi_{0}\right|e^{i\angle_{0}}$ and $\psi_{1}=\left|\psi_{1}\right|e^{i\angle_{1}}$
establishes a differential phase $\angle_{1}-\angle_{0}=\omega_{\mathrm{c}}\Delta T$,
where $\omega_{\mathrm{c}}$ is the angular frequency of the optical
carrier.
\item \emph{Measurement and degrees of freedom}. The recovering of the user
information at the receiver will be referred to as the anbit measurement
and can be carried out via two different ways: (\emph{i}) a coherent
measurement, implementable using coherent detection, or (\emph{ii})
a differential measurement, associated to a direct detection scheme
(Fig.\,2b). The former retrieves the moduli and phases of $\psi_{0,1}$
(4 real degrees of freedom) and the latter only provides information
about $\left|\psi_{0,1}\right|^{2}$ and $\angle_{1}-\angle_{0}$
(3 real degrees of freedom, in this case the global phase of $\psi_{0,1}$
cannot be recovered, see Supplementary Note 1). Hence, the number
of effective degrees of freedom (EDFs) where the user information
can be encoded depends solely on the kind of anbit measurement employed
at the receiver. Although a differential measurement provides the
lowest number of EDFs, it is the most economical anbit measurement
strategy in PIP.\newpage{}
\item \emph{Geometric representations}. An anbit with 4 EDFs (coherent measurement)
can be geometrically represented by using a polar diagram illustrating
the moduli and phases of $\psi_{0,1}$ (Fig.\,2c). An anbit with
3 EDFs (differential measurement) may be represented in the generalised
Bloch sphere (GBS), with a radius different from 1 (Fig.\,2d). 
\item \emph{Multiple anbits}. A multi-anbit gate will require to operate
in a Hilbert space \textquoteleft higher\textquoteright{} than $\mathscr{E}_{1}$.
The construction of such a Hilbert space can be carried out in APC
by using the tensor product \cite{key-P33} or the Cartesian product
\cite{key-P34}. The former will allow us to extrapolate multi-anbit
gates from QC (e.g. controlled gates). The latter will be of great
benefit to construct multi-anbit linear operations which otherwise
would exhibit a nonlinear nature using the tensor product (e.g. the
fan-in and fan-out gates, see below). In Supplementary Note 1, we
detail the main properties of the tensor and Cartesian products within
the framework of APC. 
\end{itemize}
Despite the fact that the anbit is similar to the qubit (and to its
classical counterpart, the cebit \cite{key-P24}), the following fundamental
differences should be highlighted: (\emph{i}) the anbit norm may be
different from 1 and can be modified using a non-unitary operation,
(\emph{ii}) the vector superposition of $\hat{\mathbf{e}}_{0}$ and
$\hat{\mathbf{e}}_{1}$ is preserved after an anbit measurement (Fig.\,2b),
(\emph{iii}) an anbit has 1 or 2 more EDFs than the qubit, (\emph{iv})
multiple anbits can be composed by using not only the tensor product
but also the Cartesian product, (\emph{v}) an anbit may be defined
in a one-dimensional Hilbert space (the qubit cannot be restricted
to one dimension given that a global phase is not observable \cite{key-P10}).
However, in contrast to QC, in APC we will not be able to perform
instantaneous non-local operations among different anbits (i.e., the
entanglement of multiple units of information) because the underlying
physical laws are deterministic \cite{key-P35}.

Furthermore, taking into account the vector formalism required to
describe an anbit and its mathematical similitude with a qubit (but
with the essential differences discussed above), let us introduce
at this point the use of Dirac\textquoteright s notation in order
to: (\emph{i}) simplify the mathematical calculations when designing
complex APC computing architectures and (\emph{ii}) extrapolate diverse
analysis and design strategies from QC, preserving the same notation
between both computation theories. Therefore, from now on, let us
express the anbit as $\bigl|\psi\bigr\rangle=\psi_{0}\bigl|0\bigr\rangle+\psi_{1}\bigl|1\bigr\rangle$,
with $\boldsymbol{\uppsi}\equiv\bigl|\psi\bigr\rangle$, $\hat{\mathbf{e}}_{0}\equiv\bigl|0\bigr\rangle$
and $\hat{\mathbf{e}}_{1}\equiv\bigl|1\bigr\rangle$. 

\subsubsection*{\emph{Fundamentals of combinational design: basic anbit gates}}

\noindent The second natural step to construct the computation theory
is to introduce the basic anbit operations (or gates), see Fig.\,3.
The simplest operation that can be built in APC is a gate of a single
anbit: a non-feedback (or combinational) system carrying out a transformation
between two different anbits, the input anbit $\bigl|\psi\bigr\rangle=\psi_{0}\bigl|0\bigr\rangle+\psi_{1}\bigl|1\bigr\rangle$
and the output anbit $\bigl|\varphi\bigr\rangle=\varphi_{0}\bigl|0\bigr\rangle+\varphi_{1}\bigl|1\bigr\rangle$
(Fig.\,3a). Mathematically, the gate is described via an arbitrary
mapping (or operator) $\widehat{\mathrm{F}}:\mathscr{E}_{1}\rightarrow\mathscr{E}_{1}$,
which will be assumed to be a holomorphic function for convenience.
In this way, such a mapping can be written as a power series $\widehat{\mathrm{F}}=\widehat{\mathrm{F}}^{\left(1\right)}+\widehat{\mathrm{F}}^{\left(2\right)}+\widehat{\mathrm{F}}^{\left(3\right)}+\ldots$,
with $\widehat{\mathrm{F}}^{\left(k\right)}$ accounting for the linear
($k=1$) and nonlinear ($k>1$) responses of the gate. Considering
that the PIP circuits are typically linear systems \cite{key-P12,key-P13,key-P14},
we will focus our attention on the case $\widehat{\mathrm{F}}\equiv\widehat{\mathrm{F}}^{\left(1\right)}$,
that is, in linear gates. Nonetheless, it is worth mentioning that
APC may also be constructed by using nonlinear gates (see Methods).

\newpage{}

Specifically, a single-anbit linear gate is described by a linear
operator (or endomorphism) $\widehat{\mathrm{F}}$ exhibiting the
following general properties:
\begin{itemize}
\item \emph{Uniqueness}. The input and output anbits are always related
by a unique endomorphism $\widehat{\mathrm{F}}$. This property directly
follows from the uniqueness of a linear transformation between vector
spaces \cite{key-P36}.
\item \emph{Matrix representation}. Given an orthonormal vector basis $\mathcal{B}_{1}=\left\{ \left|0\right\rangle ,\left|1\right\rangle \right\} $,
the matrix representation of $\widehat{\mathrm{F}}$ is unique and
is given by the expression (Supplementary Note 2):
\[
F=\left(\begin{array}{cc}
\bigl\langle0|\widehat{\mathrm{F}}|0\bigr\rangle & \bigl\langle0|\widehat{\mathrm{F}}|1\bigr\rangle\\
\bigl\langle1|\widehat{\mathrm{F}}|0\bigr\rangle & \bigl\langle1|\widehat{\mathrm{F}}|1\bigr\rangle
\end{array}\right).\tag{1}
\]
Hence, the gate can equivalently be described by the matrix $F$ and
the input-output relation $\bigl|\varphi\bigr\rangle=\widehat{\mathrm{F}}\bigl|\psi\bigr\rangle$
can be expressed as $\left[\bigl|\varphi\bigr\rangle\right]_{\mathcal{B}_{1}}=F\left[\bigl|\psi\bigr\rangle\right]_{\mathcal{B}_{1}}$,
where $\left[\bigl|\varphi\bigr\rangle\right]_{\mathcal{B}_{1}}=\bigl(\begin{array}{cc}
\varphi_{0} & \varphi_{1}\end{array}\bigr)^{T}$ and $\left[\bigl|\psi\bigr\rangle\right]_{\mathcal{B}_{1}}=\bigl(\begin{array}{cc}
\psi_{0} & \psi_{1}\end{array}\bigr)^{T}$ ($T$ denotes the transpose matrix).
\item \emph{Reversibility}. By definition, a gate is reversible when $\widehat{\mathrm{F}}$
is a bijective mapping (automorphism). In such circumstances, $\det\left(F\right)\neq0$
and the input anbit can be recovered from the output anbit by applying
the inverse mapping $\widehat{\mathrm{F}}^{-1}$, whose matrix representation
is $F^{-1}$. Contrariwise, a gate is irreversible when $\det\left(F\right)=0$
and the input anbit cannot be retrieved from the output anbit given
that $\cancel{\exists}\widehat{\mathrm{F}}^{-1}$. 
\item \emph{Non-locality}. The input-output relation $\bigl|\varphi\bigr\rangle=\widehat{\mathrm{F}}\bigl|\psi\bigr\rangle$
is non-local and causal. The input and output anbits are respectively
encoded by two different electric fields $\boldsymbol{\mathcal{E}}\left(\mathbf{r}_{1},t_{1}\right)$
and $\boldsymbol{\mathcal{E}}\left(\mathbf{r}_{2},t_{2}\right)$ with
$\mathbf{r}_{1}\neq\mathbf{r}_{2}$ and $t_{1}<t_{2}$ (see Methods). 
\item \emph{Classes of linear gates}. Since a linear gate can be described
by a matrix, we will use a terminology based on matrix theory \cite{key-P37}
to define different classes of operations:\linebreak{}
unitary\,gates (U-gates), general\,linear\,gates (G-gates) and
general\,matrix\,gates (M-gates).\linebreak{}
 Concretely, the U-gates will account for the linear reversible mappings
that preserve the norm of the input anbit (conservative operations)
via a unitary matrix transformation. Contrariwise, the G- and M-gates
will describe non-conservative linear operations. While, by definition,
a G-gate is always reversible, an M-gate may be reversible or irreversible,
encompassing both possibilities. Hence, a G-gate will be described
by a general linear (i.e. non-singular) matrix, whereas an M-gate
will be associated to a general complex matrix (singular or non-singular).
\item \emph{Geometric representation}. Using differential measurement (that
will be the case in most scenarios of APC), a single-anbit gate may
be geometrically interpreted as a trajectory between two different
points in the GBS (Fig.\,3a). Specifically, the kind of trajectory
depends on the class of the gate, see Supplementary Note 2 for more
details. 
\item \emph{Algebraic structure and universal matrices}. The U- and G-gates
belong to the $\mathrm{U}\left(2\right)$ and $\mathrm{GL}\left(2,\mathbb{C}\right)$
Lie groups, respectively, while an M-gate belongs to the $\mathfrak{gl}\left(2,\mathbb{C}\right)$
Lie algebra \cite{key-P37}. Using the fundamentals of these algebraic
structures \cite{key-P10,key-P23,key-P37,key-P38}, it is straightforward
to find a universal (or arbitrary) matrix in each class of gate, which
must be able to describe all the possible $2\times2$ matrix transformations
associated to the class when varying the value of its entries, encoded
by parameters. In particular, a universal matrix of a U-gate reads
as follows \cite{key-P10,key-P23}:
\[
U=e^{i\delta}R_{\hat{\mathbf{n}}}\left(\alpha\right)=e^{i\delta}\left(\begin{array}{cc}
\cos\frac{\alpha}{2}-in_{z}\sin\frac{\alpha}{2} & -\left(n_{y}+in_{x}\right)\sin\frac{\alpha}{2}\\
\left(n_{y}-in_{x}\right)\sin\frac{\alpha}{2} & \cos\frac{\alpha}{2}+in_{z}\sin\frac{\alpha}{2}
\end{array}\right),\tag{2}
\]
where $\delta\in\left[0,2\pi\right)$ is a global phase shifting and
$R_{\hat{\mathbf{n}}}\left(\alpha\right)$ is a rotation matrix accounting
for a rotation of an angle $\alpha\in\left[0,2\pi\right]$ around
an arbitrary unit vector $\hat{\mathbf{n}}=n_{x}\hat{\mathbf{x}}+n_{y}\hat{\mathbf{y}}+n_{z}\hat{\mathbf{z}}$
($n_{x,y,z}\in\left[-1,1\right]$) in the GBS (Fig.\,3b). In addition,
a possible universal matrix of a G- or M-gate is a parametric matrix
(denoted as $G$ or $M$, respectively) with the four entries described
by four independent complex numbers \cite{key-P37,key-P38}. Nevertheless,
in the former case (G-gate), the condition $\det\left(G\right)\neq0$
must be fulfilled.
\item \emph{PIP implementation}. The optical implementation (or circuit
architecture) of a U-, G- or M-gate must be able to perform the $2\times2$
matrix transformation described by its universal matrix by utilising
basic PIP devices: phase shifters (PSs), resonators, attenuators,
amplifiers, and beam splitters (combiners) such as directional couplers
(DICs) and multi-mode interferometers (MMIs). Since a universal matrix
may be implementable by\linebreak{}
different equivalent circuit architectures, we should introduce here
the concept of\linebreak{}
\emph{minimal circuit architecture} (MCA), defined as the PIP implementation
encompassing\linebreak{}
the minimum number of basic devices. Furthermore, since any PIP circuit
is fully\linebreak{}
characterized by analysing the Lorentz reciprocity and the forward-backward
(FB)\linebreak{}
symmetry \cite{key-P13}, we include an extended discussion about
these properties within the context of APC in Supplementary Note 2.
\end{itemize}
Uncovering the MCA of the U-, G-, and M-gates requires to explore
diverse matrix factorization techniques that allow us to implement
the universal matrix of each class of operation by utilising PIP technology.
After a thorough examination of the matrix theory literature\linebreak{}
\cite{key-P10,key-P34,key-P36,key-P37,key-P38,key-P39,key-P40,key-P41,key-P42},
two different factorization techniques should be taken into consideration
in our discussions: Euler\textquoteright s rotation theorem and the
singular value decomposition (SVD).

As reported in ref.\,\cite{key-P23}, a $2\times2$ universal unitary
matrix of the form given by Eq.\,(2) cannot be directly implemented
by using mainstream PIP devices because of the arbitrary nature of
$\hat{\mathbf{n}}$. Here, we can take advantage of Euler\textquoteright s
rotation theorem to factorize the $U$ matrix as a concatenation of
three rotations around two Cartesian axes of the GBS, which are implementable
via PSs, MMIs and DICs. In addition, taking into account that a U-gate
can be regarded as a $2\times2$ universal unitary signal processor,
then the MCA of a U-gate (Fig.\,3b) must be the same as the MCA of
a $2\times2$ universal unitary signal PIP processor, shown in Fig.\,4a
of ref.\,\cite{key-P23} and based on the Euler factorization $U=e^{i\delta}R_{\hat{\mathbf{n}}}\left(\alpha\right)\equiv e^{i\delta}R_{\hat{\mathbf{z}}}\left(\alpha_{3}\right)R_{\hat{\mathbf{x}}}\left(\alpha_{2}\right)R_{\hat{\mathbf{z}}}\left(\alpha_{1}\right)$.
The matrices $R_{\hat{\mathbf{z}}}\bigl(\alpha_{1,3}\bigr)$ can be
implemented by PSs integrated in parallel uncoupled waveguides and
the matrix $R_{\hat{\mathbf{x}}}\left(\alpha_{2}\right)$ may be generated
by a synchronous DIC with tunable mode-coupling coefficient $\kappa=\alpha_{2}/\left(2L\right)$,
where $L$ is the length of its arms. This MCA preserves the Lorentz
reciprocity but breaks the FB symmetry (provided that $\alpha_{1}\neq\alpha_{3}$).
Equivalent circuit architectures of a U-gate may be explored by selecting
different rotation vectors when using Euler\textquoteright s rotation
theorem. As an example, in Supplementary Note 2 it is shown a scheme
built from fixed couplers, based on the factorization $U=e^{i\delta}R_{\hat{\mathbf{z}}}\left(\alpha_{3}\right)R_{\hat{\mathbf{y}}}\left(\alpha_{2}\right)R_{\hat{\mathbf{z}}}\left(\alpha_{1}\right)$.\newpage{}

While a U-gate is a conservative transformation (given that it preserves
the norm of the input anbit or, equivalently, the power of the 2D
wave that encodes the anbit), both G- and M-gates are non-conservative
transformations. This implies that their MCAs will require to include
attenuators and amplifiers. Remarkably, a common MCA for both kind
of gates is found from the SVD \cite{key-P34,key-P36,key-P39}, which
factorizes the universal matrices of these gates as a function of
two U-gates along with a $2\times2$ diagonal matrix with positive
real entries, implementable by using tunable optical attenuators and
amplifiers (Fig.\,3c). The reciprocal (non-reciprocal) nature of
such devices preserves (breaks) the Lorentz reciprocity in the MCA.
Likewise, note that the FB symmetry is broken in the MCA when using
the circuit of Fig.\,3b to implement the U-gates. 

Although equivalent circuit architectures of the U-, G- and M-gates
can be proposed by using matrix factorizations different from Euler\textquoteright s
rotation theorem and the SVD, all of them lead to optical schemes
integrating a higher number of basic PIP devices than the structures
depicted in Fig.\,3 (see Supplementary Note 2). 

So far, we have presented the basic operations of a single anbit.
Nonetheless, bearing in mind that PIP is a reconfigurable hardware
\cite{key-P12,key-P13,key-P14}, the design of complex combinational
architectures requires to introduce an additional fundamental piece:
a \emph{controlled gate}. In APC, such a kind of gate may be defined
in the same way as in QC \cite{key-P10}. By convention, a controlled
anbit gate performs a transformation $\widehat{\mathrm{F}}$ on the
target anbits when the control anbits are equal to $\bigl|1\bigr\rangle$.
Otherwise, the target anbits remain invariant at the output. Figure\,4a
illustrates the functionality of a controlled gate with a single target
anbit $\bigl|t\bigr\rangle$ and a single control anbit $\bigl|c\bigr\rangle$.
Using the tensor product, the mathematical formalisation and properties
of a controlled gate in APC can be directly extrapolated from QC (Supplementary
Note 2), with the essential difference that a controlled gate may
be constructed from a non-unitary $\widehat{\mathrm{F}}$ mapping
in APC.

An additional crucial difference between a controlled gate in APC
and QC emerges when analysing its implementation using PIP technology.
Since the reconfigurability of a PIP circuit is realised by utilising
classical electrical control signals \cite{key-P12,key-P13,key-P14},
the implementation of a controlled anbit gate does not require the
intricate design strategies and architectures employed in optical
QC \cite{key-P10,key-P21,key-P43,key-P44} (however, these schemes
could be extrapolated to APC, if desired). As seen in Fig.\,4b, the
simplest implementation of a controlled anbit gate arises from an
electro-optic design, where the control anbit is encoded by the electrical
control signals of the PIP platform and the target anbit is encoded
by a 2D optical wave (alternatively, both control and target anbits
can be encoded by optical waves, giving rise to an all-optical architecture
requiring a higher footprint than that of the electro-optic design,
see Fig.\,4c). In this fashion, the same MCAs as those of the U-,
G-, and M-gates (Fig.\,3) may be employed to perform controlled operations
of each kind of gate. Therefore, the electro-optic architecture only
entails the definition of a mapping between the control anbit and
the electrical control signals of the PIP circuit (e.g. via software
\cite{key-P45}).

As an illustrative example, the electro-optic implementation of the
controlled-NOT (CNOT) gate is sketched in Fig.\,4d. Any suitable
mapping between the control anbit and the electrical control signals
must guarantee that the $2\times2$ unitary matrix transformations
$F=iR_{\hat{\mathbf{x}}}\left(\pi\right)$ (the Pauli matrix $\sigma_{x}$)
and $F=I$ (the identity matrix) are induced on the amplitudes of
$\bigl|t\bigr\rangle$ when $\bigl|c\bigr\rangle=\bigl|1\bigr\rangle$
and $\bigl|c\bigr\rangle=\bigl|0\bigr\rangle$, respectively. It is
worthy to note that, in contrast to the seminal optical implementation
of the quantum CNOT gate reported in ref.\,\cite{key-P43}, in APC
the implementation of this gate integrates a lower number of basic
devices and does not require to use extra (ancilla or garbage) units
of information. 

Interestingly, the concept of controlled gates can be easily extended
to the case of multiple control anbits without requiring extra devices
in the optical circuits. Accordingly, multi-controlled operations
such as the Toffoli gate (an indispensable tool to implement Boolean
functions) can be carried out in APC using the same circuit as that
of Fig.\,4d by encoding an additional control anbit in the electrical
control signals, see Supplementary Note 2. 

\subsubsection*{\emph{Fundamentals of sequential design: taming prohibited operations
in QC}}

\noindent In a combinational system such as the U-, G-, and M-gates,
the outputs depend solely on the inputs. Nevertheless, in a sequential
system, the outputs can be connected with the inputs allowing feedback
operations. Since sequential architectures are of paramount importance
in DC (e.g. to construct digital memories \cite{key-P8,key-P9}),
we wonder about the possibility of exploiting such a kind of systems
within the context of APC. 

Remarkably, in contrast to QC, feedback schemes will be allowed in
APC thanks to the feasibility of performing summation (fan-in) and
cloning (fan-out) of anbits using PIP circuits. These are prohibited
operations within the realm of QC that will however play a fundamental
role to construct any sequential architecture in APC. Concretely,
both fan-in and fan-out anbit gates can be implemented via the PIP
circuit depicted in Fig.\,5a, which preserves the Lorentz reciprocity
and the FB symmetry (provided that the amplifiers have a reciprocal
and FB symmetric behaviour). This scheme transforms the input anbits
$\bigl|\psi\bigr\rangle$ and $\bigl|\varphi\bigr\rangle$ into the
output anbits $\bigl|\psi+\varphi\bigr\rangle=\bigl(\psi_{0}+\varphi_{0}\bigr)\bigl|0\bigr\rangle+\bigl(\psi_{1}+\varphi_{1}\bigr)\bigl|1\bigr\rangle$
and $\bigl|\psi-\varphi\bigr\rangle=\bigl(\psi_{0}-\varphi_{0}\bigr)\bigl|0\bigr\rangle+\bigl(\psi_{1}-\varphi_{1}\bigr)\bigl|1\bigr\rangle$.
Thus, setting $\bigl|\varphi\bigr\rangle=\bigl|\mathbf{0}\bigr\rangle=0\bigl|0\bigr\rangle+0\bigl|1\bigr\rangle$
(the null vector of $\mathscr{E}_{1}$) we will carry out a fan-out
operation of the anbit $\bigl|\psi\bigr\rangle$ (a perfect cloning)
and taking $\bigl|\varphi\bigr\rangle\neq\bigl|\mathbf{0}\bigr\rangle$
we will perform a fan-in operation of the anbits $\bigl|\psi\bigr\rangle$
and $\bigl|\varphi\bigr\rangle$. Moreover, in order to guarantee
a linear behaviour, both fan-in and fan-out gates should be defined
by using the Cartesian product, which allows us to independently transform
the anbit amplitudes $\psi_{0}$, $\psi_{1}$, $\varphi_{0}$ and
$\varphi_{1}$ (conversely, the tensor product leads to multi-anbit
nonlinear operations, see Supplementary Note 3, including a more in-depth
discussion about the mathematical properties and optical implementation
of these gates). 

Figure 5b shows the simplest sequential architecture that can be built
in APC, integrated by both fan-in and fan-out gates along with two
M-gates ($\widehat{\mathrm{M}}_{1}$ and $\widehat{\mathrm{M}}_{2}$)
to complete the feedback loop. The analysis of the input-output relation
$\bigl|\varphi\bigr\rangle=\widehat{\mathrm{M}}_{\mathrm{eq}}\bigl|\psi\bigr\rangle$
indicates that this sequential scheme is equivalent to a combinational
M-gate described by the matrix representation\linebreak{}
$M_{\mathrm{eq}}=\bigl(I-M_{1}M_{2}\bigr)^{-1}M_{1}$. Hence, the
existence of $M_{\mathrm{eq}}$ is closely linked to the condition
$\det\bigl(I-M_{1}M_{2}\bigr)\neq0$. Contrariwise, the loop cannot
be built because the matrix $I-M_{1}M_{2}$ is singular. In Supplementary
Note 3, we provide further information about the analysis and properties
of this structure.

Although the same single-anbit operation $\widehat{\mathrm{M}}_{\mathrm{eq}}$
can be implemented via the MCA of an M-gate (Fig.\,3c), the potential
of this basic sequential scheme relies on the fact that it establishes
the fundamental strategies to analyse and design more complex sequential
architectures in APC. For instance, the intricate multi-anbit combinational
scheme shown in Fig.\,5c can elegantly be simplified with the equivalent
sequential architecture depicted in Fig.\,5d, composed by a lower
number of gates. Surprisingly, both systems have the same input-output
relation (see Supplementary Note 3). 

\subsection*{Discussion}

\noindent These results lay the theoretical foundations of APC, a
new computing paradigm conceived to exploit the full potential of
PIP technology and, consequently, leading to the emergence of an entire
field of research within computational science and photonics. In addition,
APC can be regarded as a new optical design toolbox which blazes a
trail for manufacturing advanced photonic computing architectures
that can team-up with digital electronic processors to unlock in the
near- and middle-term the serious limitations imposed by the demise
of Moore\textquoteright s and Dennard\textquoteright s laws. 

Bearing in mind that this work is completely devoted to establishing
the fundamentals of APC, a roadmap must be specified to further develop
this computation theory in forthcoming works (Fig.\,1d). Firstly,
the fundamentals of combinational design should be extended to the
case of multiple anbits for both U- , G-, and M-gates. We may expect
that the MCA of these multi-anbit gates can also be employed to implement
controlled gates with multiple target anbits (encoding the control
anbits in the electrical control signals of the PIP platform). Secondly,
the fundamentals of sequential design should be further developed
to the case of multiple anbits, embracing the research of fan-in,
fan-out and feedback operations. Given that a digital memory is built
from a multi-bit sequential architecture in DC \cite{key-P9}, the
study of multi-anbit feedback schemes could be of paramount importance
to revisit the concept of memory within the scope of APC. Outstandingly,
the capacity to scale both combinational and sequential architectures
to multiple anbits is inherently related to the feasibility to scale
the PIP circuits integrating multiple waveguides \cite{key-P12,key-P13,key-P14}
in combination with the exploitation of WDM technique \cite{key-P19}
to perform massive parallel computing of anbits. Thirdly, APC must
be completed by conceiving a specific gamut of analog search algorithms
to speed up the solution of a representative set of deterministic
polynomial time (P) and non-deterministic polynomial time (NP) computational
problems (here, it is not possible to rigorously specify such a set
of problems because it previously requires to completely develop the
multi-anbit operations. In any case, as an example, two possible P
and NP problems that could be evaluated are the maximum cardinality
matching \cite{key-P46} and the integer factorization problem \cite{key-P47},
respectively). The time, resources and energy required in APC to solve
these computational problems must be compared with the time, resources
and energy required in DC, QC and NC using the general methodologies
of computational science \cite{key-P10,key-P48}.

Compared with these existing computational models, APC relaxes some
of their theoretical and technological limitations, see Table 1. While
in APC we have the possibility of defining both linear and nonlinear
gates, DC and QC can only be constructed from nonlinear and linear
gates, respectively \cite{key-P9,key-P10}. Moreover, NC embraces
both linear and nonlinear operations, but the global transformation
induced on the units of information in a neural network is nonlinear
\cite{key-P49}. In contrast, in APC there exists the possibility
of exclusively performing linear or nonlinear operations (or a combination
of both). A similar remark applies to the reversible and irreversible
nature of the operations. Both design possibilities can be found in
APC via the U-, G- and M-gates, a feature that is not shared by the
other computation theories (in DC reversible operations are not efficient
since ancilla and garbage bits are required \cite{key-P50}, in QC
all operations are reversible, and in NC the multi-dimensional transformation
of a neural network is irreversible, inherited from its nonlinear
nature). 

\newpage{}

On the other hand, an essential difference between DC and APC relies
on the fact that a combinational APC architecture can take advantage
of both forward and backward propagations of light to compute the
double of units of information (the same (a different) multi-anbit
transformation will be induced in each propagation direction when
the circuit preserves (breaks) the FB symmetry). This property also
applies to QC and NC when implementing the corresponding computational
architectures via PIP technology \cite{key-P6,key-P44}.

The additional properties shown in Table 1 highlight common differences
of APC, DC and NC with QC. Although QC provides the outstanding feature
of performing instantaneous non-local operations (via the entanglement
of qubits, a characteristic that is not shared by the other computation
theories), APC can be readily implemented by current PIP technology
operating at room temperature. Along this line, it is worth mentioning
that APC can also be implemented in any technological platform enabling
analog signal processing based on matrix transformations (e.g. optical
metamaterials \cite{key-P17,key-P18}, electronics \cite{key-P27,key-P51,key-P52}
and acoustics \cite{key-P53}).

Likewise, being APC a computation theory using classical waves, we
may expect that this new computing paradigm has more tolerance to
external noise than QC. This subsequently implies that we will require
a lower number of extra units of information than in QC to detect
and correct the data errors, simplifying the scalability of APC architectures
to a large number of anbits.

An additional intriguing feature of APC arises from the general nature
of its mathematical framework, inherited from the versatile properties
of the anbit and the basic gates, allowing to implement (at least
partially) other computing paradigms using APC architectures (see
Supplementary Note 5). 

Given that any computation theory has associated an information theory,
APC also leads to an additional field of research: the \emph{Analog
Programmable-Photonic Information} (API). Concretely, API will be
focus on the study of anbit modulation formats (Supplementary Note
1), analysis of noise, error detection and correction strategies,
entropy analysis, data compression, and cryptography techniques. Both
APC and API have the potential to spark a crucial impact on fundamental
and applied research, as well as on our information society.

\subsection*{Methods}

\subsubsection*{\emph{Space-encoding modulation}}

\noindent The description of this modulation format (similar to the
path-encoding strategy employed in optical QC \cite{key-P54}) can
be done by specifying the electric field strength encoding the anbit
of Fig.\,2a. According to the usual features of the optical waveguides
employed in PIP \cite{key-P13}, we may assume that the parallel waveguides
of Fig.\,2a have a negligible inter-waveguide mode-coupling and operate
in the paraxial and single-mode regimes. Hence, a space-encoded anbit
is characterized by an electric field of the form:
\begin{align*}
\boldsymbol{\mathcal{E}}\left(\mathbf{r},t\right) & \simeq\sum_{k=0}^{1}\textrm{Re}\left\{ \psi_{k}\left(t-\beta_{k}^{\left(1\right)}r_{\parallel}\right)\hat{\mathbf{e}}_{k}\left(r_{\perp,1},r_{\perp,2},\omega_{\mathrm{c}}\right)e^{i\omega_{\mathrm{c}}t}e^{-i\beta_{k}^{\left(0\right)}r_{\parallel}}\right\} \\
 & =\sum_{k=0}^{1}\left|\psi_{k}\left(t-\beta_{k}^{\left(1\right)}r_{\parallel}\right)\right|\hat{\mathbf{e}}_{k}\left(r_{\perp,1},r_{\perp,2},\omega_{\mathrm{c}}\right)\cos\left(\omega_{\mathrm{c}}t-\beta_{k}^{\left(0\right)}r_{\parallel}+\angle_{k}\right),\tag{3}
\end{align*}
where the anbit amplitudes $\psi_{0}=\left|\psi_{0}\right|e^{i\angle_{0}}$
and $\psi_{1}=\left|\psi_{1}\right|e^{i\angle_{1}}$ play the role
of the optical wave packets (or complex envelopes), $\omega_{\mathrm{c}}$
is the angular frequency of the optical carrier,\linebreak{}
$\mathbf{r}=r_{\perp,1}\hat{\mathbf{r}}_{\perp,1}+r_{\perp,2}\hat{\mathbf{r}}_{\perp,2}+r_{\parallel}\hat{\mathbf{r}}_{\parallel}$
is the vector position written as a function of its transverse ($r_{\perp,1}$,
$r_{\perp,2}$) and longitudinal ($r_{\parallel}$) components, and
$\hat{\mathbf{e}}_{k}$ and $\beta_{k}\left(\omega\right)\simeq\beta_{k}^{\left(0\right)}+\left(\omega-\omega_{\mathrm{c}}\right)\beta_{k}^{\left(1\right)}$
are respectively the normalised mode profile and the propagation constant
of the fundamental mode in waveguide $k$ (being $\beta_{k}^{\left(0\right)}=\beta_{k}\left(\omega=\omega_{\mathrm{c}}\right)$,
$\beta_{k}^{\left(1\right)}=\mathrm{d}\beta_{k}\left(\omega=\omega_{\mathrm{c}}\right)/\mathrm{d}\omega$
and omitting the dispersive terms $\beta_{k}^{\left(n\geq2\right)}$
in the Taylor series expansion of $\beta_{k}\left(\omega\right)$).
In particular, $\hat{\mathbf{e}}_{k}$ must satisfy the condition
\cite{key-P55}:
\[
\iint_{-\infty}^{\infty}\hat{\mathbf{e}}_{k}\times\hat{\mathbf{h}}_{k}^{\ast}\cdot\hat{\mathbf{r}}_{\parallel}\mathrm{d}r_{\perp,1}\mathrm{d}r_{\perp,2}=2,\tag{4}
\]
being $\hat{\mathbf{h}}_{k}$ the normalised mode profile of the magnetic
field strength. Equation (4) guarantees that the optical power propagated
by the fundamental modes of both waveguides can be calculated as $\mathcal{P}=\left|\psi_{0}\right|^{2}+\left|\psi_{1}\right|^{2}$.

Using Eq.\,(3), it is straightforward to describe the electric field
strength at the input $\boldsymbol{\mathcal{E}}\left(\mathbf{r}_{1},t_{1}\right)$
and at the output $\boldsymbol{\mathcal{E}}\left(\mathbf{r}_{2},t_{2}\right)$
of the single-anbit gate depicted in Fig.\,3a, which must be particularised
at two different vector positions $\mathbf{r}_{1}\neq\mathbf{r}_{2}$
and time instants $t_{1}\neq t_{2}$. Taking into account the causal
response of the materials employed in PIP \cite{key-P13}, then it
follows that $t_{1}<t_{2}$. Hence, the input-output relation of the
gate is non-local and causal.

\subsubsection*{\emph{Nonlinear anbit gates}}

\noindent Single-anbit nonlinear gates can be implemented in PIP,
e.g., by means of the Pockels and Kerr effects, which allow to carry
out second- and third-order nonlinear anbit transformations, respectively.
For instance, stimulating the self-phase modulation effect in two
parallel uncoupled waveguides (similar to those of depicted in Fig.\,2a),
a nonlinear anbit operation of the form $\widehat{\mathrm{F}}\bigl|\psi\bigr\rangle=\psi_{0}\exp\bigl(-i\gamma\bigl|\psi_{0}\bigr|^{2}L_{\mathrm{eff}}\bigr)\bigl|0\bigr\rangle+\psi_{1}\exp\bigl(-i\gamma\bigl|\psi_{1}\bigr|^{2}L_{\mathrm{eff}}\bigr)\bigl|1\bigr\rangle$
may be obtained ($\gamma$ and $L_{\mathrm{eff}}$ are nonlinear parameters
of the waveguides \cite{key-P56}). In the same vein, optical devices
such as nonlinear directional couplers \cite{key-P57} and ring resonators
\cite{key-P58} can also be employed to exploit third-order nonlinearities
in silicon platforms.

The most general definition of a single-anbit gate (including both
linear and nonlinear contributions) is given by the expression $\widehat{\mathrm{F}}\bigl|\psi\bigr\rangle:=f_{0}\bigl(\psi_{0},\psi_{1}\bigr)\bigl|0\bigr\rangle+f_{1}\bigl(\psi_{0},\psi_{1}\bigr)\bigl|1\bigr\rangle$,
with $f_{0}$ and $f_{1}$ belonging to $\mathcal{F}\left(\mathbb{C}^{2},\mathbb{C}\right)$.
Thus, $\widehat{\mathrm{F}}$ will induce a nonlinear transformation
on the input anbit when the functions $f_{0,1}$ have a nonlinear
behaviour. Therefore, using holomorphic $f_{0,1}$ functions in a
neighbourhood of a reference point $\bigl(\psi_{0,\mathrm{ref}},\psi_{1,\mathrm{ref}}\bigr)\in\mathbb{C}^{2}$,
we will be able to build a nonlinear response of the desired order.

The main drawback of operating with nonlinear anbit gates relies on
the fact that we cannot deal with a matrix formalism. Nevertheless,
the mathematical description of the above nonlinear anbit operation
can be simplified by performing a Taylor series expansion of $f_{0,1}$.
To this end, let us introduce the vectors $\mathbf{z}:=\psi_{0}\hat{\mathbf{z}}_{0}+\psi_{1}\hat{\mathbf{z}}_{1}$
and $\mathbf{z}_{\mathrm{ref}}:=\psi_{0,\mathrm{ref}}\hat{\mathbf{z}}_{0}+\psi_{1,\mathrm{ref}}\hat{\mathbf{z}}_{1}$
belonging to the vector space $\mathcal{Z}=\mathrm{span}\left\{ \hat{\mathbf{z}}_{0},\hat{\mathbf{z}}_{1}\right\} $
(isomorphic to $\mathbb{C}^{2}$) and being $\hat{\mathbf{z}}_{0,1}$
complex orthonormal vectors. In this way, we can write $\widehat{\mathrm{F}}=\sum_{n=1}^{\infty}\widehat{\mathrm{F}}^{\left(n\right)}$
where:
\[
\widehat{\mathrm{F}}^{\left(n\right)}\bigl|\psi\bigr\rangle=\frac{1}{n!}\mathrm{d}^{n}\left.f_{0}\right\rfloor _{\mathbf{z}_{\mathrm{ref}}}\left(\mathbf{z}\right)\bigl|0\bigr\rangle+\frac{1}{n!}\mathrm{d}^{n}\left.f_{1}\right\rfloor _{\mathbf{z}_{\mathrm{ref}}}\left(\mathbf{z}\right)\bigl|1\bigr\rangle,\tag{5}
\]
is the $n$-th order nonlinear response of the gate with:
\[
\left.\mathrm{d}^{n}f_{0}\right\rfloor _{\mathbf{z}_{\mathrm{ref}}}\left(\mathbf{z}\right)=\sum_{i_{1},\ldots,i_{n}\in\left\{ 0,1\right\} }\psi_{i_{1}}\cdots\psi_{i_{n}}\frac{\partial^{n}f_{0}\bigl(\mathbf{z}_{\mathrm{ref}}\bigr)}{\partial\hat{\mathbf{z}}_{i_{1}}\cdots\partial\hat{\mathbf{z}}_{i_{n}}},\tag{6}
\]
and similar for $\left.\mathrm{d}^{n}f_{1}\right\rfloor _{\mathbf{z}_{\mathrm{ref}}}\left(\mathbf{z}\right)$.
Obviously, this nonlinear mathematical framework should be further
extended in forthcoming contributions by defining diverse classes
of nonlinear anbit gates, encompassing both combinational and sequential
computational architectures. 

\subsection*{Author contributions}

\noindent A. Macho Ortiz originally conceived the idea of APC and
API. A. Macho Ortiz and J. Capmany developed the theory. D. P\'erez
L\'opez, J. Aza\~{n}a and J. Capmany supervised the work. All authors
contributed to write the manuscript.

\newpage{}

\noindent \begin{center}
\newpage{}
\par\end{center}

\noindent \begin{center}
\includegraphics[width=13.8cm,height=20cm,keepaspectratio]{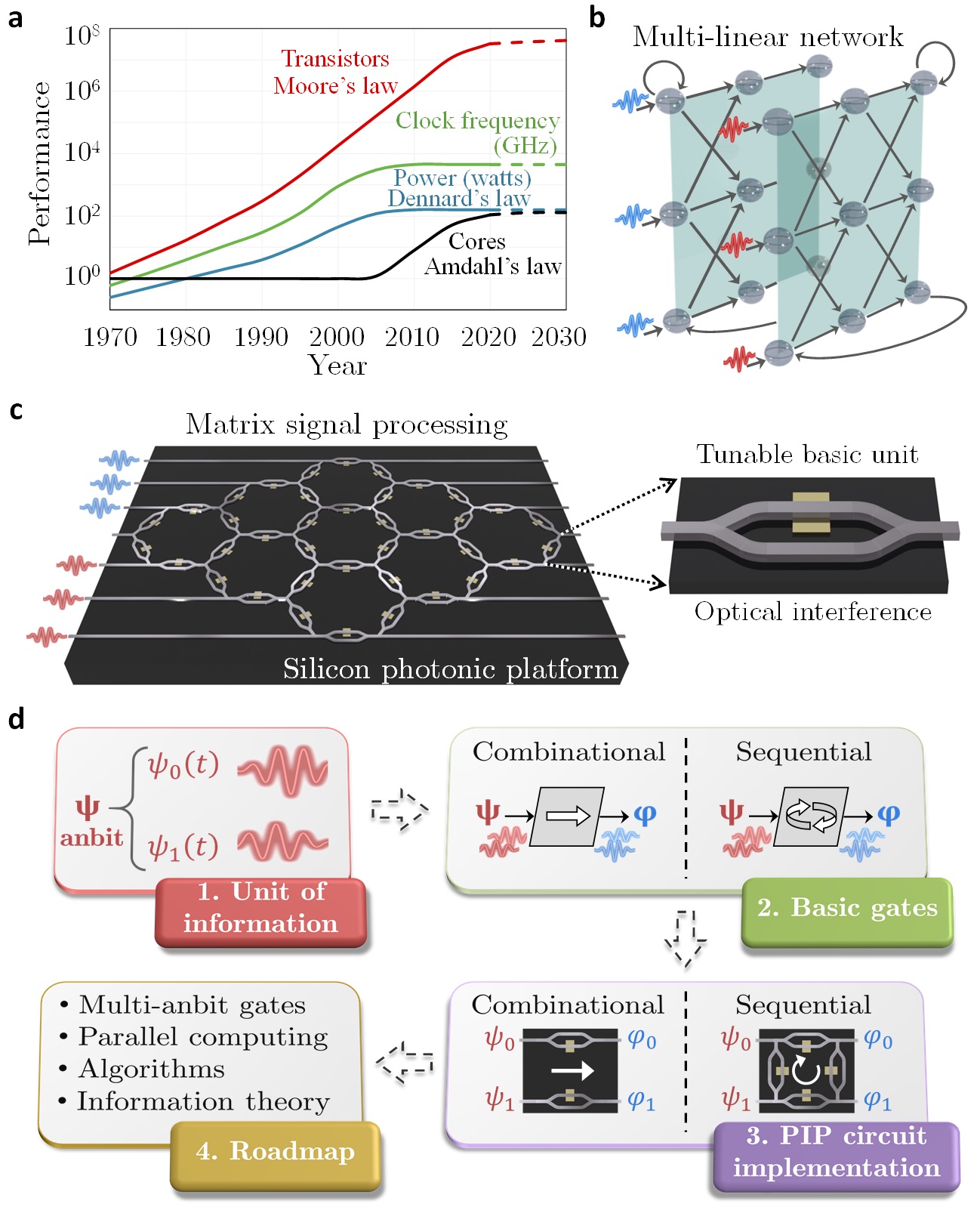}
\par\end{center}

\noindent \textbf{\small{}Fig.\,1 Limitations of digital\,electronics
and new\,computation\,theory proposed in this work.}\linebreak{}
\textbf{\small{}a}{\small{} Historical evolution and perspective of
the main performance parameters of digital electronics \cite{key-P3}.
}\textbf{\small{}b}{\small{} Distributed, parallel and adaptive network
performing multi-linear analog operations via real-time matrix transformations
of the input signals, a scenario where digital electronic paradigm
shows significant mathematical and technological limitations \cite{key-P5,key-P6}.
}\textbf{\small{}c}{\small{} Programmable integrated photonic (PIP)
circuit integrated into a silicon photonic platform. This system-on-chip
technology carries out parallel reconfigurable matrix transformations
on the analog input signals using optical interference as a fundamental
physical principle. }\textbf{\small{}d}{\small{} Flowchart of the
steps required to construct the new computation theory, termed Analog
Programmable-Photonic Computation (APC) and implementable with PIP
technology. APC revolves around the idea of performing analog operations
on a new unit of information, the analog bit (anbit), evolving the
concept of optical signal processing shown in (}\textbf{\small{}c}{\small{})
into true optical computing.}{\small \par}
\noindent \begin{center}
\includegraphics[width=11cm,height=20cm,keepaspectratio]{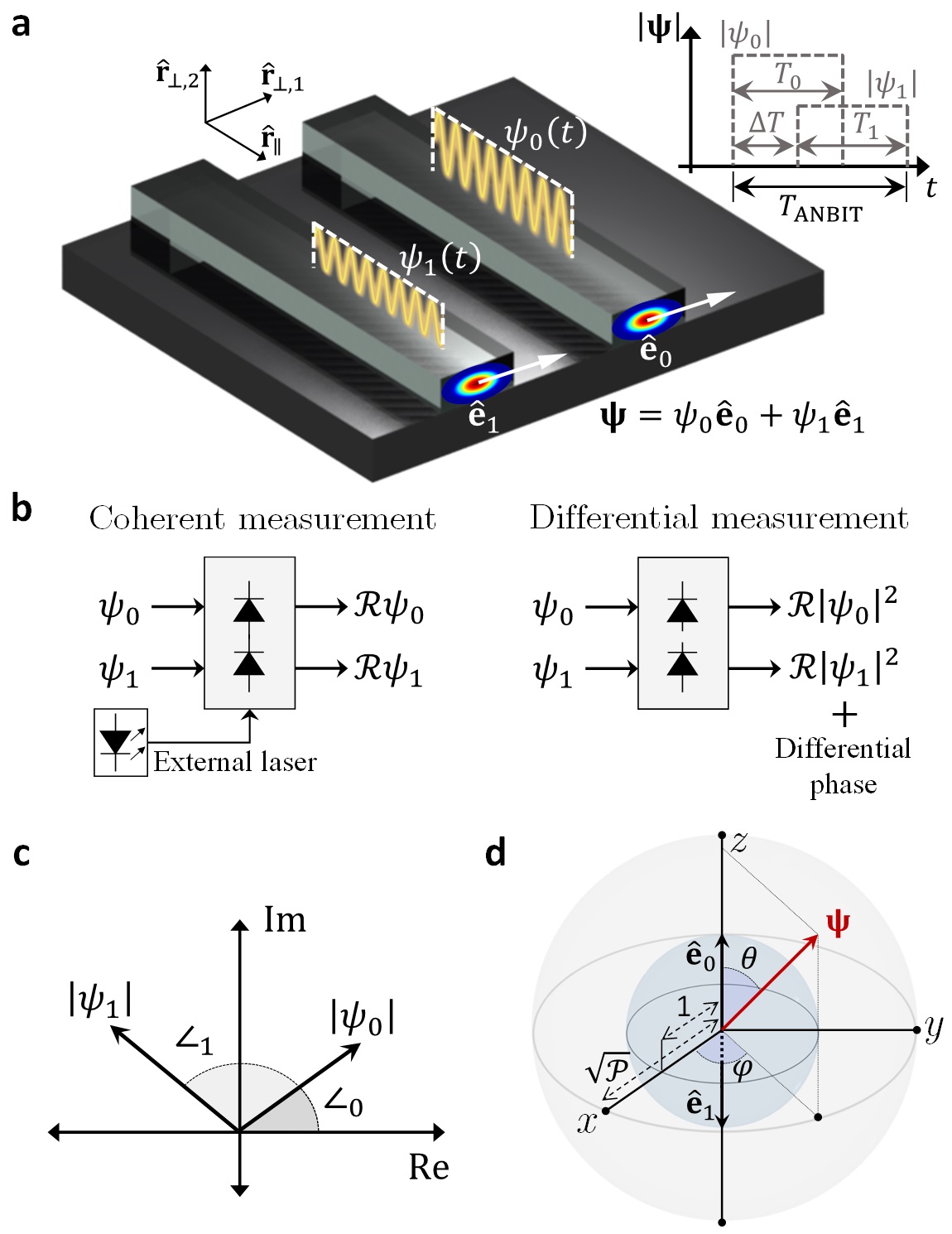}
\par\end{center}

\noindent \textbf{\small{}Fig.\,2 The analog bit. a}{\small{} Physical
implementation of an anbit $\boldsymbol{\uppsi}\left(t\right)=\psi_{0}\left(t\right)\hat{\mathbf{e}}_{0}+\psi_{1}\left(t\right)\hat{\mathbf{e}}_{1}$
using PIP technology and a space-encoding modulation (see Methods).
The anbit amplitudes $\psi_{0,1}=\left|\psi_{0,1}\right|e^{i\angle_{0,1}}$
are encoded by the optical wave packets propagated in the fundamental
modes $\hat{\mathbf{e}}_{0,1}$ of two uncoupled waveguides. }\textbf{\small{}b}{\small{}
Different classes of anbit measurement using coherent or direct detection
at the optical receiver. In the former case, an anbit of the form
$\mathcal{R}\psi_{0}\hat{\mathbf{e}}_{0}+\mathcal{R}\psi_{1}\hat{\mathbf{e}}_{1}$
is measured, where $\mathcal{R}$ is the responsivity of the photodiodes.
In the latter case, an anbit of the form $\mathcal{R}\left|\psi_{0}\right|^{2}\hat{\mathbf{e}}_{0}+\mathcal{R}\left|\psi_{1}\right|^{2}e^{i\left(\angle_{1}-\angle_{0}\right)}\hat{\mathbf{e}}_{1}$
is retrieved (Supplementary Note 1). }\textbf{\small{}c}{\small{}
Geometric representation of an anbit with 4 effective degrees of freedom
(EDFs) using a polar diagram in the complex plane. }\textbf{\small{}d}{\small{}
Geometric representation of an anbit with 3 EDFs in the generalised
Bloch sphere (GBS). In such a situation, the anbit $\boldsymbol{\uppsi}$
can be rewritten as $\boldsymbol{\uppsi}=\sqrt{\mathcal{P}}\bigl[\cos\bigl(\theta/2\bigr)\hat{\mathbf{e}}_{0}+e^{i\varphi}\sin\bigl(\theta/2\bigr)\hat{\mathbf{e}}_{1}\bigr]$,
see Supplementary Note 1.}{\small \par}
\noindent \begin{center}
\includegraphics[width=11.5cm,height=20cm,keepaspectratio]{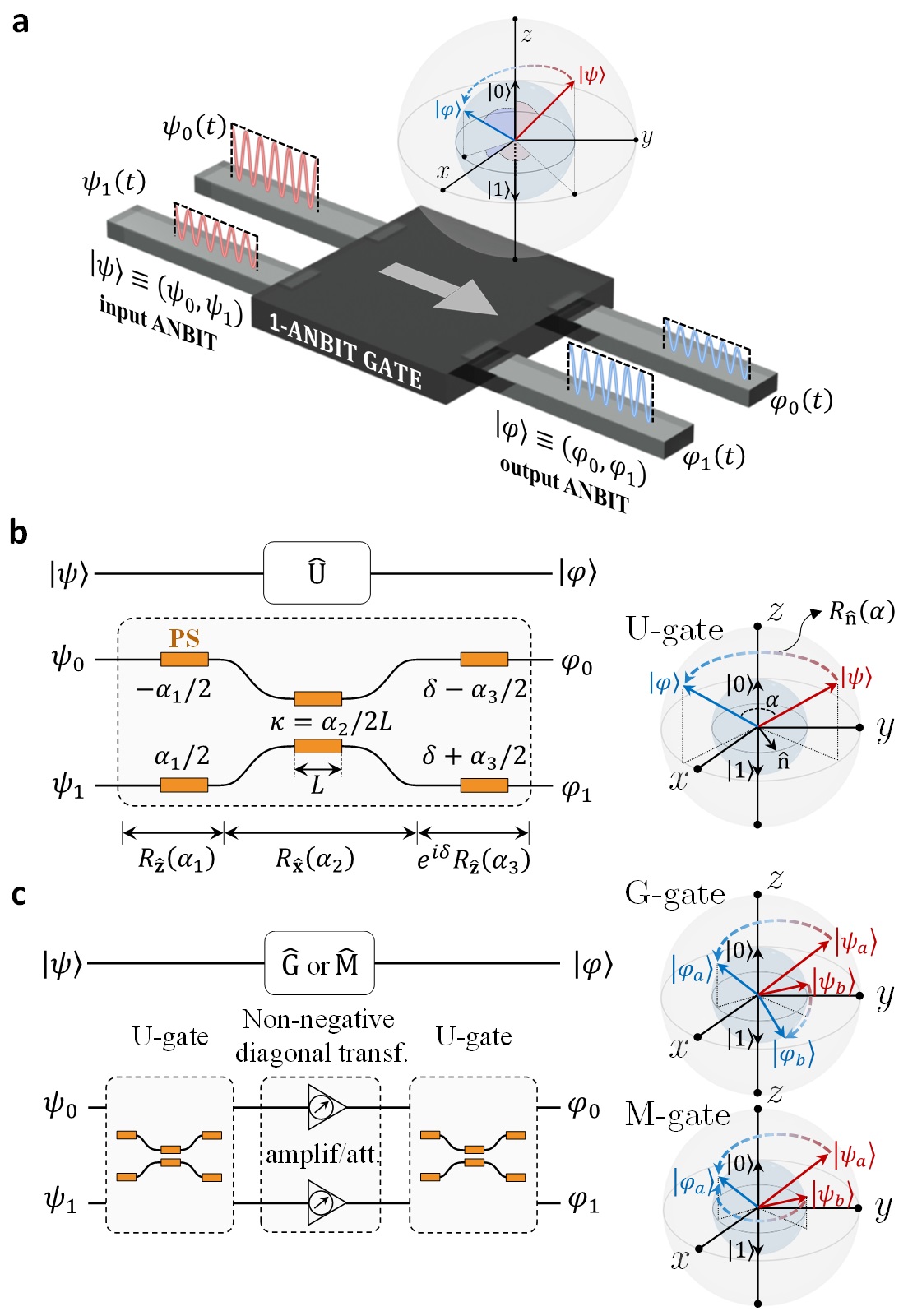}
\par\end{center}

\noindent \textbf{\small{}Fig.\,3 Basic combinational anbit gates.}{\small{}
}\textbf{\small{}a}{\small{} A combinational single-anbit gate is
a non-feedback system performing a transformation between two different
2D vector functions: the input anbit $\bigl|\psi\bigr\rangle=\psi_{0}\left(t\right)\bigl|0\bigr\rangle+\psi_{1}\left(t\right)\bigl|1\bigr\rangle$
and the output anbit $\bigl|\varphi\bigr\rangle=\varphi_{0}\left(t\right)\bigl|0\bigr\rangle+\varphi_{1}\left(t\right)\bigl|1\bigr\rangle$.
Using differential anbit measurement, the gate can be geometrically
represented as a trajectory between two different points in the GBS.
}\textbf{\small{}b}{\small{} Minimal circuit architecture (MCA) of
a U-gate, implementing the universal unitary matrix of Eq.\,(2) via
the Euler factorization $U=e^{i\delta}R_{\hat{\mathbf{n}}}\left(\alpha\right)\equiv e^{i\delta}R_{\hat{\mathbf{z}}}\left(\alpha_{3}\right)R_{\hat{\mathbf{x}}}\left(\alpha_{2}\right)R_{\hat{\mathbf{z}}}\left(\alpha_{1}\right)$.
The U-gate generates a rotation around an arbitrary unit vector $\hat{\mathbf{n}}$
of the GBS, preserving the norm of the input anbit. }\textbf{\small{}c}{\small{}
MCA of a G- and M-gate, based on the singular value decomposition.
While a G-gate is a reversible operation (two different input anbits
$\bigl|\psi_{a}\bigr\rangle$ and $\bigl|\psi_{b}\bigr\rangle$ are
always transformed into two different output anbits $\bigl|\varphi_{a}\bigr\rangle$
and $\bigl|\varphi_{b}\bigr\rangle$), an M-gate may be an irreversible
operation (two different input anbits $\bigl|\psi_{a}\bigr\rangle$
and $\bigl|\psi_{b}\bigr\rangle$ may generate the same output anbit
$\bigl|\varphi_{a}\bigr\rangle$).}{\small \par}
\noindent \begin{center}
\includegraphics[width=12.5cm,height=20cm,keepaspectratio]{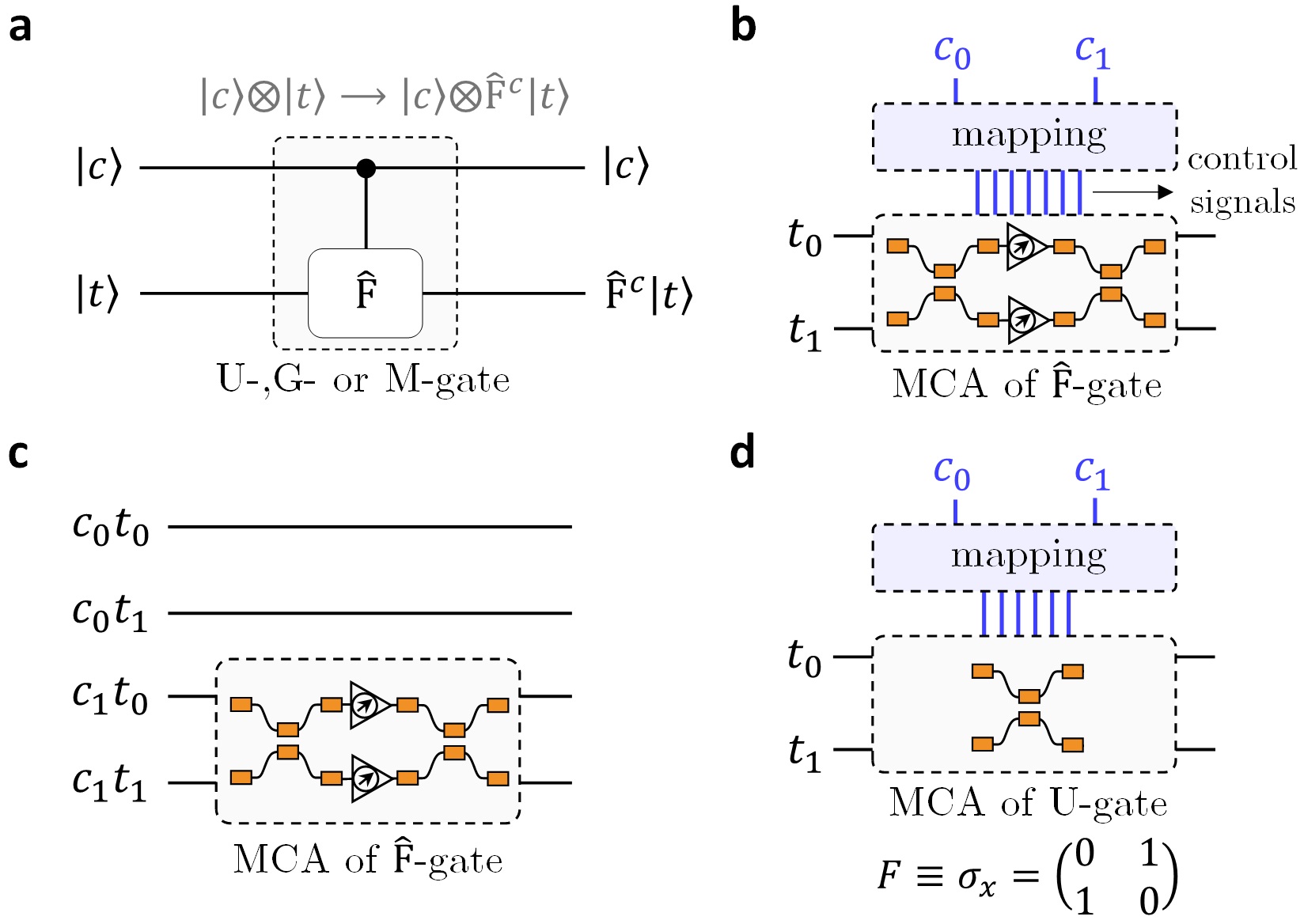}
\par\end{center}

\noindent \textbf{\small{}Fig.\,4 Controlled anbit gates.}{\small{}
}\textbf{\small{}a}{\small{} Functional scheme of a controlled gate
with a single control anbit $\bigl|c\bigr\rangle\in\left\{ \bigl|0\bigr\rangle,\bigl|1\bigr\rangle\right\} $
and a single target anbit $\bigl|t\bigr\rangle$. Inspired in a controlled
quantum gate \cite{key-P10}, the operation $\widehat{\mathrm{F}}$
(a U-, G-, or M-gate) is applied to $\bigl|t\bigr\rangle$ when $\bigl|c\bigr\rangle=\bigl|1\bigr\rangle$
or, otherwise, $\bigl|t\bigr\rangle$ remains invariant at the output.
}\textbf{\small{}b}{\small{} Electro-optic design of a controlled
gate. The PIP circuit must implement the MCA of the $\widehat{\mathrm{F}}$-gate
(here we depict the MCA of an M-gate to cover the general case), whose
basic devices are controlled by electrical signals (blue lines) mapped
with the electrical amplitudes of $\bigl|c\bigr\rangle=c_{0}\bigl|0\bigr\rangle+c_{1}\bigl|1\bigr\rangle$,
e.g., via software \cite{key-P45}. The optical inputs encode the
amplitudes of $\bigl|t\bigr\rangle=t_{0}\bigl|0\bigr\rangle+t_{1}\bigl|1\bigr\rangle$
(black lines). }\textbf{\small{}c}{\small{} All-optical design of
a controlled gate. The optical inputs encode the amplitudes of $\bigl|c\bigr\rangle\otimes\bigl|t\bigr\rangle$,
where $\otimes$ is the tensor product.}\textbf{\small{} d}{\small{}
Electro-optic implementation of the controlled-NOT anbit gate. The
PIP circuit is the MCA of a U-gate since $F=\sigma_{x}$ is a unitary
matrix.}{\small \par}
\noindent \begin{center}
\newpage{}
\par\end{center}

\noindent \begin{center}
\includegraphics[width=15cm,height=20cm,keepaspectratio]{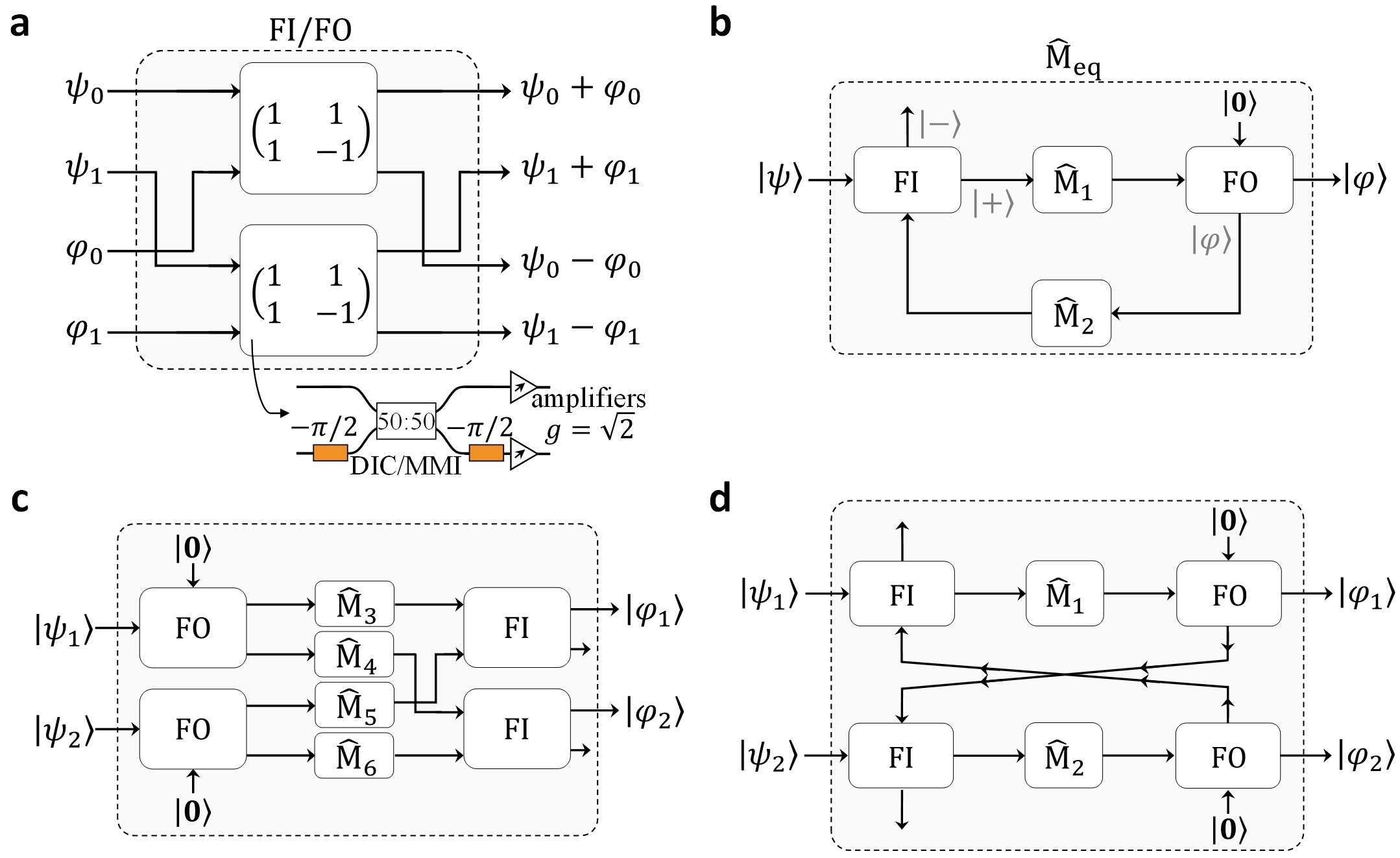}
\par\end{center}

\noindent \textbf{\small{}Fig.\,5 Sequential anbit architectures.
a}{\small{} Optical implementation using PIP technology of both fan-in
(FI) and fan-out (FO) anbit gates. The FI operation maps the input
$\bigl|\psi\bigr\rangle\times\bigl|\varphi\bigr\rangle$ into the
output $\bigl|\psi+\varphi\bigr\rangle\times\bigl|\psi-\varphi\bigr\rangle$,
where $\times$ is the Cartesian product. The FO operation performs
a perfect cloning of $\bigl|\psi\bigr\rangle$ when $\varphi_{0}=\varphi_{1}=0$,
i.e., taking $\bigl|\varphi\bigr\rangle=\bigl|\mathbf{0}\bigr\rangle$,
where $\bigl|\mathbf{0}\bigr\rangle=0\bigl|0\bigr\rangle+0\bigl|1\bigr\rangle$
is the null anbit. }\textbf{\small{}b}{\small{} Sequential}\linebreak{}
{\small{}computational architecture of a single anbit, composed by
both FI and FO gates along with 2 single-anbit M-gates ($\widehat{\mathrm{M}}_{1}$
and $\widehat{\mathrm{M}}_{2}$). }\textbf{\small{}c}{\small{} Multi-anbit
combinational architecture composed by 4 single-anbit M-gates, 2 FI
gates and 2 FO gates. }\textbf{\small{}d}{\small{} Equivalent multi-anbit
sequential architecture, integrating}\linebreak{}
{\small{}2 single-anbit M-gates, 2 FI gates and 2 FO gates.}{\small \par}
\noindent \begin{center}
\newpage{}
\par\end{center}

\begin{center}
{\small{}}%
\begin{tabular}{ccccc}
\toprule 
\textbf{\small{}Properties} & \textbf{\small{}DC} & \textbf{\small{}QC} & \textbf{\small{}NC} & \textbf{\small{}APC}\tabularnewline
\midrule
\midrule 
{\small{}Linear computation} & \textcolor{red}{\small{}$\boldsymbol{\times}$} & \textcolor{blue}{\small{}$\boldsymbol{\checkmark}$} & \textcolor{red}{\small{}$\boldsymbol{\times}$} & \textcolor{blue}{\small{}$\boldsymbol{\checkmark}$}\tabularnewline
\midrule 
{\small{}Nonlinear computation} & \textcolor{blue}{\small{}$\boldsymbol{\checkmark}$} & \textcolor{red}{\small{}$\boldsymbol{\times}$} & \textcolor{blue}{\small{}$\boldsymbol{\checkmark}$} & \textcolor{blue}{\small{}$\boldsymbol{\checkmark}$}\tabularnewline
\midrule 
{\small{}Reversible operations} & \textcolor{red}{\small{}$\boldsymbol{\times}$} & \textcolor{blue}{\small{}$\boldsymbol{\checkmark}$} & \textcolor{red}{\small{}$\boldsymbol{\times}$} & \textcolor{blue}{\small{}$\boldsymbol{\checkmark}$}\tabularnewline
\midrule 
{\small{}Irreversible operations} & \textcolor{blue}{\small{}$\boldsymbol{\checkmark}$} & \textcolor{red}{\small{}$\boldsymbol{\times}$} & \textcolor{blue}{\small{}$\boldsymbol{\checkmark}$} & \textcolor{blue}{\small{}$\boldsymbol{\checkmark}$}\tabularnewline
\midrule 
{\small{}Forward-backward propagation} & \textcolor{red}{\small{}$\boldsymbol{\times}$} & \textcolor{blue}{\small{}$\boldsymbol{\checkmark}$} & \textcolor{blue}{\small{}$\boldsymbol{\checkmark}$} & \textcolor{blue}{\small{}$\boldsymbol{\checkmark}$}\tabularnewline
\midrule 
{\small{}Parallel computing} & \textcolor{blue}{\small{}$\boldsymbol{\checkmark}$} & \textcolor{blue}{\small{}$\boldsymbol{\checkmark}$} & \textcolor{blue}{\small{}$\boldsymbol{\checkmark}$} & \textcolor{blue}{\small{}$\boldsymbol{\checkmark}$}\tabularnewline
\midrule 
{\small{}Summation, cloning and feedback} & \textcolor{blue}{\small{}$\boldsymbol{\checkmark}$} & \textcolor{red}{\small{}$\boldsymbol{\times}$} & \textcolor{blue}{\small{}$\boldsymbol{\checkmark}$} & \textcolor{blue}{\small{}$\boldsymbol{\checkmark}$}\tabularnewline
\midrule 
{\small{}Instantaneous non-locality} & \textcolor{red}{\small{}$\boldsymbol{\times}$} & \textcolor{blue}{\small{}$\boldsymbol{\checkmark}$} & \textcolor{red}{\small{}$\boldsymbol{\times}$} & \textcolor{red}{\small{}$\boldsymbol{\times}$}\tabularnewline
\midrule 
{\small{}Implementation with current technology} & \textcolor{blue}{\small{}$\boldsymbol{\checkmark}$} & \textcolor{red}{\small{}$\boldsymbol{\times}$} & \textcolor{blue}{\small{}$\boldsymbol{\checkmark}$} & \textcolor{blue}{\small{}$\boldsymbol{\checkmark}$}\tabularnewline
\midrule 
{\small{}Operation at room temperature} & \textcolor{blue}{\small{}$\boldsymbol{\checkmark}$} & \textcolor{red}{\small{}$\boldsymbol{\times}$} & \textcolor{blue}{\small{}$\boldsymbol{\checkmark}$} & \textcolor{blue}{\small{}$\boldsymbol{\checkmark}$}\tabularnewline
\midrule 
{\small{}Tolerance to external perturbations} & \textcolor{blue}{\small{}$\boldsymbol{\checkmark}$} & \textcolor{red}{\small{}$\boldsymbol{\times}$} & \textcolor{blue}{\small{}$\boldsymbol{\checkmark}$} & \textcolor{blue}{\small{}$\boldsymbol{\checkmark}$}\tabularnewline
\bottomrule
\end{tabular}
\par\end{center}{\small \par}

\noindent \textbf{\small{}Table 1.}{\small{} Qualitative comparison
of APC }\emph{\small{}vs}{\small{} digital computation (DC), quantum
computation (QC) and neuromorphic computation (NC).}{\small \par}
\noindent \begin{center}
\newpage{}
\par\end{center}

\noindent \begin{center}
\textbf{\LARGE{}Supplementary Information}
\par\end{center}{\LARGE \par}

\vspace{1cm}

\tableofcontents{}

\newpage{}

\section*{Supplementary Note 1: the analog bit\label{sec:1}}

\addcontentsline{toc}{section}{Supplementary Note 1: the analog bit}

\noindent In this section, we provide further information about the
analog bit (or anbit for short) including supplementary notes about:
the anbit modulation formats, the anbit measurement, the generalised
Bloch sphere (GBS), the anbit amplitudes, and the mathematical properties
of the tensor and Cartesian products within the context of Analog
Programmable-Photonic Computation (APC).

\subsection*{1.1 Anbit modulation formats}

\addcontentsline{toc}{subsection}{1.1 Anbit modulation formats}

\noindent The optical implementation of an anbit $\boldsymbol{\uppsi}\left(t\right)=\psi_{0}\left(t\right)\hat{\mathbf{e}}_{0}+\psi_{1}\left(t\right)\hat{\mathbf{e}}_{1}$
can be carried out by exploiting the different degrees of freedom
of light (space, mode, polarisation, frequency and time), unveiling
a gamut of \emph{anbit modulation formats}.

\paragraph*{Space-encoding modulation (SEM).}

This modulation technique (described in Fig.\,2a and Methods of the
main text) is the simplest strategy to implement an anbit in programmable
integrated photonics (PIP) using current technology.

\paragraph*{Mode-encoding modulation (MEM).}

Instead of implementing $\boldsymbol{\uppsi}\left(t\right)$ by encoding
$\hat{\mathbf{e}}_{0}$ and $\hat{\mathbf{e}}_{1}$ in the fundamental
modes of two different single-mode waveguides, we can associate $\hat{\mathbf{e}}_{0}$
and $\hat{\mathbf{e}}_{1}$ to two different guided modes of a single
waveguide operating in the multi-mode regime (Supplementary Figure
1a). Therefore, a mode-encoded anbit is characterized by an electric
field strength of the form (paraxial regime is assumed):
\begin{equation}
\boldsymbol{\mathcal{E}}\left(\mathbf{r},t\right)\simeq\sum_{k=0}^{1}\textrm{Re}\left\{ \psi_{k}\left(t-\beta_{k}^{\left(1\right)}r_{\parallel}\right)\hat{\mathbf{e}}_{k}\left(r_{\perp,1},r_{\perp,2},\omega_{\mathrm{c}}\right)e^{i\omega_{\mathrm{c}}t}e^{-i\beta_{k}^{\left(0\right)}r_{\parallel}}\right\} ,\label{eq:1.1.1}
\end{equation}
where the anbit amplitudes $\psi_{0}$ and $\psi_{1}$ are encoded
by the optical wave packets (or complex envelopes) propagated by the
guided modes, $\omega_{\mathrm{c}}$ is the angular frequency of the
optical carrier, $\mathbf{r}=r_{\perp,1}\hat{\mathbf{r}}_{\perp,1}+r_{\perp,2}\hat{\mathbf{r}}_{\perp,2}+r_{\parallel}\hat{\mathbf{r}}_{\parallel}$
is the vector position written as a function of its transverse ($r_{\perp,1}$,
$r_{\perp,2}$) and longitudinal ($r_{\parallel}$) components, and
$\hat{\mathbf{e}}_{k}$ and $\beta_{k}\left(\omega\right)\simeq\beta_{k}^{\left(0\right)}+\left(\omega-\omega_{\mathrm{c}}\right)\beta_{k}^{\left(1\right)}$
are respectively the normalised mode profile and the propagation constant
of the $k$-th guided mode (being $\beta_{k}^{\left(0\right)}=\beta_{k}\left(\omega=\omega_{\mathrm{c}}\right)$,
$\beta_{k}^{\left(1\right)}=\mathrm{d}\beta_{k}\left(\omega=\omega_{\mathrm{c}}\right)/\mathrm{d}\omega$
and omitting the dispersive terms $\beta_{k}^{\left(n\geq2\right)}$
in the Taylor series expansion of $\beta_{k}\left(\omega\right)$).
This modulation format requires to use mode-division multiplexing
devices and circuits to generate, transform and measure the anbits.

\paragraph*{Polarisation-encoding modulation (PEM).}

While the above modulation formats only make use of a single transverse
component of the electric field (in the $\hat{\mathbf{r}}_{\perp,1}$
direction), a polarisation-encoded anbit exploits the two transverse
components $\hat{\mathbf{e}}_{0}$ and $\hat{\mathbf{e}}_{1}$ of
the fundamental mode propagated by a single-mode waveguide (Supplementary
Figure 1b). In this vein, the electric field strength propagating
a polarisation-encoded anbit is found to be (paraxial conditions):
\begin{equation}
\boldsymbol{\mathcal{E}}\left(\mathbf{r},t\right)\simeq\sum_{k=0}^{1}\textrm{Re}\left\{ \psi_{k}\left(t-\beta^{\left(1\right)}r_{\parallel}\right)\hat{\mathbf{e}}_{k}\left(r_{\perp,1},r_{\perp,2},\omega_{\mathrm{c}}\right)e^{i\omega_{\mathrm{c}}t}e^{-i\beta^{\left(0\right)}r_{\parallel}}\right\} ,\label{eq:1.1.2}
\end{equation}
with the same propagation constant $\beta\left(\omega\right)\simeq\beta^{\left(0\right)}+\left(\omega-\omega_{\mathrm{c}}\right)\beta^{\left(1\right)}$
for both transverse components of the fundamental mode considering
a lowly-birefringent waveguide \cite{key-S1}. This modulation format
entails the utilisation of polarisation devices and systems to generate,
transform and measure the anbits.

\paragraph*{Frequency-encoding modulation (FEM).}

The frequency domain can also be employed to implement $\boldsymbol{\uppsi}\left(t\right)$
by encoding $\hat{\mathbf{e}}_{0}$ and $\hat{\mathbf{e}}_{1}$ in
the same fundamental mode of a given single-mode waveguide, but stimulated
at two different angular frequencies $\omega_{0}$ and $\omega_{1}$.
The spectral separation between both optical carriers must avoid the
frequency overlapping between the Fourier transforms of $\psi_{0}\left(t\right)$
and $\psi_{1}\left(t\right)$, namely $\widetilde{\psi}_{0}\left(\omega\right)$
and $\widetilde{\psi}_{1}\left(\omega\right)$ (Supplementary Figure
1c). Specifically, the electric field strength associated to a frequency-encoded
anbit is (paraxial regime):
\begin{equation}
\boldsymbol{\mathcal{E}}\left(\mathbf{r},t\right)\simeq\sum_{k=0}^{1}\textrm{Re}\left\{ \psi_{k}\left(t-\beta_{k}^{\left(1\right)}r_{\parallel}\right)\hat{\mathbf{e}}\left(r_{\perp,1},r_{\perp,2},\omega_{k}\right)e^{i\omega_{k}t}e^{-i\beta_{k}^{\left(0\right)}r_{\parallel}}\right\} ,\label{eq:1.1.3}
\end{equation}
with $\beta_{k}^{\left(0\right)}=\beta\left(\omega=\omega_{k}\right)$
and $\beta_{k}^{\left(1\right)}=\mathrm{d}\beta\left(\omega=\omega_{k}\right)/\mathrm{d}\omega$,
being $\beta\left(\omega\right)$ the propagation constant of the
fundamental mode. Note that the vector basis $\mathcal{B}_{1}=\left\{ \hat{\mathbf{e}}_{0},\hat{\mathbf{e}}_{1}\right\} $
of the anbit is implemented with the normalised profile $\hat{\mathbf{e}}$
of the fundamental mode stimulated at the angular frequencies $\omega_{0}$
and $\omega_{1}$, respectively. The Fourier transform of $\boldsymbol{\mathcal{E}}$
takes the form:
\begin{equation}
\widetilde{\boldsymbol{\mathcal{E}}}\left(\mathbf{r},\omega\right)\simeq\sum_{k=0}^{1}\textrm{Re}\left\{ \widetilde{\psi}_{k}\left(\omega-\omega_{k}\right)\hat{\mathbf{e}}\left(r_{\perp,1},r_{\perp,2},\omega_{k}\right)e^{-i\beta_{k}^{\left(0\right)}r_{\parallel}}e^{-i\left(\omega-\omega_{k}\right)\beta_{k}^{\left(1\right)}r_{\parallel}}\right\} .\label{eq:1.1.4}
\end{equation}
Here, wavelength-division multiplexing devices and circuits will be
required to generate, transform and measure the anbits.

\paragraph*{Time-encoding modulation (TEM).}

The time domain is an additional degree of freedom of light that can
also be explored to implement an anbit in optics. The basic idea is
to generate a complex envelope of the form $\mathcal{A}\left(t\right)=\psi_{0}\left(t-t_{0}\right)+\psi_{1}\left(t-t_{1}\right)$
that will be propagated by the fundamental mode of a single-mode waveguide.
In particular, the time delay $\Delta\tau=\left|t_{1}-t_{0}\right|$
must avoid the temporal overlapping between $\psi_{0}$ and $\psi_{1}$
(Supplementary Figure 1d). Here, the vector basis $\mathcal{B}_{1}=\left\{ \hat{\mathbf{e}}_{0},\hat{\mathbf{e}}_{1}\right\} $
of the anbit is implemented with the normalised profile $\hat{\mathbf{e}}$
of the fundamental mode stimulated at $t=t_{0}$ and $t=t_{1}$, respectively.
For sufficient short propagation distances where the chromatic dispersion
of the waveguide can be neglected, the electric field strength of
a time-encoded anbit can be approximated as (paraxial regime):
\begin{align}
\boldsymbol{\mathcal{E}}\left(\mathbf{r},t\right) & \simeq\textrm{Re}\left\{ \mathcal{A}\left(t-\beta^{\left(1\right)}r_{\parallel}\right)\hat{\mathbf{e}}\left(r_{\perp,1},r_{\perp,2},\omega_{\mathrm{c}}\right)e^{i\omega_{\mathrm{c}}t}e^{-i\beta^{\left(0\right)}r_{\parallel}}\right\} \nonumber \\
 & =\sum_{k=0}^{1}\textrm{Re}\left\{ \psi_{k}\left(t-t_{k}-\beta^{\left(1\right)}r_{\parallel}\right)\hat{\mathbf{e}}\left(r_{\perp,1},r_{\perp,2},\omega_{\mathrm{c}}\right)e^{i\omega_{\mathrm{c}}t}e^{-i\beta^{\left(0\right)}r_{\parallel}}\right\} .\label{eq:1.1.5}
\end{align}
However, for sufficient propagation distances, the chromatic dispersion
will generate fluctuations in $\Delta\tau$ that could give rise to
an undesirable temporal overlapping between $\psi_{0}$ and $\psi_{1}$.
Consequently, this propagation impairment is the main drawback of
this modulation format, which would require an external synchronization
system to control the fluctuations of $\Delta\tau$ within the whole
computational system.
\noindent \begin{center}
\includegraphics[width=15cm,height=10.2cm,keepaspectratio]{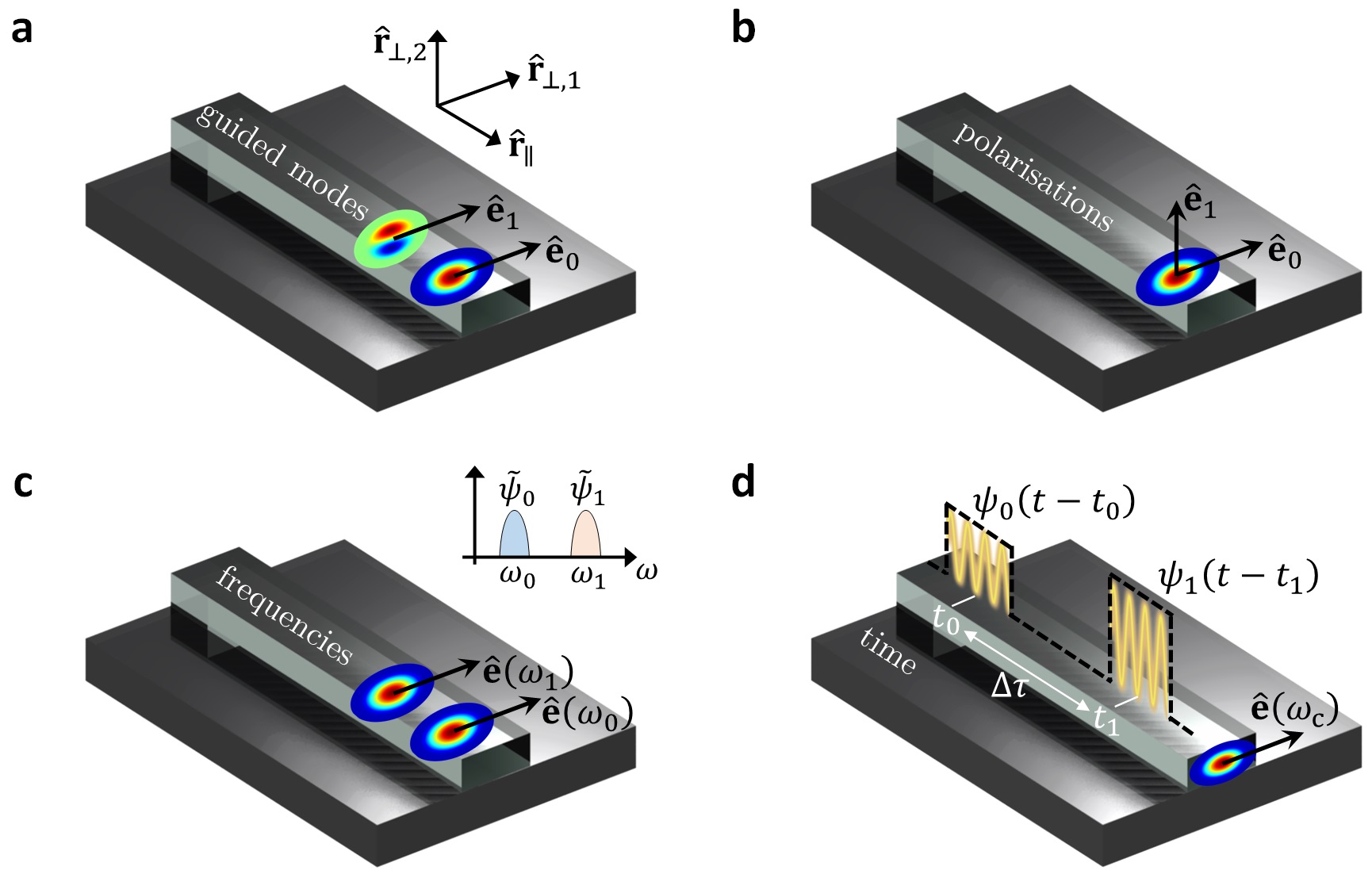}
\par\end{center}

\noindent \textbf{\small{}Supplementary Figure 1.}{\small{} Anbit
modulation formats.}\textbf{\small{} a}{\small{} Mode-encoding modulation.
}\textbf{\small{}b}{\small{} Polarisation-encoding modulation. }\textbf{\small{}c}{\small{}
Frequency-encoding modulation. }\textbf{\small{}d}{\small{} Time-encoding
modulation. Each modulation format exploits a different degree of
freedom of light to encode an anbit.\label{Fig.S1}}{\small \par}

\paragraph*{Pulse shape.}

In a similar way to digital information theory \cite{key-S2,key-S3},
we can engineer the temporal shape of the complex envelopes (encoding
$\psi_{0}\left(t\right)$ and $\psi_{1}\left(t\right)$) to improve
the tolerance of the above anbit modulation formats to the different
propagation impairments within a PIP circuit (optical filtering, optical
crosstalk, chromatic dispersion, polarisation-mode dispersion, and
nonlinear Kerr effects, among others \cite{key-S4}). In particular,
using rectangular or quasi-rectangular shapes in $\psi_{0}\left(t\right)$
and $\psi_{1}\left(t\right)$ (e.g., via super-Gaussian pulses), we
may explore the introduction of a duty cycle (the ratio of pulse width
of $\left|\psi_{0}\left(t\right)\right|$ and $\left|\psi_{1}\left(t\right)\right|$
over the total time intervals $T_{0}$ and $T_{1}$ where $\psi_{0}\left(t\right)$
and $\psi_{1}\left(t\right)$ are respectively defined) to study the
pulse-to-pulse interaction between different anbits. In such a scenario,
we should compare non-return-to-zero (NRZ) versus return-to-zero (RZ)
pulse shapes (Supplementary Figure 2). This will lead to different
versions of each modulation format (e.g., NRZ-SEM, 33\%RZ-SEM, 50\%RZ-SEM...).
Likewise, arbitrary shapes of $\psi_{0}\left(t\right)$ and $\psi_{1}\left(t\right)$
(such as prolate spheroidal wave functions \cite{key-S5}) can also
be investigated in each modulation format to improve the tolerance
of the anbit to the aforementioned propagation impairments, although
it is out of the scope of this work.

\paragraph*{Other modulation formats.}

The implementation of APC in other technologies different from PIP
(e.g., in electronics or acoustics) will require to develop new modulation
formats that allow us to physically generate the anbit in the corresponding
hardware platforms.
\noindent \begin{center}
\includegraphics[width=15cm,height=7cm,keepaspectratio]{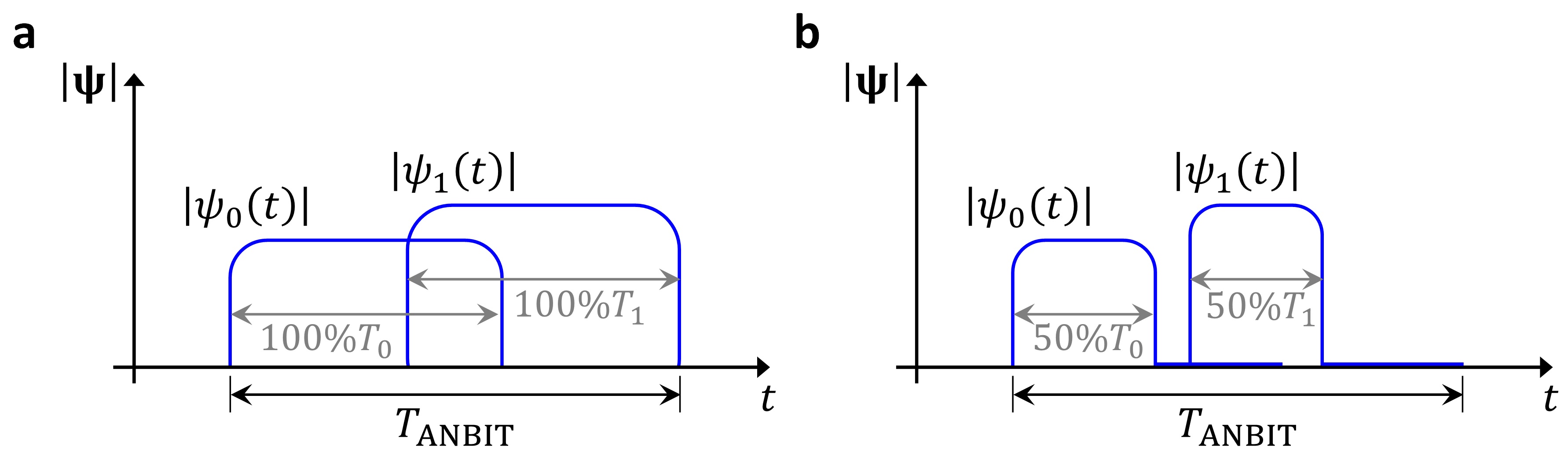}
\par\end{center}

\noindent \textbf{\small{}Supplementary Figure 2.}{\small{} Different
temporal shapes of rectangular (or quasi-rectangular) anbit amplitudes.
}\textbf{\small{}a}{\small{} Non-return-to-zero (NRZ), where the pulse
width covers the whole duration of $\left|\psi_{0}\left(t\right)\right|$
and $\left|\psi_{1}\left(t\right)\right|$. }\textbf{\small{}b}{\small{}
Return-to-zero (RZ), where the pulse width encompasses only a percentage
(50\% in this figure) of the time intervals $T_{0}$ and $T_{1}$
over which $\left|\psi_{0}\left(t\right)\right|$ and $\left|\psi_{1}\left(t\right)\right|$
are respectively defined.\label{Fig.S2}}{\small \par}

\subsection*{1.2 Anbit measurement\label{subsec:1.2-Anbit-measurement}}

\addcontentsline{toc}{subsection}{1.2 Anbit measurement}

\noindent Each modulation format requires a different optical architecture
to perform a coherent or a differential anbit measurement. Here, we
only discuss the architectures associated to both classes of anbit
measurement within the context of a SEM, the simplest modulation format
to implement APC using current PIP technology. 

In general, any anbit measurement can be mathematically described
as an anbit mapping $\boldsymbol{\uppsi}\rightarrow\boldsymbol{\upvarphi}$,
which transforms the input (optical) anbit: 
\begin{equation}
\boldsymbol{\uppsi}=\psi_{0}\hat{\mathbf{e}}_{0}+\psi_{1}\hat{\mathbf{e}}_{1}=\left|\psi_{0}\right|e^{i\angle_{0}}\hat{\mathbf{e}}_{0}+\left|\psi_{1}\right|e^{i\angle_{1}}\hat{\mathbf{e}}_{1},\label{eq:1.2.1}
\end{equation}
into an output (electrical) anbit $\boldsymbol{\upvarphi}=\varphi_{0}\hat{\mathbf{e}}_{0}+\varphi_{1}\hat{\mathbf{e}}_{1}$.
The main difference between a coherent and a differential measurement
is that each kind of measurement establishes a different mapping between
$\boldsymbol{\uppsi}$ and $\boldsymbol{\upvarphi}$, as detailed
below. In this context, for the sake of simplicity, we will assume
that any (coherent or differential) anbit measurement will be carried
out at the receiver by using the same orthonormal vector basis $\mathcal{B}_{1}=\left\{ \hat{\mathbf{e}}_{0},\hat{\mathbf{e}}_{1}\right\} $
as that of the transmitter.

\paragraph*{Coherent anbit measurement.}

Here, the 4 real degrees of freedom of $\boldsymbol{\uppsi}$ (the
moduli and phases of $\psi_{0,1}$) can be completely recovered, leading
to 4 \emph{effective degrees of freedom} (EDFs) where the user information
can be encoded and retrieved. Using a coherent optical receiver, e.g.,
combining two quadrature homodyne receivers (Supplementary Figure
3), the phase and quadrature photocurrents obtained are (see pp.\,66-68
of ref.\,\cite{key-S6}):
\begin{align}
I_{\mathrm{I},k} & =\mathcal{R}\left|\psi_{k}\right|\cos\angle_{k},\label{eq:1.2.2}\\
I_{\mathrm{Q},k} & =\mathcal{R}\left|\psi_{k}\right|\sin\angle_{k},\label{eq:1.2.3}
\end{align}
where $k\in\left\{ 0,1\right\} $ and $\mathcal{R}$ is the responsivity
of the photodiodes. Next, applying a simple signal processing routine
to the above photocurrents using the expressions:
\begin{align}
\left|\psi_{k}\right| & =\frac{1}{\mathcal{R}}\sqrt{I_{\mathrm{I},k}^{2}+I_{\mathrm{Q},k}^{2}},\ \ \ \ \angle_{k}=\arctan\frac{I_{\mathrm{Q},k}}{I_{\mathrm{I},k}},\label{eq:1.2.4}
\end{align}
we will be able to retrieve the moduli and phases of $\psi_{0,1}$.
Consequently, we can safely assume that the output (electrical) anbit
is of the form $\boldsymbol{\upvarphi}=\mathcal{R}\left|\psi_{0}\right|e^{i\angle_{0}}\hat{\mathbf{e}}_{0}+\mathcal{R}\left|\psi_{1}\right|e^{i\angle_{1}}\hat{\mathbf{e}}_{1}\equiv\mathcal{R}\boldsymbol{\uppsi}$.
\noindent \begin{center}
\includegraphics[width=14cm,height=10cm,keepaspectratio]{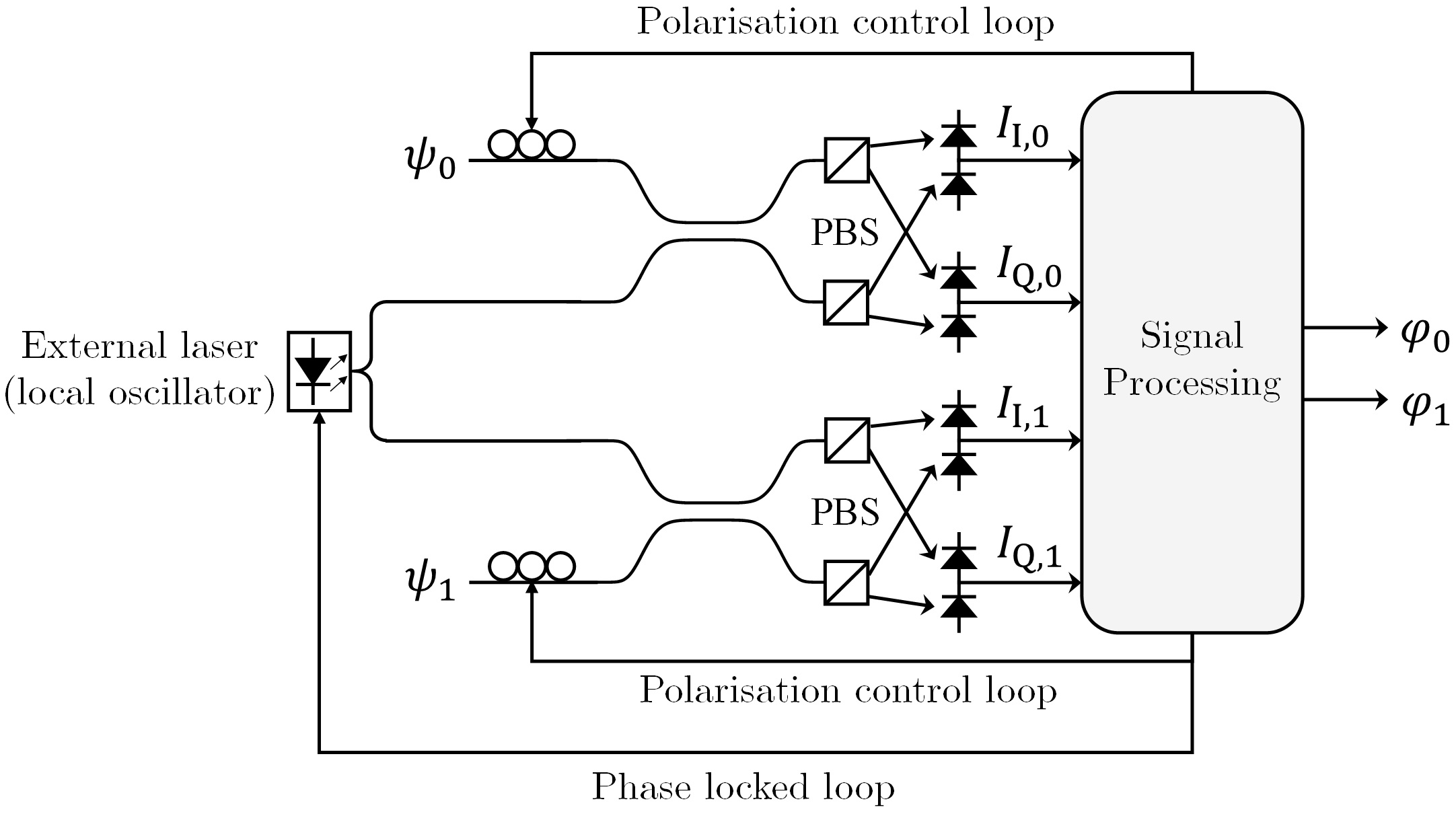}
\par\end{center}

\noindent \textbf{\small{}Supplementary Figure 3.}{\small{} Structure
of the optical receiver proposed to perform a coherent anbit measurement.
The system is composed by two quadrature homodyne receivers (PBS:
polarisation beam splitter) \cite{key-S6}.}{\small \par}

\paragraph*{Differential anbit measurement.}

Using two different photodiodes with the same responsivity $\mathcal{R}$
to measure $\psi_{0}$ and $\psi_{1}$ via direct detection (see Fig.\,2b),
we will obtain two independent photocurrents: $I_{0}=\mathcal{R}\left|\psi_{0}\right|^{2}$
and $I_{1}=\mathcal{R}\left|\psi_{1}\right|^{2}$. The time delay
between both photocurrents will be the same as the time delay $\Delta T$
between $\psi_{0}\left(t\right)$ and $\psi_{1}\left(t\right)$ depicted
in Fig.\,2a. As mentioned in the main text, $\Delta T$ provides
information about the differential phase between $\psi_{0}$ and $\psi_{1}$
via the relation $\angle_{1}-\angle_{0}=\omega_{\mathrm{c}}\Delta T$,
where $\omega_{\mathrm{c}}$ is the angular frequency of the optical
carrier. Consequently, the output anbit is found to be $\boldsymbol{\upvarphi}=\mathcal{R}\left|\psi_{0}\right|^{2}\hat{\mathbf{e}}_{0}+\mathcal{R}\left|\psi_{1}\right|^{2}e^{i\left(\angle_{1}-\angle_{0}\right)}\hat{\mathbf{e}}_{1}$. 

As seen, this anbit measurement technique only provides 3 EDFs where
the user information can be encoded and retrieved: $\left|\psi_{0}\right|$,
$\left|\psi_{1}\right|$ and $\angle_{1}-\angle_{0}$. Here, the phase
$\angle_{0}$ cannot be recovered. Consequently, this (global) phase
term can be omitted to describe $\boldsymbol{\uppsi}$, which may
be restated as:
\begin{equation}
\boldsymbol{\uppsi}=\left|\psi_{0}\right|\hat{\mathbf{e}}_{0}+\left|\psi_{1}\right|e^{i\left(\angle_{1}-\angle_{0}\right)}\hat{\mathbf{e}}_{1}.\label{eq:1.2.5}
\end{equation}
Note that Supplementary Equations \ref{eq:1.2.1} and \ref{eq:1.2.5}
generate the same output anbit $\boldsymbol{\upvarphi}$ in a differential
measurement. For this reason, both expressions are considered to be
equivalent to describe the input anbit. Alternatively, the above equation
can be rewritten as:
\begin{equation}
\boldsymbol{\uppsi}=\sqrt{\mathcal{P}}\left(\cos\frac{\theta}{2}\hat{\mathbf{e}}_{0}+e^{i\varphi}\sin\frac{\theta}{2}\hat{\mathbf{e}}_{1}\right),\label{eq:1.2.6}
\end{equation}
by identifying $\sqrt{\mathcal{P}}\cos\left(\theta/2\right)\equiv\left|\psi_{0}\right|$,
$\sqrt{\mathcal{P}}\sin\left(\theta/2\right)\equiv\left|\psi_{1}\right|$,
$\varphi\equiv\angle_{1}-\angle_{0}$ and $\mathcal{P}\equiv\left|\psi_{0}\right|^{2}+\left|\psi_{1}\right|^{2}$.
In particular, Supplementary Equation \ref{eq:1.2.6} will allow us
to geometrically represent an anbit with 3 EDFs in the generalised
Bloch sphere (see below). 

On the other hand, a particular scenario associated to a differential
measurement should be mentioned. The case where the differential phase
$\angle_{1}-\angle_{0}$ is not encoded by a time delay $\Delta T$
between $\psi_{0}$ and $\psi_{1}$ at the transmitter. This can be
done, for instance, modulating $\psi_{0}$ and $\psi_{1}$ with two
different optical carriers having the same angular frequency $\omega_{\mathrm{c}}$
but exhibiting a different phase $\angle_{0}\neq\angle_{1}$ at $t=0$.
In such a situation, the differential phase $\angle_{1}-\angle_{0}$
cannot be recovered with a differential anbit measurement because
the time delay between the photocurrents is found to be null. Here,
a differential measurement will only recover the information encoded
by $\left|\psi_{0}\right|$ and $\left|\psi_{1}\right|$ (2 EDFs).
This remark highlights the importance of the parameter $\Delta T$
depicted in Fig.\,2a of the paper.

\subsection*{1.3 Geometric representation: the generalised Bloch Sphere (GBS)}

\addcontentsline{toc}{subsection}{1.3 Geometric representation: the generalised Bloch Sphere (GBS)}

\noindent The construction of the GBS is actually more complex than
a mere analogy of Supplementary Equation \ref{eq:1.2.6} with the
spherical coordinates $\bigl(\sqrt{\mathcal{P}},\theta,\varphi\bigr)$
in the vector space $\mathbb{R}^{3}$. Specifically, the construction
of the GBS can be regarded as a geometrical transformation of the
hypersphere $S^{2}\bigl(\sqrt{\mathcal{P}}\bigr)\subset\mathbb{R}^{3}$
into the hypersphere $S^{3}\bigl(\sqrt{\mathcal{P}}\bigr)\subset\mathbb{C}^{2}$
\cite{key-S7,key-S8}. Supplementary Figure 4 illustrates such a transformation,
summarised in the following steps:
\begin{enumerate}
\item Let us start from an anbit with 3 EDFs described by Supplementary
Equation \ref{eq:1.2.5}. Here, we can identify the 3-tuple $\bigl(\left|\psi_{1}\right|\cos\left(\angle_{1}-\angle_{0}\right),\left|\psi_{1}\right|\sin\left(\angle_{1}-\angle_{0}\right),\left|\psi_{0}\right|\bigr)$
with the Cartesian coordinates $\left(x,y,z\right)$ of the hypersphere
$S^{2}\bigl(\sqrt{\mathcal{P}}\bigr)\subset\mathbb{R}^{3}$ with $z>0$
and radius: 
\begin{equation}
\sqrt{x^{2}+y^{2}+z^{2}}=\sqrt{\left|\psi_{0}\right|^{2}+\left|\psi_{1}\right|^{2}}=\sqrt{\mathcal{P}},\label{eq:1.3.1}
\end{equation}
see Supplementary Figure 4a. Hence, Supplementary Equation \ref{eq:1.2.5}
can be rewritten as:
\begin{equation}
\boldsymbol{\uppsi}=z\hat{\mathbf{e}}_{0}+\left(x+iy\right)\hat{\mathbf{e}}_{1}.\label{eq:1.3.2}
\end{equation}
Here, note that $\hat{\mathbf{e}}_{0}\equiv\hat{\mathbf{z}}$ and
$\hat{\mathbf{e}}_{1}\equiv\hat{\mathbf{x}}$. Moving from Cartesian
to spherical coordinates $\bigl(\sqrt{\mathcal{P}},\theta,\varphi\bigr)$
via the transformations: 
\begin{align}
x & =\sqrt{\mathcal{P}}\sin\theta\cos\varphi,\label{eq:1.3.3}\\
y & =\sqrt{\mathcal{P}}\sin\theta\sin\varphi,\label{eq:1.3.4}\\
z & =\sqrt{\mathcal{P}}\cos\theta,\label{eq:1.3.5}
\end{align}
the anbit can equivalently be expressed as:
\begin{equation}
\boldsymbol{\uppsi}=\sqrt{\mathcal{P}}\left(\cos\theta\hat{\mathbf{e}}_{0}+e^{i\varphi}\sin\theta\hat{\mathbf{e}}_{1}\right),\label{eq:1.3.6}
\end{equation}
with $\theta\in\left[0,\pi/2\right]$ and $\varphi\in\left[0,2\pi\right)$.
\item In order to extend the geometrical representation of $\boldsymbol{\uppsi}$
to the whole sphere (i.e. including the points with $z<0$), we should
perform the coordinate transformation $\theta^{\prime}:=2\theta$,
which maps the Cartesian coordinates $\left(x,y,z\right)$ into the
coordinates $\left(x^{\prime},y^{\prime},z^{\prime}\right)$ depicted
in Supplementary Figure 4b. Hence, Supplementary Equation \ref{eq:1.3.6}
becomes:
\begin{equation}
\boldsymbol{\uppsi}=\sqrt{\mathcal{P}}\left(\cos\frac{\theta}{2}^{\prime}\hat{\mathbf{e}}_{0}+e^{i\varphi}\sin\frac{\theta}{2}^{\prime}\hat{\mathbf{e}}_{1}\right),\label{eq:1.3.7}
\end{equation}
with $\theta^{\prime}\in\left[0,\pi\right]$ and $\varphi\in\left[0,2\pi\right)$.
However, note that a coordinate transformation cannot perform a geometrical
transformation on its own.
\item In order to transform $S^{2}\bigl(\sqrt{\mathcal{P}}\bigr)\subset\mathbb{R}^{3}$
into $S^{3}\bigl(\sqrt{\mathcal{P}}\bigr)\subset\mathbb{C}^{2}$,
we must carry out the relabelling $\theta^{\prime}\rightarrow\theta$,
which is equivalent to reinterpret the coordinates $\left(x^{\prime},y^{\prime},z^{\prime}\right)$
as the original Cartesian coordinates $\left(x,y,z\right)$. As a
result, $\hat{\mathbf{e}}_{1}$ now appears located at the south pole
of the new sphere: the GBS (Supplementary Figure 4c). Finally, applying
this relabelling to Supplementary Equation \ref{eq:1.3.7}, we obtain
the sought Supplementary Equation \ref{eq:1.2.6}.
\end{enumerate}
\noindent \begin{center}
\includegraphics[width=15.5cm,height=9cm,keepaspectratio]{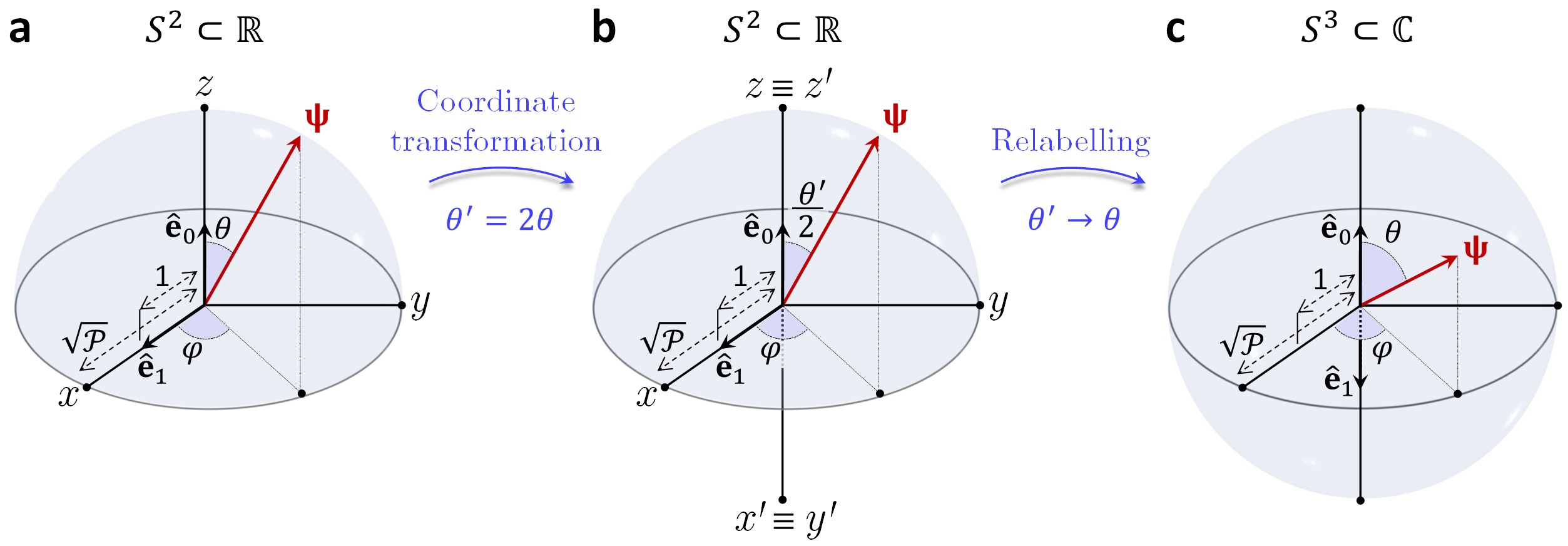}
\par\end{center}

\noindent \textbf{\small{}Supplementary Figure 4.}{\small{} Geometrical
construction of the generalised Bloch sphere (GBS) to represent an
anbit with 3 effective degrees of freedom (Supplementary Equation
}\ref{eq:1.2.6}{\small{}). }\textbf{\small{}a}{\small{},}\textbf{\small{}b}{\small{}
Transformation of the hypersphere $S^{2}\bigl(\sqrt{\mathcal{P}}\bigr)\subset\mathbb{R}^{3}$
into }\textbf{\small{}c}{\small{} the hypersphere $S^{3}\bigl(\sqrt{\mathcal{P}}\bigr)\subset\mathbb{C}^{2}$,
the GBS.}{\small \par}

\paragraph*{Final remark: Dirac's notation.}

As seen from the GBS, it is clear the mathematical similitude between
an anbit and a quantum bit (qubit), but with the basic differences
discussed in the paper. In this sense and to be coherent with the
main text at the end of subsection ``\emph{Unit of information: the
analog bit}'', let us introduce at this point the use of Dirac\textquoteright s
notation from now on. Therefore, let us recast the anbit as $\bigl|\psi\bigr\rangle=\psi_{0}\bigl|0\bigr\rangle+\psi_{1}\bigl|1\bigr\rangle$
by performing the identifications $\boldsymbol{\uppsi}\equiv\bigl|\psi\bigr\rangle$,
$\hat{\mathbf{e}}_{0}\equiv\bigl|0\bigr\rangle$ and $\hat{\mathbf{e}}_{1}\equiv\bigl|1\bigr\rangle$.
As we will see through the next sections, Dirac\textquoteright s notation
will allow us to simplify the mathematical framework of APC. 

\subsection*{1.4 Anbit amplitudes: component isomorphism\label{subsec:1.4}}

\addcontentsline{toc}{subsection}{1.4 Anbit amplitudes: component isomorphism}

\noindent Consider the single-anbit vector space $\mathscr{E}_{1}=\mathrm{span}(\mathcal{B}_{1})$,
where $\mathcal{B}_{1}=\left\{ \bigl|u\bigr\rangle,\bigl|w\bigr\rangle\right\} $
is a vector basis (that may be the canonical orthonormal basis $\left\{ \bigl|0\bigr\rangle,\bigl|1\bigr\rangle\right\} $
or a different basis). For any anbit $\bigl|\psi\bigr\rangle\in\mathscr{E}_{1}$,
$\exists\bigl(\psi_{u},\psi_{w}\bigr)\in\mathbb{C}^{2}$ (referred
to as the anbit amplitudes) satisfying that $\bigl|\psi\bigr\rangle=\psi_{u}\bigl|u\bigr\rangle+\psi_{w}\bigl|w\bigr\rangle$.
Using an algebraic terminology, $\psi_{u}$ and $\psi_{w}$ are specifically
the components of the vector $\bigl|\psi\bigr\rangle$ associated
to the basis $\mathcal{B}_{1}$, which can be calculated using the
inner product of $\mathscr{E}_{1}$ as follows \cite{key-S9}:
\begin{align}
\psi_{u}=\mathrm{proj}_{\mathrm{span}\left\{ \bigl|u\bigr\rangle\right\} }\left(\bigl|\psi\bigr\rangle\right)=\frac{\bigl\langle u|\psi\bigr\rangle}{\bigl\langle u|u\bigr\rangle}, & \ \ \ \ \ \psi_{w}=\mathrm{proj}_{\mathrm{span}\left\{ \bigl|w\bigr\rangle\right\} }\left(\bigl|\psi\bigr\rangle\right)=\frac{\bigl\langle w|\psi\bigr\rangle}{\bigl\langle w|w\bigr\rangle}.\label{eq:1.4.1}
\end{align}
In this context, a useful analytical tool that allows us to provide
a matrix nature to the concept of component is the \emph{component
isomorphism}, defined as the mapping $\left[\cdot\right]_{\mathcal{B}_{1}}:\mathscr{E}_{1}\rightarrow M_{2\times1}\left(\mathbb{C}\right)$:
\begin{equation}
\left[\bigl|\psi\bigr\rangle\right]_{\mathcal{B}_{1}}:=\left(\begin{array}{c}
\psi_{u}\\
\psi_{w}
\end{array}\right).\label{eq:1.4.2}
\end{equation}
In particular, this mapping is employed in APC to describe the input-output
relation of the anbit gates using matrices instead of linear operators
(see the description of equation 1 in the main text and Supplementary
Note 2 on p.\,\pageref{par:Matrix-representation.}).

\subsection*{1.5 Multiple anbits: tensor product \emph{vs} Cartesian product}

\addcontentsline{toc}{subsection}{1.5 Multiple anbits: tensor product \textit{vs} Cartesian product}

\noindent In order to operate with multiple anbits, we must construct
a vector space ``higher'' than the single-anbit vector space $\mathscr{E}_{1}$.
This can be done by using the tensor product or the Cartesian product.
In this subsection, we detail the main properties of both operations
within the context of APC.

\paragraph*{Tensor product.}

The definition and mathematical properties of the tensor product can
be found rigorously detailed in Ch.\,2 of ref.\,\cite{key-S10}.
Here, we will only highlight the following remarks:
\begin{itemize}
\item The notation employed in APC will be the same as the notation used
in quantum computing (QC) to describe a system with multiple qubits.
The $n$-anbit vector space $\mathscr{E}_{n}$, constructed from $\mathscr{E}_{1}$
using the tensor product, is denoted as:
\begin{equation}
\mathscr{E}_{n}=\underset{n-1\textrm{ times}}{\underbrace{\mathscr{E}_{1}\otimes\mathscr{E}_{1}\otimes\ldots\otimes\mathscr{E}_{1}}}\equiv\mathscr{E}_{1}^{\otimes\left(n-1\right)}.\label{eq:1.5.1}
\end{equation}
\item A vector belonging to $\mathscr{E}_{n}$ ($\mathscr{E}_{1}$) will
usually be denoted by using an uppercase (lowercase) Greek letter,
e.g., $\bigl|\Psi\bigr\rangle=\bigl|\psi_{1}\bigr\rangle\otimes\bigl|\psi_{2}\bigr\rangle\otimes\ldots\otimes\bigl|\psi_{n}\bigr\rangle$,
with $\bigl|\Psi\bigr\rangle\in\mathscr{E}_{n}$ and $\bigl|\psi_{k}\bigr\rangle\in\mathscr{E}_{1}$,
$\forall k\in\left\{ 1,\ldots,n\right\} $. Here, we can use a more
economical notation by describing the above expression of the form
$\bigl|\Psi\bigr\rangle=\bigl|\psi_{1},\psi_{2},\ldots,\psi_{n}\bigr\rangle$.
\item The dimension of such a vector space is $\dim\bigl(\mathscr{E}_{n}\bigr)=\dim\bigl(\mathscr{E}_{1}\bigr)^{n}$.
\item The scalar-vector product is ($\lambda\in\mathbb{C}$, $\bigl|\Psi\bigr\rangle\in\mathscr{E}_{n}$):
\begin{align}
\lambda\bigl|\Psi\bigr\rangle & =\lambda\bigl|\psi_{1}\bigr\rangle\otimes\bigl|\psi_{2}\bigr\rangle\otimes\ldots\otimes\bigl|\psi_{n}\bigr\rangle\nonumber \\
 & =\bigl|\psi_{1}\bigr\rangle\otimes\lambda\bigl|\psi_{2}\bigr\rangle\otimes\ldots\otimes\bigl|\psi_{n}\bigr\rangle\nonumber \\
 & =\bigl|\psi_{1}\bigr\rangle\otimes\bigl|\psi_{2}\bigr\rangle\otimes\ldots\otimes\lambda\bigl|\psi_{n}\bigr\rangle.\label{eq:1.5.2}
\end{align}
\item The null vector of $\mathscr{E}_{n}$ is constructed by applying $n-1$
times the tensor product to the null vector $\bigl|\mathbf{0}\bigr\rangle:=0\bigl|0\bigr\rangle+0\bigl|1\bigr\rangle$
of $\mathscr{E}_{1}$: $\bigl|\mathbf{0}\bigr\rangle^{\otimes\left(n-1\right)}$.
For the sake of simplicity, we will also denote the null vector of
$\mathscr{E}_{n}$ as $\bigl|\mathbf{0}\bigr\rangle$ and the context
should avoid the confusion.
\item The dual space of $\mathscr{E}_{1}$ is $\mathscr{L}\bigl(\mathscr{E}_{1},\mathbb{C}\bigr)$:
the set of homomorphisms $\bigl\langle\varphi\bigr|$ (described by
a ``bra'') that transform an anbit (described by a ``ket'') $\bigl|\psi\bigr\rangle\in\mathscr{E}_{1}$
into the complex number $\bigl\langle\varphi|\psi\bigr\rangle$ by
using the standard complex inner product of $\mathscr{E}_{1}$. Along
this line, note that the dual space of $\mathscr{E}_{n}=\mathscr{E}_{1}^{\otimes\left(n-1\right)}$
is $\mathscr{L}\bigl(\mathscr{E}_{1}^{\otimes\left(n-1\right)},\mathbb{C}\bigr)=\mathscr{L}\bigl(\mathscr{E}_{1},\mathbb{C}\bigr)^{\otimes\left(n-1\right)}$,
that is, the set of homomorphisms $\bigl\langle\Phi\bigr|$ constructed
from the tensor product of $n$ different homomorphisms $\bigl\langle\varphi_{1}\bigr|,\bigl\langle\varphi_{2}\bigr|,\ldots,\bigl\langle\varphi_{n}\bigr|$
belonging to $\mathscr{L}\bigl(\mathscr{E}_{1},\mathbb{C}\bigr)$:
$\bigl\langle\Phi\bigr|=\bigl\langle\varphi_{1}\bigr|\otimes\bigl\langle\varphi_{2}\bigr|\otimes\ldots\otimes\bigl\langle\varphi_{n}\bigr|$.
Here, we can use a more economical notation rewriting the above expression
as $\bigl\langle\Phi\bigr|=\bigl\langle\varphi_{1},\varphi_{2},\ldots,\varphi_{n}\bigr|$.
\item The inner product of $\mathscr{E}_{n}$ is defined by combining the
inner product of $\mathscr{E}_{1}$ along with the multiplication
operation of $\mathbb{C}$. As an example, the inner product of $\bigl|\Psi\bigr\rangle=\bigl|\psi_{1}\bigr\rangle\otimes\bigl|\psi_{2}\bigr\rangle$
and $\bigl|\Phi\bigr\rangle=\bigl|\varphi_{1}\bigr\rangle\otimes\bigl|\varphi_{2}\bigr\rangle$
belonging to $\mathscr{E}_{2}$ is defined as:
\begin{equation}
\bigl\langle\Psi|\Phi\bigr\rangle:=\bigl\langle\psi_{1}|\varphi_{1}\bigr\rangle\cdot\bigl\langle\psi_{2}|\varphi_{2}\bigr\rangle.\label{eq:1.5.3}
\end{equation}
\item The vector space $\mathscr{E}_{n}$ is a Hilbert space.
\item The canonical vector basis of $\mathscr{E}_{n}$ is built by applying
the tensor product to the canonical vector basis of $\mathscr{E}_{1}$.
As an example, the canonical vector basis of $\mathscr{E}_{2}$ is:
\begin{equation}
\mathcal{B}_{2}=\left\{ \bigl|k\bigr\rangle\otimes\bigl|l\bigr\rangle\right\} _{k,l\in\left\{ 0,1\right\} }=\left\{ \bigl|0,0\bigr\rangle,\bigl|0,1\bigr\rangle,\bigl|1,0\bigr\rangle,\bigl|1,1\bigr\rangle\right\} .\label{eq:1.5.4}
\end{equation}
Using the inner product given by Supplementary Equation \ref{eq:1.5.3},
it follows that $\mathcal{B}_{2}$ is orthonormal.
\item The components of a given ket $\bigl|\psi_{1},\psi_{2},\ldots,\psi_{n}\bigr\rangle$
belonging to $\mathscr{E}_{n}$ with vector basis $\mathcal{B}_{n}$
can be calculated from the components (or anbit amplitudes) of the
individuals kets $\bigl|\psi_{1,\ldots,n}\bigr\rangle$ belonging
to $\mathscr{E}_{1}$ with vector basis $\mathcal{B}_{1}$ by using
the Kronecker product $\otimes_{\mathrm{K}}$: 
\begin{equation}
\left[\bigl|\psi_{1},\psi_{2},\ldots,\psi_{n}\bigr\rangle\right]_{\mathcal{B}_{n}}=\left[\bigl|\psi_{1}\bigr\rangle\right]_{\mathcal{B}_{1}}\otimes_{\mathrm{K}}\left[\bigl|\psi_{2}\bigr\rangle\right]_{\mathcal{B}_{1}}\otimes_{\mathrm{K}}\ldots\otimes_{\mathrm{K}}\left[\bigl|\psi_{n}\bigr\rangle\right]_{\mathcal{B}_{1}},\label{eq:1.5.5}
\end{equation}
where $\left[\cdot\right]_{\mathcal{B}_{n}}$ and $\left[\cdot\right]_{\mathcal{B}_{1}}$are
the component isomorphisms associated to the bases $\mathcal{B}_{n}$
and $\mathcal{B}_{1}$, respectively (see p.\,\pageref{subsec:1.4}).
Supplementary Equation \ref{eq:1.5.5} applies to any basis $\mathcal{B}_{n}$
and $\mathcal{B}_{1}$ (not necessarily being the canonical bases).
For instance, consider two different anbits $\bigl|\psi\bigr\rangle=\psi_{0}\bigl|0\bigr\rangle+\psi_{1}\bigl|1\bigr\rangle$
and $\bigl|\varphi\bigr\rangle=\varphi_{0}\bigl|0\bigr\rangle+\varphi_{1}\bigl|1\bigr\rangle$.
The tensor product is:
\begin{align}
\bigl|\psi\bigr\rangle\otimes\bigl|\varphi\bigr\rangle & =\left(\psi_{0}\bigl|0\bigr\rangle+\psi_{1}\bigl|1\bigr\rangle\right)\otimes\left(\varphi_{0}\bigl|0\bigr\rangle+\varphi_{1}\bigl|1\bigr\rangle\right)\nonumber \\
 & =\psi_{0}\varphi_{0}\bigl|0,0\bigr\rangle+\psi_{0}\varphi_{1}\bigl|0,1\bigr\rangle+\psi_{1}\varphi_{0}\bigl|1,0\bigr\rangle+\psi_{1}\varphi_{1}\bigl|1,1\bigr\rangle.\label{eq:1.5.6}
\end{align}
Therefore, the components of $\bigl|\psi\bigr\rangle\otimes\bigl|\varphi\bigr\rangle\in\mathscr{E}_{2}$
with vector basis $\mathcal{B}_{2}$ given by Supplementary Equation
\ref{eq:1.5.4} can be calculated as:
\begin{equation}
\left[\bigl|\psi\bigr\rangle\otimes\bigl|\varphi\bigr\rangle\right]_{\mathcal{B}_{2}}=\left(\begin{array}{c}
\psi_{0}\varphi_{0}\\
\psi_{0}\varphi_{1}\\
\psi_{1}\varphi_{0}\\
\psi_{1}\varphi_{1}
\end{array}\right)=\left(\begin{array}{c}
\psi_{0}\\
\psi_{1}
\end{array}\right)\otimes_{\mathrm{K}}\left(\begin{array}{c}
\varphi_{0}\\
\varphi_{1}
\end{array}\right)\equiv\left[\bigl|\psi\bigr\rangle\right]_{\mathcal{B}_{1}}\otimes_{\mathrm{K}}\left[\bigl|\varphi\bigr\rangle\right]_{\mathcal{B}_{1}}.\label{eq:1.5.7}
\end{equation}
Supplementary Equation \ref{eq:1.5.7} can be applied, e.g., to obtain
the anbit amplitudes associated to the optical inputs depicted in
Fig.\,4c of the paper.
\item The tensor product is non-commutative. The 1:1 correspondence between
$\otimes$ and $\otimes_{\mathrm{K}}$ (emerged from the component
isomorphism) allows us to infer that $\otimes$ inherits the non-commutative
property from $\otimes_{\mathrm{K}}$. Using the above example, it
is straightforward to verify that $\bigl|\psi\bigr\rangle\otimes\bigl|\varphi\bigr\rangle\neq\bigl|\varphi\bigr\rangle\otimes\bigl|\psi\bigr\rangle$
(when $\bigl|\psi\bigr\rangle\neq\bigl|\varphi\bigr\rangle$):
\begin{align}
\bigl|\psi\bigr\rangle\otimes\bigl|\varphi\bigr\rangle\overset{1:1}{\longleftrightarrow}\left[\bigl|\psi\bigr\rangle\otimes\bigl|\varphi\bigr\rangle\right]_{\mathcal{B}_{2}} & =\left[\bigl|\psi\bigr\rangle\right]_{\mathcal{B}_{1}}\otimes_{\mathrm{K}}\left[\bigl|\varphi\bigr\rangle\right]_{\mathcal{B}_{1}}\nonumber \\
 & \neq\left[\bigl|\varphi\bigr\rangle\right]_{\mathcal{B}_{1}}\otimes_{\mathrm{K}}\left[\bigl|\psi\bigr\rangle\right]_{\mathcal{B}_{1}}=\left[\bigl|\varphi\bigr\rangle\otimes\bigl|\psi\bigr\rangle\right]_{\mathcal{B}_{2}}\overset{1:1}{\longleftrightarrow}\bigl|\varphi\bigr\rangle\otimes\bigl|\psi\bigr\rangle.\label{eq:1.5.8}
\end{align}
\end{itemize}

\paragraph*{Cartesian product.}

The definition and mathematical properties of the Cartesian product
(equivalent to the direct sum when is applied to a finite number of
vector spaces) are rigorously detailed in refs.\,\cite{key-S9,key-S11}.
Here, we will only highlight the following remarks:
\begin{itemize}
\item The $n$-anbit vector space $\mathscr{E}_{n}$, constructed from $\mathscr{E}_{1}$
using the Cartesian product, is denoted as:
\begin{equation}
\mathscr{E}_{n}=\underset{n-1\textrm{ times}}{\underbrace{\mathscr{E}_{1}\times\mathscr{E}_{1}\times\ldots\times\mathscr{E}_{1}}}\equiv\mathscr{E}_{1}^{\times\left(n-1\right)}.\label{eq:1.5.9}
\end{equation}
\item A vector belonging to $\mathscr{E}_{n}$ ($\mathscr{E}_{1}$) will
usually be denoted by using an uppercase (lowercase) Greek letter,
e.g., $\bigl|\Psi\bigr\rangle=\bigl|\psi_{1}\bigr\rangle\times\bigl|\psi_{2}\bigr\rangle\times\ldots\times\bigl|\psi_{n}\bigr\rangle$,
with $\bigl|\Psi\bigr\rangle\in\mathscr{E}_{n}$ and $\bigl|\psi_{k}\bigr\rangle\in\mathscr{E}_{1}$,
$\forall k\in\left\{ 1,\ldots,n\right\} $. Here, we can also use
a classical notation by describing the above expression via an $n$-tuple
$\bigl|\Psi\bigr\rangle=\bigl(\bigl|\psi_{1}\bigr\rangle,\bigl|\psi_{2}\bigr\rangle,\ldots,\bigl|\psi_{n}\bigr\rangle\bigr)$.
\item The dimension of such a vector space is $\dim\bigl(\mathscr{E}_{n}\bigr)=n\dim\bigl(\mathscr{E}_{1}\bigr)$.
\item The scalar-vector product is ($\lambda\in\mathbb{C}$, $\bigl|\Psi\bigr\rangle\in\mathscr{E}_{n}$):
\begin{align}
\lambda\bigl|\Psi\bigr\rangle & =\lambda\bigl|\psi_{1}\bigr\rangle\times\lambda\bigl|\psi_{2}\bigr\rangle\times\ldots\times\lambda\bigl|\psi_{n}\bigr\rangle\equiv\bigl(\lambda\bigl|\psi_{1}\bigr\rangle,\lambda\bigl|\psi_{2}\bigr\rangle,\ldots,\lambda\bigl|\psi_{n}\bigr\rangle\bigr).\label{eq:1.5.10}
\end{align}
\item The null vector of $\mathscr{E}_{n}$ is constructed by applying $n-1$
times the Cartesian product to the null vector $\bigl|\mathbf{0}\bigr\rangle$
of $\mathscr{E}_{1}$: $\bigl|\mathbf{0}\bigr\rangle^{\times\left(n-1\right)}$.
For the sake of simplicity, we will also denote the null vector of
$\mathscr{E}_{n}$ with the ket $\bigl|\mathbf{0}\bigr\rangle$ and
the context should avoid the confusion.
\item The dual space of $\mathscr{E}_{n}=\mathscr{E}_{1}^{\times\left(n-1\right)}$
is $\mathscr{L}\bigl(\mathscr{E}_{1}^{\times\left(n-1\right)},\mathbb{C}\bigr)=\mathscr{L}\bigl(\mathscr{E}_{1},\mathbb{C}\bigr)^{\times\left(n-1\right)}$,
that is, the set of homomorphisms $\bigl\langle\Phi\bigr|$ constructed
from the Cartesian product of $n$ different homomorphisms $\bigl\langle\varphi_{1}\bigr|,\bigl\langle\varphi_{2}\bigr|,\ldots,\bigl\langle\varphi_{n}\bigr|$
belonging to $\mathscr{L}\bigl(\mathscr{E}_{1},\mathbb{C}\bigr)$:
$\bigl\langle\Phi\bigr|=\bigl\langle\varphi_{1}\bigr|\times\bigl\langle\varphi_{2}\bigr|\times\ldots\times\bigl\langle\varphi_{n}\bigr|\equiv\bigl(\bigl\langle\varphi_{1}\bigr|,\bigl\langle\varphi_{2}\bigr|,\ldots,\bigl\langle\varphi_{n}\bigr|\bigr)$.
\item The inner product of $\mathscr{E}_{n}$ is defined by combining the
inner product of $\mathscr{E}_{1}$ along with the addition operation
of $\mathbb{C}$. As an example, the inner product of $\bigl|\Psi\bigr\rangle=\bigl|\psi_{1}\bigr\rangle\times\bigl|\psi_{2}\bigr\rangle$
and $\bigl|\Phi\bigr\rangle=\bigl|\varphi_{1}\bigr\rangle\times\bigl|\varphi_{2}\bigr\rangle$
belonging to $\mathscr{E}_{2}$ is defined as:
\begin{equation}
\bigl\langle\Psi|\Phi\bigr\rangle:=\bigl\langle\psi_{1}|\varphi_{1}\bigr\rangle+\bigl\langle\psi_{2}|\varphi_{2}\bigr\rangle.\label{eq:1.5.11}
\end{equation}
\item The vector space $\mathscr{E}_{n}$ is a Hilbert space.
\item The canonical vector basis of $\mathscr{E}_{n}$ is built by combining
the canonical vector basis of $\mathscr{E}_{1}$ along with the null
anbit of $\mathscr{E}_{1}$. As an illustrative example, the canonical
vector basis of $\mathscr{E}_{2}$ is found to be:
\begin{equation}
\mathcal{B}_{2}=\left\{ \left(\bigl|0\bigr\rangle,\bigl|\mathbf{0}\bigr\rangle\right),\left(\bigl|1\bigr\rangle,\bigl|\mathbf{0}\bigr\rangle\right),\left(\bigl|\mathbf{0}\bigr\rangle,\bigl|0\bigr\rangle\right),\left(\bigl|\mathbf{0}\bigr\rangle,\bigl|1\bigr\rangle\right)\right\} .\label{eq:1.5.12}
\end{equation}
Using the inner product given by Supplementary Equation \ref{eq:1.5.11},
it follows that $\mathcal{B}_{2}$ is orthonormal.
\item The components of a given ket $\bigl|\psi_{1}\bigr\rangle\times\bigl|\psi_{2}\bigr\rangle\times\ldots\times\bigl|\psi_{n}\bigr\rangle$
belonging to $\mathscr{E}_{n}$ with vector basis $\mathcal{B}_{n}$
can be calculated from the components (or anbit amplitudes) of the
individuals kets $\bigl|\psi_{1,\ldots,n}\bigr\rangle$ belonging
to $\mathscr{E}_{1}$ with vector basis $\mathcal{B}_{1}$ as follows:
\begin{equation}
\left[\bigl|\psi_{1}\bigr\rangle\times\bigl|\psi_{2}\bigr\rangle\times\ldots\times\bigl|\psi_{n}\bigr\rangle\right]_{\mathcal{B}_{n}}=\left(\begin{array}{c}
\left[\bigl|\psi_{1}\bigr\rangle\right]_{\mathcal{B}_{1}}\\
\left[\bigl|\psi_{2}\bigr\rangle\right]_{\mathcal{B}_{1}}\\
\cdots\\
\left[\bigl|\psi_{n}\bigr\rangle\right]_{\mathcal{B}_{1}}
\end{array}\right),\label{eq:1.5.13}
\end{equation}
where $\left[\cdot\right]_{\mathcal{B}_{n}}$ and $\left[\cdot\right]_{\mathcal{B}_{1}}$are
the component isomorphisms associated to the bases $\mathcal{B}_{n}$
and $\mathcal{B}_{1}$, respectively. Supplementary Equation \ref{eq:1.5.13}
applies to any basis $\mathcal{B}_{n}$ and $\mathcal{B}_{1}$ (not
necessarily being the canonical bases). For instance, consider two
different anbits $\bigl|\psi\bigr\rangle=\psi_{0}\bigl|0\bigr\rangle+\psi_{1}\bigl|1\bigr\rangle$
and $\bigl|\varphi\bigr\rangle=\varphi_{0}\bigl|0\bigr\rangle+\varphi_{1}\bigl|1\bigr\rangle$.
The Cartesian product is:
\begin{align}
\bigl|\psi\bigr\rangle\times\bigl|\varphi\bigr\rangle\equiv\bigl(\bigl|\psi\bigr\rangle,\bigl|\varphi\bigr\rangle\bigr) & =\bigl(\sum_{k=0}^{1}\psi_{k}\bigl|k\bigr\rangle,\sum_{l=0}^{1}\varphi_{l}\bigl|l\bigr\rangle\bigr)\nonumber \\
 & =\sum_{k=0}^{1}\psi_{k}\bigl(\bigl|k\bigr\rangle,\bigl|\mathbf{0}\bigr\rangle\bigr)+\sum_{l=0}^{1}\varphi_{l}\bigl(\bigl|\mathbf{0}\bigr\rangle,\bigl|l\bigr\rangle\bigr).\label{eq:1.5.14}
\end{align}
Therefore, the components of $\bigl|\psi\bigr\rangle\times\bigl|\varphi\bigr\rangle\in\mathscr{E}_{2}$
with vector basis $\mathcal{B}_{2}$ given by Supplementary Equation
\ref{eq:1.5.12} can be calculated as:
\begin{equation}
\left[\bigl|\psi\bigr\rangle\times\bigl|\varphi\bigr\rangle\right]_{\mathcal{B}_{2}}=\left(\begin{array}{c}
\psi_{0}\\
\psi_{1}\\
\varphi_{0}\\
\varphi_{1}
\end{array}\right)\equiv\left(\begin{array}{c}
\left[\bigl|\psi\bigr\rangle\right]_{\mathcal{B}_{1}}\\
\left[\bigl|\varphi\bigr\rangle\right]_{\mathcal{B}_{1}}
\end{array}\right).\label{eq:1.5.15}
\end{equation}
Supplementary Equation \ref{eq:1.5.15} can be applied, e.g., to obtain
the anbit amplitudes associated to the optical inputs shown in Fig.\,5a
of the paper.
\item The Cartesian product is non-commutative: $\bigl|\psi\bigr\rangle\times\bigl|\varphi\bigr\rangle\neq\bigl|\varphi\bigr\rangle\times\bigl|\psi\bigr\rangle$
(when $\bigl|\psi\bigr\rangle\neq\bigl|\varphi\bigr\rangle$).
\end{itemize}

\paragraph*{Final remarks.}

Consider two different anbits $\bigl|\psi\bigr\rangle=\psi_{0}\bigl|0\bigr\rangle+\psi_{1}\bigl|1\bigr\rangle$
and $\bigl|\varphi\bigr\rangle=\varphi_{0}\bigl|0\bigr\rangle+\varphi_{1}\bigl|1\bigr\rangle$.
It is worthy to note that the tensor product mixes the information
encoded by the anbit amplitudes (see Supplementary Equation \ref{eq:1.5.7}).
In contrast, the Cartesian product preserves this information (see
Supplementary Equation \ref{eq:1.5.15}) allowing us to independently
transform the anbit amplitudes $\psi_{0}$, $\psi_{1}$, $\varphi_{0}$
and $\varphi_{1}$ in a multi-anbit gate (e.g. the fan-in gate, see
Fig.\,5a of the paper).

On the other hand, as commented in the main text, there are some specific
multi-anbit gates (such as the fan-in and fan-out gates) that exhibit
a linear (nonlinear) nature when using the Cartesian (tensor) product
to construct them, see Supplementary Note 3 on p.\,\pageref{Cartesian vs tensor product in FI}
for more details.

\newpage{}

\section*{Supplementary Note 2: combinational design\label{sec:2}}

\addcontentsline{toc}{section}{Supplementary Note 2: combinational design}

\noindent In this section, we include additional information about
the fundamental pieces of combinational APC architectures: single-anbit
linear gates and controlled gates.

\subsection*{2.1 General properties of single-anbit linear gates}

\addcontentsline{toc}{subsection}{2.1 General properties of single-anbit linear gates}

\noindent A single-anbit linear gate is described by a linear operator
(or endomorphism) $\widehat{\mathrm{F}}:\mathscr{E}_{1}\rightarrow\mathscr{E}_{1}$
carrying out a transformation between two different anbits: the input
anbit $\bigl|\psi\bigr\rangle=\psi_{0}\bigl|0\bigr\rangle+\psi_{1}\bigl|1\bigr\rangle$
and the output anbit $\bigl|\varphi\bigr\rangle=\varphi_{0}\bigl|0\bigr\rangle+\varphi_{1}\bigl|1\bigr\rangle$.
In the next lines, we include supplementary notes about the general
properties of $\widehat{\mathrm{F}}$ reported in the main text.

\paragraph*{Matrix representation.\label{par:Matrix-representation.}}

Consider the canonical (orthonormal) vector basis $\mathcal{B}_{1}=\left\{ \left|0\right\rangle ,\left|1\right\rangle \right\} $
of $\mathscr{E}_{1}$. Using the fundamentals of linear algebra \cite{key-S12},
it is well-known that any \emph{linear} mapping of $\mathscr{E}_{1}$
has associated the following (and unique) matrix representation:
\begin{equation}
F=M_{\mathcal{B}_{1}}^{\mathcal{B}_{1}}\bigl(\widehat{\mathrm{F}}\bigr)=\left(\begin{array}{cc}
\bigl[\widehat{\mathrm{F}}\bigl|0\bigr\rangle\bigr]_{\mathcal{B}_{1}} & \bigl[\widehat{\mathrm{F}}\bigl|1\bigr\rangle\bigr]_{\mathcal{B}_{1}}\end{array}\right),\label{eq:2.1.1}
\end{equation}
where $\left[\cdot\right]_{\mathcal{B}_{1}}$ is the component isomorphism
detailed on p.\,\pageref{subsec:1.4}. Using Supplementary Equation\,\ref{eq:1.4.1}
to calculate the components of $\widehat{\mathrm{F}}\left|0\right\rangle $
and $\widehat{\mathrm{F}}\left|1\right\rangle $ associated to the
basis $\mathcal{B}_{1}$, we find that:
\begin{align}
\bigl[\widehat{\mathrm{F}}\bigl|0\bigr\rangle\bigr]_{\mathcal{B}_{1}}=\left(\begin{array}{c}
\bigl\langle0|\widehat{\mathrm{F}}|0\bigr\rangle\\
\bigl\langle1|\widehat{\mathrm{F}}|0\bigr\rangle
\end{array}\right) & ,\ \ \ \ \ \bigl[\widehat{\mathrm{F}}\bigl|1\bigr\rangle\bigr]_{\mathcal{B}_{1}}=\left(\begin{array}{c}
\bigl\langle0|\widehat{\mathrm{F}}|1\bigr\rangle\\
\bigl\langle1|\widehat{\mathrm{F}}|1\bigr\rangle
\end{array}\right).\label{eq:2.1.2}
\end{align}
Combining Supplementary Equations \ref{eq:2.1.1} and \ref{eq:2.1.2},
we obtain equation 1 of the paper, reproduced here for clarity:
\begin{equation}
F=\left(\begin{array}{cc}
\bigl\langle0|\widehat{\mathrm{F}}|0\bigr\rangle & \bigl\langle0|\widehat{\mathrm{F}}|1\bigr\rangle\\
\bigl\langle1|\widehat{\mathrm{F}}|0\bigr\rangle & \bigl\langle1|\widehat{\mathrm{F}}|1\bigr\rangle
\end{array}\right).\label{eq:2.1.3}
\end{equation}

Moreover, applying the isomorphism $\left[\cdot\right]_{\mathcal{B}_{1}}$
to the input-output relation $\bigl|\varphi\bigr\rangle=\widehat{\mathrm{F}}\bigl|\psi\bigr\rangle$,
we find that $\bigl[\bigl|\varphi\bigr\rangle\bigr]_{\mathcal{B}_{1}}=\bigl[\widehat{\mathrm{F}}\bigl|\psi\bigr\rangle\bigr]_{\mathcal{B}_{1}}$where:
\begin{equation}
\bigl[\widehat{\mathrm{F}}\bigl|\psi\bigr\rangle\bigr]_{\mathcal{B}_{1}}=\left(\begin{array}{c}
\bigl\langle0|\widehat{\mathrm{F}}|\psi\bigr\rangle\\
\bigl\langle1|\widehat{\mathrm{F}}|\psi\bigr\rangle
\end{array}\right)=\left(\begin{array}{c}
\psi_{0}\bigl\langle0|\widehat{\mathrm{F}}|0\bigr\rangle+\psi_{1}\bigl\langle0|\widehat{\mathrm{F}}|1\bigr\rangle\\
\psi_{0}\bigl\langle1|\widehat{\mathrm{F}}|0\bigr\rangle+\psi_{1}\bigl\langle1|\widehat{\mathrm{F}}|1\bigr\rangle
\end{array}\right)=F\left(\begin{array}{c}
\psi_{0}\\
\psi_{1}
\end{array}\right).\label{eq:2.1.4}
\end{equation}
Hence, $\bigl|\varphi\bigr\rangle=\widehat{\mathrm{F}}\bigl|\psi\bigr\rangle$
can equivalently be expressed via the matrix relation $\left[\bigl|\varphi\bigr\rangle\right]_{\mathcal{B}_{1}}=F\left[\bigl|\psi\bigr\rangle\right]_{\mathcal{B}_{1}}$,
with $\left[\bigl|\varphi\bigr\rangle\right]_{\mathcal{B}_{1}}=\bigl(\begin{array}{cc}
\varphi_{0} & \varphi_{1}\end{array}\bigr)^{T}$ and $\left[\bigl|\psi\bigr\rangle\right]_{\mathcal{B}_{1}}=\bigl(\begin{array}{cc}
\psi_{0} & \psi_{1}\end{array}\bigr)^{T}$ ($T$ denotes the transpose matrix). As commented above on p.\,\pageref{subsec:1.4},
the component isomorphism allows us to describe the input-output relation
of the anbit gates using matrices instead of linear operators.

\paragraph*{Ket-bra representation.\label{par:Ket-bra-representation.}}

Keeping in mind that $M_{\mathcal{B}_{1}}^{\mathcal{B}_{1}}\left(\cdot\right)$
is an isomorphism \cite{key-S12}, it should be noted that we can
always recover the input from the output in Supplementary Equation\,\ref{eq:2.1.1}.
In this vein, the linear operator $\widehat{\mathrm{F}}$ can be calculated
from the matrix $F$ as follows:
\begin{equation}
\widehat{\mathrm{F}}=\sum_{n,m}F_{nm}\bigl|n\bigr\rangle\bigl\langle m\bigr|,\label{eq:2.1.5}
\end{equation}
where $n,m\in\left\{ 0,1\right\} $ and $F_{nm}$ is the entry of
$F$ located at the $\left(n+1\right)$-th row and $\left(m+1\right)$-th
column. Supplementary Equation \ref{eq:2.1.5} will be referred to
as the \emph{ket-bra representation} of the gate. This equation may
be derived from Supplementary Equation \ref{eq:2.1.3}, which can
be recast of the form $F_{nm}=\bigl\langle n|\widehat{\mathrm{F}}|m\bigr\rangle$.
Using this expression, we infer that $F_{nm}\bigl|n\bigr\rangle\bigl\langle m\bigr|=\bigl|n\bigr\rangle\bigl\langle n|\widehat{\mathrm{F}}|m\bigr\rangle\bigl\langle m\bigr|$.
Applying the double summation $\sum_{n,m}$ at both sides and taking
into account that the vector basis $\mathcal{B}_{1}$ satisfies the
closure relation $\sum_{n}\bigl|n\bigr\rangle\bigl\langle n|=\widehat{1}$
(where $\widehat{1}$ is the identity operator) \cite{key-S10}, then
it directly follows Supplementary Equation \ref{eq:2.1.5}.

\paragraph*{Geometric representation.}

Using differential measurement, the input anbit $\bigl|\psi\bigr\rangle$
and the output anbit $\bigl|\varphi\bigr\rangle$ can be geometrically
represented in the GBS as two different points (Fig.\,3a). If the
anbit transformation $\widehat{\mathrm{F}}$ is a U-gate, both points
are connected by a trajectory describing a rotation of an angle $\alpha$
around an arbitrary unit vector $\hat{\mathbf{n}}$ (Fig.\,3b). Such
a trajectory preserves the radius in the GBS. However, if $\widehat{\mathrm{F}}$
is a G- or M-gate, there is no specific \emph{kind} of trajectory
connecting both points. This can be inferred from the singular value
decomposition (SVD), which factorizes $F$ as a function of two U-gates
($U_{1}$ and $U_{2}$) along with a $2\times2$ diagonal matrix $D$
with non-negative real entries \cite{key-S9,key-S12,key-S13}: $F=U_{2}DU_{1}$.
The matrices $U_{1}$ and $U_{2}$ can be respectively described by
two different rotations of angles $\alpha_{1}$ and $\alpha_{2}$
around two arbitrary unit vectors $\hat{\mathbf{n}}_{1}$ and $\hat{\mathbf{n}}_{2}$:
$U_{1}=e^{i\delta_{1}}R_{\hat{\mathbf{n}}_{1}}\left(\alpha_{1}\right)$
and $U_{2}=e^{i\delta_{2}}R_{\hat{\mathbf{n}}_{1}}\left(\alpha_{2}\right)$
(the global phases $\delta_{1}$ and $\delta_{2}$ are not observable
in the GBS). The diagonal matrix:
\begin{equation}
D=\left(\begin{array}{cc}
d_{1} & 0\\
0 & d_{2}
\end{array}\right),\label{eq:2.1.6}
\end{equation}
with $d_{1}\geq d_{2}\geq0$, modifies $\left|\psi_{0}\right|$ and
$\left|\psi_{1}\right|$ but, unfortunately, this parametric matrix
cannot be associated to a specific class of trajectory. As an example,
taking $d_{2}=0$, $D$ projects $\left|\psi\right\rangle $ onto
the positive $z$-axis of the GBS annihilating the anbit amplitude
$\psi_{1}$. However, if $d_{1}=d_{2}=1$, then $D$ preserves the
location of $\left|\psi\right\rangle $. Consequently, a G- or M-gate
cannot be represented by a universal kind of trajectory. Each numerical
case leads to a different trajectory in the GBS. 

\paragraph*{Lorentz reciprocity and forward-backward (FB) symmetry.}

By definition, an optical circuit is said to be reciprocal if and
only if the Lorentz reciprocity theorem is satisfied in such a system
\cite{key-S4,key-S14,key-S15}. By definition, an optical circuit
is said to be symmetric (or, equivalently, it is said that preserves
the FB symmetry) if and only if the system has the same transfer matrices
associated to the forward and backward light propagation directions
\cite{key-S4}. Specifically, the reciprocity and FB symmetry of an
optical circuit may be theoretically analysed via the scattering and
transfer matrices \cite{key-S4,key-S16}.

In the following, let us discuss some remarks and logical implications
between the $F$ matrix of a single-anbit linear gate and the scattering
and transfer matrices associated with an optical implementation of
such a gate (for the sake of clarity, the reader should know the definition
and basic properties of the scattering matrix ($S$), the forward
and backward transfer matrices ($T_{\mathrm{f}}$, $T_{\mathrm{b}}$),
and the reduced forward and backward transfer matrices ($\widetilde{T}_{\mathrm{f}}$,
$\widetilde{T}_{\mathrm{b}}$) of an optical circuit, see Ch.\,2
of ref.\,\cite{key-S4} for more details): 
\begin{enumerate}
\item The Lorentz reciprocity and the FB symmetry are not properties of
a gate. These are optical properties exclusively associated with a
specific PIP implementation of a gate. Therefore, a gate could be
implemented by different PIP circuits performing the same $F$ matrix
transformation, but exhibiting opposite optical properties.
\item An optical circuit preserving (breaking) the FB symmetry will induce
the same (a different) anbit transformation in each propagation direction. 
\item If an optical circuit of a gate preserves the FB symmetry ($T_{\mathrm{f}}=T_{\mathrm{b}}$),
then the circuit is reciprocal ($S=S^{T}$). The converse is not true
\cite{key-S4,key-S15}. 
\item The scattering matrix of a $2\times2$ \emph{non-reflective} optical
system is found to be \cite{key-S4}:
\begin{equation}
S=\left(\begin{array}{c|c}
0 & \widetilde{T}_{\mathrm{b}}\\
\hline \widetilde{T}_{\mathrm{f}} & 0
\end{array}\right)\in M_{4}\left(\mathbb{C}\right),\label{eq:2.1.7}
\end{equation}
with $\widetilde{T}_{\mathrm{f}},\widetilde{T}_{\mathrm{b}}\in M_{2}\left(\mathbb{C}\right)$.
Assuming that light reflection can usually be neglected in a PIP circuit
\cite{key-S4,key-S17}, we will consider that the \emph{ideal} optical
implementation of a gate is non-reflective. Hence, the scattering
matrix of any optical implementation of a gate can ideally be assumed
of the form:
\begin{equation}
S=\left(\begin{array}{c|c}
0 & \widetilde{T}_{\mathrm{b}}\\
\hline F & 0
\end{array}\right),\label{eq:2.1.8}
\end{equation}
with $\widetilde{T}_{\mathrm{f}}\equiv F$. The form of the matrix
$\widetilde{T}_{\mathrm{b}}$ directly depends on the specific optical
architecture of the gate.
\item If the circuit of a gate is reciprocal, we know that $S=S^{T}$ \cite{key-S4}.
Accordingly, $\widetilde{T}_{\mathrm{b}}\equiv F^{T}$ and the scattering
matrix becomes:
\begin{equation}
S=\left(\begin{array}{c|c}
0 & F^{T}\\
\hline F & 0
\end{array}\right).\label{eq:2.1.9}
\end{equation}
\item If the circuit of a gate is reciprocal and the gate satisfies the
condition $F=F^{T}$, then the circuit preserves the FB symmetry ($\widetilde{T}_{\mathrm{f}}=\widetilde{T}_{\mathrm{b}}\equiv F$)
and the scattering matrix is reduced to:
\begin{equation}
S=\left(\begin{array}{c|c}
0 & F\\
\hline F & 0
\end{array}\right).\label{eq:2.1.10}
\end{equation}
Using the contrapositive statement, if the circuit of a gate breaks
the FB symmetry ($\widetilde{T}_{\mathrm{f}}\neq\widetilde{T}_{\mathrm{b}}$),
then it follows that the Lorentz reciprocity is broken \emph{or} the
gate is described by an asymmetric matrix ($F\neq F^{T}$). Both situations
can take place simultaneously.
\item There is no logical implication between the concepts of circuital
reciprocity and gate reversibility. As commented above, the reciprocity
is a functional property of an optical implementation of a gate, while
the reversibility is a mathematical property of the endomorphism defining
the gate (discussed in the main text). In the same vein, there is
no logical implication between the concepts of FB symmetry and gate
reversibility.
\end{enumerate}
\newpage{}

\subsection*{2.2 Equivalent circuit architectures of single-anbit linear gates}

\addcontentsline{toc}{subsection}{2.2 Equivalent circuit architectures of single-anbit linear gates}

\paragraph*{U-gates.}

Equivalent optical implementations of a single-anbit U-gate can be
proposed by selecting different rotation vectors in Euler\textquoteright s
rotation theorem to factorize the $2\times2$ universal unitary matrix
given by equation 2 of the paper \cite{key-S18}. For instance, instead
of performing the factorization $U=e^{i\delta}R_{\hat{\mathbf{z}}}\left(\alpha_{3}\right)R_{\hat{\mathbf{x}}}\left(\alpha_{2}\right)R_{\hat{\mathbf{z}}}\left(\alpha_{1}\right)$,
whose optical implementation is shown in Fig.\,3b, we can select
the unit vector $\hat{\mathbf{y}}$ in the second rotation:
\begin{equation}
U=e^{i\delta}R_{\hat{\mathbf{z}}}\left(\alpha_{3}\right)R_{\hat{\mathbf{y}}}\left(\alpha_{2}\right)R_{\hat{\mathbf{z}}}\left(\alpha_{1}\right)=e^{i\delta}\left(\begin{array}{cc}
\cos\frac{\alpha_{2}}{2}e^{-i\left(\frac{\alpha_{3}+\alpha_{1}}{2}\right)} & -\sin\frac{\alpha_{2}}{2}e^{-i\left(\frac{\alpha_{3}-\alpha_{1}}{2}\right)}\\
\sin\frac{\alpha_{2}}{2}e^{i\left(\frac{\alpha_{3}-\alpha_{1}}{2}\right)} & \cos\frac{\alpha_{2}}{2}e^{i\left(\frac{\alpha_{3}+\alpha_{1}}{2}\right)}
\end{array}\right).\label{eq:2.2.1}
\end{equation}
Supplementary Figure 5 shows the $2\times2$ optical system whose
(reduced) forward transfer matrix $\widetilde{T}_{\mathrm{f}}$ is
given by the above equation. In contrast to the architecture depicted
in Fig.\,3b, which requires a tunable symmetric directional coupler
(DIC), this scheme is based on fixed couplers (50:50 beam splitters
implemented via multi-mode interferometers (MMIs) or fixed symmetric
DICs). Although the fabrication of a fixed coupler is simpler than
that of a tunable coupler, the main drawback of this implementation
of a U-gate is that it will approximately require the double footprint
than that of the circuit of Fig.\,3b because of a twice number of
couplers must be integrated in this architecture.
\noindent \begin{center}
\includegraphics[width=11.5cm,height=7cm,keepaspectratio]{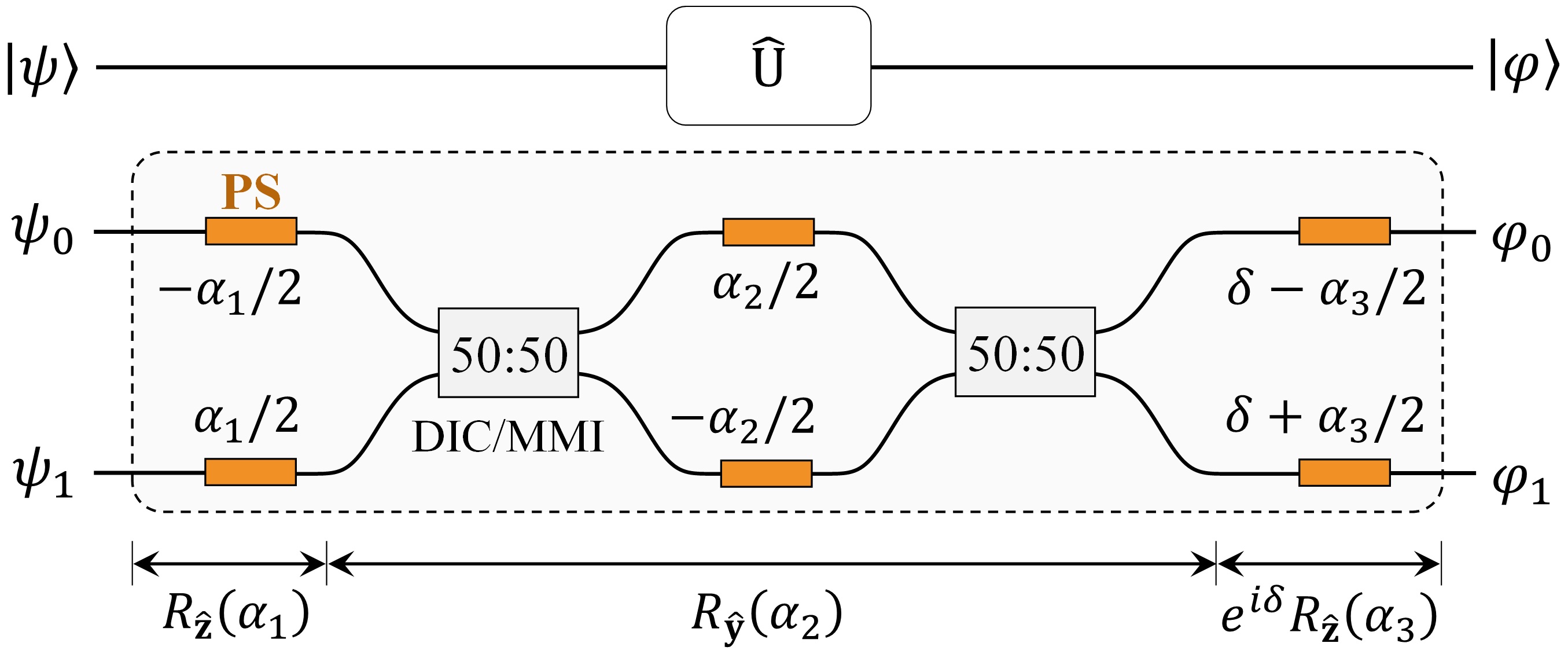}
\par\end{center}

\noindent \textbf{\small{}Supplementary Figure 5.}{\small{} Optical
circuit of a U-gate implementing the universal unitary matrix of equation
2 of the paper via Euler's factorization given by Supplementary Equation
\ref{eq:2.2.1}.\\ }{\small \par}

On the other hand, it is natural to wonder about the possibility of
building a more compact architecture of a U-gate (a $2\times2$ universal
unitary system) than the scheme proposed in Fig.\,3b by using a different
factorization from Euler\textquoteright s rotation theorem. Remarkably,
this question has been recently studied in ref.\,\cite{key-S18}
within the context of unitary signal PIP processors. As reported in
this reference, alternative $2\times2$ unitary factorization techniques
can be found in the mathematical literature (e.g. the cosine-sine
decomposition), but all of them lead to optical schemes of a $2\times2$
universal unitary system integrating a higher number of basic devices
than in our proposal.

\paragraph*{G-gates.}

An optical implementation of a single-anbit G-gate (different from
but equivalent to the architecture depicted in Fig.\,3c of the paper,
based on the SVD) can be found by using \emph{Mostow's decomposition}
\cite{key-S19,key-S20}. This theorem establishes that a non-singular
matrix $G$ can be factorized as a function of a unitary matrix $U$,
a real anti-symmetric matrix $A$ ($A=-A^{T}$) and a real symmetric
matrix $B$ ($B=B^{T}$) following the expression:

\newpage{}

\begin{equation}
G=Ue^{iA}e^{B}.\label{eq:2.2.2}
\end{equation}
The main handicap of Mostow's decomposition is that the exponential
matrices cannot be directly implemented by employing mainstream PIP
devices. In order to unveil the optical architecture of Supplementary
Equation \ref{eq:2.2.2}, it should be noted that $E_{1}=e^{iA}$
and $E_{2}=e^{B}$ are Hermitian (but non-unitary) matrices:
\begin{align}
E_{1}^{\dagger} & =\left(e^{iA}\right)^{\dagger}=e^{-iA^{\dagger}}=e^{-iA^{T}}=e^{iA}=E_{1},\label{eq:2.2.3}\\
E_{2}^{\dagger} & =\left(e^{B}\right)^{\dagger}=e^{B^{\dagger}}=e^{B^{T}}=e^{B}=E_{2},\label{eq:2.2.4}
\end{align}
with $E_{1}^{\dagger}E_{1}=E_{1}^{2}=e^{i2A}\neq I$ and $E_{2}^{\dagger}E_{2}=E_{2}^{2}=e^{2B}\neq I$
($\dagger$ denotes the conjugate transpose matrix). Therefore, $E_{1}$
and $E_{2}$ can be factorized by using the spectral decomposition
as $E_{1}=U_{1}\Lambda_{1}U_{1}^{-1}$ and $E_{2}=U_{2}\Lambda_{2}U_{2}^{-1}$,
where $U_{1,2}$ are matrices built from linearly independent eigenvectors
of $E_{1,2}$ and $\Lambda_{1,2}$ are diagonal matrices whose entries
are the corresponding (real) eigenvalues \cite{key-S9,key-S12}. Moreover,
since $\mathcal{B}_{1}=\left\{ \left|0\right\rangle ,\left|1\right\rangle \right\} $
is an orthonormal basis, then it follows that $U_{1,2}$ are unitary
matrices ($U_{1,2}^{-1}=U_{1,2}^{\dagger}$). Hence, Supplementary
Equation \ref{eq:2.2.2} can be recast as: 
\begin{equation}
G=\bigl(UU_{1}\bigr)\Lambda_{1}\bigl(U_{1}^{\dagger}U_{2}\bigr)\Lambda_{2}U_{2}^{\dagger}.\label{eq:2.2.5}
\end{equation}
As seen, we have factorized $G$ as a function of \emph{three} unitary
matrices $UU_{1}$, $U_{1}^{\dagger}U_{2}$, $U_{2}^{\dagger}$ and
\emph{two} diagonal matrices $\Lambda_{1}$, $\Lambda_{2}$. Interestingly,
this factorization is implementable using basic PIP devices, see Supplementary
Figure 6.
\noindent \begin{center}
\includegraphics[width=14cm,height=7cm,keepaspectratio]{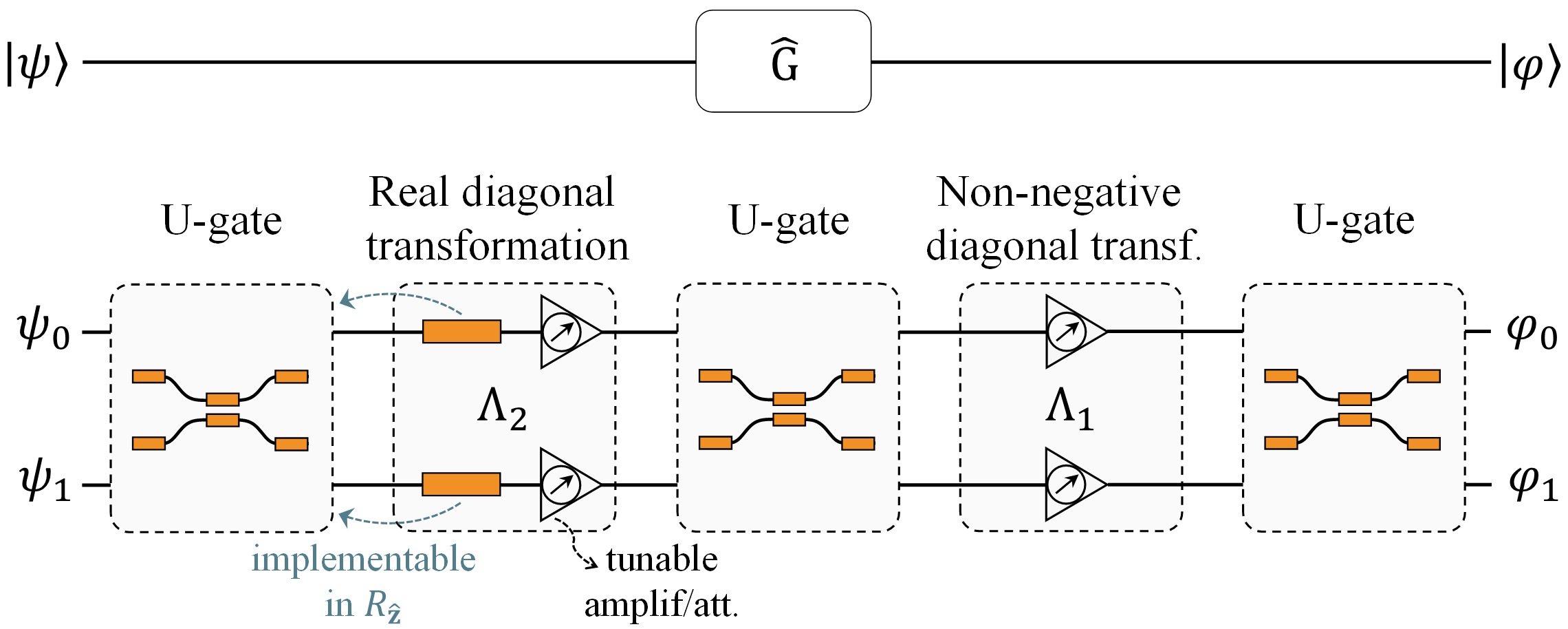}\textbf{\small{}}\\
\textbf{\small{}Supplementary Figure 6.}{\small{} Optical circuit
of a G-gate based on Mostow's factorization.\\ }
\par\end{center}{\small \par}

An additional remark about this architecture should be discussed.
While the first diagonal matrix of the circuit ($\Lambda_{2}$) may
have positive and negative real entries (the negative sign is implemented
by phase shifters (PSs) of $\pi$ rad that can alternatively be implemented
by the PSs of the final $R_{\hat{\mathbf{z}}}$ rotation in the first
U-gate), the second diagonal matrix ($\Lambda_{1}$) has non-negative
real entries. 

For completeness, let us demonstrate the non-negative nature of the
entries of $\Lambda_{1}$. Since $A$ is real anti-symmetric, then
it follows that $A$ must be of the form:
\begin{equation}
A=\left(\begin{array}{cc}
0 & a\\
-a & 0
\end{array}\right),\label{eq:2.2.6}
\end{equation}
with $a\in\mathbb{R}$. The (non-degenerate) eigenvalues of $A$ are
found to be $\mu_{1,2}=\pm ia$. Here, since the geometric and algebraic
multiplicity are the same, $A$ is diagonalizable. This implies that
$A$ can be factorized by using the spectral decomposition. Hence,
$A=U_{1}D_{1}U_{1}^{-1}$ with $D_{1}=\textrm{diag}\left(\mu_{1},\mu_{2}\right)$
and:
\begin{equation}
e^{iA}=e^{iU_{1}D_{1}U_{1}^{-1}}=U_{1}e^{iD_{1}}U_{1}^{-1}\equiv U_{1}\Lambda_{1}U_{1}^{-1},\label{eq:2.2.7}
\end{equation}
where the eigenvalues of $e^{iA}$, which are the entries of $\Lambda_{1}=e^{iD_{1}}$,
are found to be:
\begin{equation}
\lambda_{1,2}=e^{i\mu_{1,2}}=e^{\mp a}>0.\label{eq:2.2.8}
\end{equation}

\paragraph*{M-gates.}

A single-anbit M-gate is described by a complex matrix belonging to
$M_{2}\left(\mathbb{C}\right)$:
\begin{equation}
M=\left(\begin{array}{cc}
m_{11} & m_{12}\\
m_{21} & m_{22}
\end{array}\right).\label{eq:2.2.9}
\end{equation}
Remarkably, $M_{2}\left(\mathbb{C}\right)$ is the complex Lie algebra
$\mathfrak{gl}\left(2,\mathbb{C}\right)$ \cite{key-S7,key-S21},
with vector basis composed by the Pauli matrices $\left\{ \sigma_{k}\right\} _{k=0}^{3}$:
\begin{equation}
\sigma_{k}:=\left(\begin{array}{cc}
\delta_{k0}+\delta_{k3} & \delta_{k1}-i\delta_{k2}\\
\delta_{k1}+i\delta_{k2} & \delta_{k0}-\delta_{k3}
\end{array}\right),\label{eq:2.2.10}
\end{equation}
where $\delta_{kl}$ is the Kronecker delta ($\delta_{kl}=1$ with
$k=l$ and $\delta_{kl}=0$ with $k\neq l$). This implies that any
M-gate can be described as a linear combination of the Pauli matrices
(see p.\,427 of ref.\,\cite{key-S10}):
\begin{equation}
M=\sum_{k=0}^{3}\alpha_{k}\sigma_{k},\label{eq:2.2.11}
\end{equation}
where $\alpha_{0}=(m_{11}+m_{22})/2$, $\alpha_{1}=(m_{12}+m_{21})/2$,
$\alpha_{2}=i(m_{12}-m_{21})/2$ and $\alpha_{3}=(m_{11}-m_{22})/2$.
Consequently, an optical circuit implementing Supplementary Equation
\ref{eq:2.2.11} will give rise to an equivalent scheme to the architecture
of an M-gate depicted in Fig.\,3c of the paper, based on the SVD. 

To this end, let us express the Pauli matrices as a function of the
rotation matrices of the Bloch sphere (which can be implemented via
PIP technology because are unitary matrices \cite{key-S18}):
\begin{align}
\sigma_{0}=I,\ \ \ \sigma_{1}=iR_{\hat{\mathbf{x}}}\left(\pi\right), & \ \ \ \sigma_{2}=iR_{\hat{\mathbf{y}}}\left(\pi\right),\ \ \ \sigma_{3}=iR_{\hat{\mathbf{z}}}\left(\pi\right).\label{eq:2.2.12}
\end{align}
In this fashion, Supplementary Equation \ref{eq:2.2.11} becomes:
\begin{equation}
M=\alpha_{0}I+i\alpha_{1}R_{\hat{\mathbf{x}}}\left(\pi\right)+i\alpha_{2}R_{\hat{\mathbf{y}}}\left(\pi\right)+i\alpha_{3}R_{\hat{\mathbf{z}}}\left(\pi\right).\label{eq:2.2.13}
\end{equation}
The optical implementation of the above equation using PIP technology
only requires (see Supplementary Figure 7): (\emph{i}) the minimal
circuit architecture (MCA) of the U-gates to generate the rotation
matrices (see Fig.\,3b of the paper), (\emph{ii}) PSs and tunable
optical attenuators and amplifiers to generate the terms $\alpha_{0}$
and $i\alpha_{1,2,3}$, and (\emph{iii}) fan-in and fan-out gates
to obtain the linear combination of all terms (see Supplementary Note
3 for more details about these multi-anbit gates). As seen, this structure
is more complex than the architecture of an M-gate based on the SVD
(Fig.\,3c) given that it involves a higher number of basic devices,
requires ancilla anbits and generates gargabe anbits.
\noindent \begin{center}
\includegraphics[width=14cm,height=10cm,keepaspectratio]{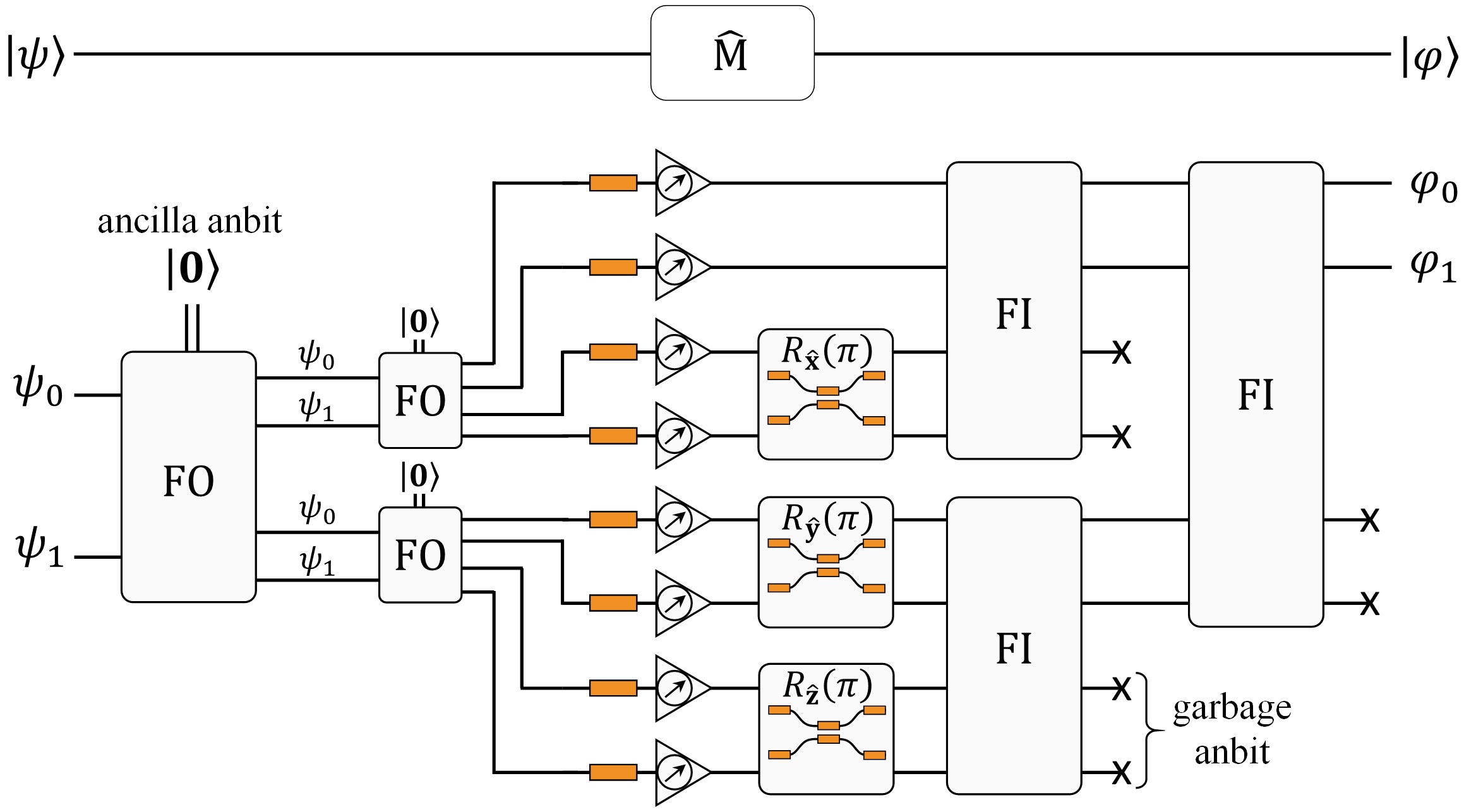}
\par\end{center}

\noindent \textbf{\small{}Supplementary Figure 7.}{\small{} Optical
circuit of an M-gate based on the factorization described by Supplementary
Equation }\ref{eq:2.2.13}{\small{}. The gates fan-in (FI) and fan-out
(FO) are described in Fig.\,5a of the main text and in Supplementary
Note 3, see p.}\,\pageref{sec:3}{\small{}.}{\small \par}

\subsection*{2.3 Controlled gates}

\addcontentsline{toc}{subsection}{2.3 Controlled gates}

\noindent In any computation theory, a controlled gate is an operation
of \emph{multiple} units of information given that the goal is to
perform a transformation on the target units of information (e.g.
the target anbits, $\bigl|t\bigr\rangle$) depending on the value
of the control units of information (e.g. the control anbits, $\bigl|c\bigr\rangle$). 

In APC, the definition of a multi-anbit gate (in this case, a controlled
gate) will depend on the kind of operation employed to describe the
multiple input and output anbits. As commented in the main text, using
the \emph{tensor product} to describe $\bigl|t\bigr\rangle$ and $\bigl|c\bigr\rangle$,
the mathematical framework of a controlled anbit gate can be directly
extrapolated from a controlled quantum gate (but differing in some
fundamental aspects). 

Here, we develop in detail the theory of such a kind of gates within
the context of APC highlighting the main differences with QC.

\subsubsection*{2.3.1 Single control anbit}

\addcontentsline{toc}{subsubsection}{2.3.1  Single control anbit}

\paragraph*{Definition and formalism.}

Let us start by considering a single-anbit linear gate (a U-, G-,
or M-gate) performing a transformation $\widehat{\mathrm{F}}$ on
an input anbit $\bigl|t\bigr\rangle$: the target anbit. The controlled
version of the $\widehat{\mathrm{F}}$-gate (termed as controlled-$\widehat{\mathrm{F}}$
gate) with a single control anbit $\bigl|c\bigr\rangle\in\left\{ \bigl|0\bigr\rangle,\bigl|1\bigr\rangle\right\} $
is a linear operator $\widehat{\mathrm{F}}_{\mathrm{C}}$ defined
via the tensor product as:
\begin{equation}
\widehat{\mathrm{F}}_{\mathrm{C}}:=\widehat{\mathrm{1}}\otimes\widehat{\mathrm{F}}^{c},\label{eq:2.3.1}
\end{equation}
which transforms the input $\bigl|c,t\bigr\rangle$ into the output:
\begin{equation}
\widehat{\mathrm{F}}_{\mathrm{C}}\bigl|c,t\bigr\rangle=\bigl(\widehat{\mathrm{1}}\otimes\widehat{\mathrm{F}}^{c}\bigr)\bigl|c,t\bigr\rangle=\bigl|c\bigr\rangle\otimes\widehat{\mathrm{F}}^{c}\bigl|t\bigr\rangle.\label{eq:2.3.2}
\end{equation}
The transformation $\widehat{\mathrm{F}}$ is applied to $\bigl|t\bigr\rangle$
when $\bigl|c\bigr\rangle=\bigl|1\bigr\rangle$ or, otherwise, $\bigl|t\bigr\rangle$
remains invariant at the output. Figure 4a of the paper shows the
symbolic representation of $\widehat{\mathrm{F}}_{\mathrm{C}}$, the
same as that of a controlled quantum gate for the sake of simplicity.
However, in contrast to QC, $\widehat{\mathrm{F}}_{\mathrm{C}}$ will
be a non-unitary operation when $\widehat{\mathrm{F}}\widehat{\mathrm{F}}^{\dagger}\neq\widehat{1}$,
that is, when $\widehat{\mathrm{F}}$ is a G- or M-gate (see below
the algebraic structure in properties).

In any case, the mathematical formalism of a controlled anbit gate
is the same as that of a controlled quantum gate. Since we work with
two input and output anbits, the underlying Hilbert space is $\mathscr{E}_{2}=\mathscr{E}_{1}\otimes\mathscr{E}_{1}$
and the canonical (orthonormal) basis is given by Supplementary Equation\,\ref{eq:1.5.4},
reproduced here for clarity:
\begin{equation}
\mathcal{B}_{2}=\left\{ \bigl|k\bigr\rangle\otimes\bigl|l\bigr\rangle\right\} _{k,l\in\left\{ 0,1\right\} }=\left\{ \bigl|0,0\bigr\rangle,\bigl|0,1\bigr\rangle,\bigl|1,0\bigr\rangle,\bigl|1,1\bigr\rangle\right\} .\label{eq:2.3.3}
\end{equation}
Hence, the matrix representation of $\widehat{\mathrm{F}}_{\mathrm{C}}$
associated to $\mathcal{B}_{2}$ is:
\begin{align}
F_{\mathrm{C}} & =M_{\mathcal{B}_{2}}^{\mathcal{B}_{2}}\bigl(\widehat{\mathrm{F}}_{\mathrm{C}}\bigr)=\bigl(\begin{array}{ccc}
\cdots & \bigl[\widehat{\mathrm{F}}_{\mathrm{C}}\bigl|k,l\bigr\rangle\bigr]_{\mathcal{B}_{2}} & \cdots\end{array}\bigr)=\left(\begin{array}{c|c}
I & 0\\
\hline 0 & F
\end{array}\right),\label{eq:2.3.4}
\end{align}
where $F_{\mathrm{C}}\in M_{4}\left(\mathbb{C}\right)$, $I$ is the
$2\times2$ identity matrix and $F\in M_{2}\left(\mathbb{C}\right)$
is the matrix representation of $\widehat{\mathrm{F}}$ associated
to the canonical basis $\mathcal{B}_{1}=\left\{ \left|0\right\rangle ,\left|1\right\rangle \right\} $
of $\mathscr{E}_{1}$, see Supplementary Equation \ref{eq:2.1.1}.
Along this line, the ket-bra representation of $\widehat{\mathrm{F}}_{\mathrm{C}}$
can be calculated from the entries of $F_{\mathrm{C}}$ as follows
(see previously p.\,\pageref{par:Ket-bra-representation.}, where
the ket-bra representation of a gate is introduced):
\begin{align}
\widehat{\mathrm{F}}_{\mathrm{C}} & =\bigl|0,0\bigr\rangle\bigl\langle0,0\bigr|+\bigl|0,1\bigr\rangle\bigl\langle0,1\bigr|\nonumber \\
 & +F_{00}\bigl|1,0\bigr\rangle\bigl\langle1,0\bigr|+F_{01}\bigl|1,0\bigr\rangle\bigl\langle1,1\bigr|+F_{10}\bigl|1,1\bigr\rangle\bigl\langle1,0\bigr|+F_{11}\bigl|1,1\bigr\rangle\bigl\langle1,1\bigr|\nonumber \\
 & =\bigl|0\bigr\rangle\bigl\langle0\bigr|\otimes\left(\bigl|0\bigr\rangle\bigl\langle0\bigr|+\bigl|1\bigr\rangle\bigl\langle1\bigr|\right)\nonumber \\
 & +\bigl|1\bigr\rangle\bigl\langle1\bigr|\otimes\left(F_{00}\bigl|0\bigr\rangle\bigl\langle0\bigr|+F_{01}\bigl|0\bigr\rangle\bigl\langle1\bigr|+F_{10}\bigl|1\bigr\rangle\bigl\langle0\bigr|+F_{11}\bigl|1\bigr\rangle\bigl\langle1\bigr|\right)\nonumber \\
 & =\bigl|0\bigr\rangle\bigl\langle0\bigr|\otimes\widehat{1}+\bigl|1\bigr\rangle\bigl\langle1\bigr|\otimes\widehat{\mathrm{F}}.\label{eq:2.3.5}
\end{align}
In contrast to Supplementary Equation \ref{eq:2.3.1}, the ket-bra
representation of $\widehat{\mathrm{F}}_{\mathrm{C}}$ can be applied
to an arbitrary control anbit $\bigl|c\bigr\rangle=c_{0}\bigl|0\bigr\rangle+c_{1}\bigl|1\bigr\rangle\notin\left\{ \bigl|0\bigr\rangle,\bigl|1\bigr\rangle\right\} $.
In such a scenario, using Supplementary Equation \ref{eq:2.3.5},
$\widehat{\mathrm{F}}_{\mathrm{C}}$ generates the output:
\begin{equation}
\widehat{\mathrm{F}}_{\mathrm{C}}\bigl|c,t\bigr\rangle=\bigl\langle0|c\bigr\rangle\bigl|0,t\bigr\rangle+\bigl\langle1|c\bigr\rangle\bigl|1\bigr\rangle\otimes\widehat{\mathrm{F}}\bigl|t\bigr\rangle=c_{0}\bigl|0,t\bigr\rangle+c_{1}\bigl|1\bigr\rangle\otimes\widehat{\mathrm{F}}\bigl|t\bigr\rangle.\label{eq:2.3.6}
\end{equation}
Note that this input-output relation is more general than Supplementary
Equation \ref{eq:2.3.2}.

On the other hand, although it is out of the scope of this work, it
is worthy to note that a controlled anbit gate can alternatively be
defined by using the \emph{Cartesian product}. This possibility cannot
be found in QC. Nevertheless, using the Cartesian product we cannot
extrapolate the mathematical framework of the controlled quantum gates
to the controlled anbit gates. In any case, for completeness, let
us discuss this possibility. Here, the basic idea is to define a controlled
anbit gate as follows: the transformation $\widehat{\mathrm{F}}$
is applied to $\bigl|t\bigr\rangle$ when $\bigl|c\bigr\rangle$ is
the null anbit $\bigl|\mathbf{0}\bigr\rangle$ or, otherwise, $\bigl|t\bigr\rangle$
remains invariant at the output. In this way, using this definition
and the canonical vector basis of $\mathscr{E}_{2}=\mathscr{E}_{1}\times\mathscr{E}_{1}$,
given by Supplementary Equation \ref{eq:1.5.12}, it is straightforward
to demonstrate that we obtain the same matrix representation of $\widehat{\mathrm{F}}_{\mathrm{C}}$
as in Supplementary Equation \ref{eq:2.3.4}. 

\paragraph*{Properties.\label{par:Properties.}}

The controlled anbit gate illustrated in Fig.\,4a of the main text
exhibits the following properties:
\begin{enumerate}
\item \emph{Endomorphism}. The linear operator $\widehat{\mathrm{F}}_{\mathrm{C}}$
is an endomorphism of the Hilbert space $\mathscr{E}_{2}=\mathscr{E}_{1}\otimes\mathscr{E}_{1}$.
\item \emph{EDF transformation}. If $\bigl|c\bigr\rangle=\bigl|1\bigr\rangle$,
then $\widehat{\mathrm{F}}_{\mathrm{C}}$ modifies the EDFs of $\bigl|t\bigr\rangle$
according to $\widehat{\mathrm{F}}$ (a U-, G-, or M-gate). For instance,
assuming a target anbit with 3 EDFs, if $\bigl|c\bigr\rangle=\bigl|1\bigr\rangle$
and $\widehat{\mathrm{F}}$ is a U-gate, then $\widehat{\mathrm{F}}_{\mathrm{C}}$
transforms these EDFs by rotating $\bigl|t\bigr\rangle$ in the GBS.
\item \emph{Reversibility}. From Supplementary Equation \ref{eq:2.3.4},
it is direct to verify that: 
\begin{equation}
\det\bigl(F_{\mathrm{C}}\bigr)=\det\bigl(F\bigr).\label{eq:2.3.7}
\end{equation}
Consequently, the controlled-$\widehat{\mathrm{F}}$ gate is reversible
if and only if the $\widehat{\mathrm{F}}$-gate is reversible ($\det\bigl(F\bigr)\neq0$).
In such a case, $\widehat{\mathrm{F}}_{\mathrm{C}}$ is an automorphism.
The inverse controlled gate (inverse automorphism) is described by
the linear operator $\widehat{\mathrm{F}}_{\mathrm{C}}^{-1}$, whose
matrix representation associated to $\mathcal{B}_{2}$ is:
\begin{align}
F_{\mathrm{C}}^{-1} & =M_{\mathcal{B}_{2}}^{\mathcal{B}_{2}}\bigl(\widehat{\mathrm{F}}_{\mathrm{C}}^{-1}\bigr)=\bigl[M_{\mathcal{B}_{2}}^{\mathcal{B}_{2}}\bigl(\widehat{\mathrm{F}}_{\mathrm{C}}\bigr)\bigr]^{-1}=\left(\begin{array}{c|c}
I & 0\\
\hline 0 & F^{-1}
\end{array}\right).\label{eq:2.3.8}
\end{align}
In addition, if the controlled-$\widehat{\mathrm{F}}$ gate is reversible,
then $\widehat{\mathrm{F}}_{\mathrm{C}}\bigl(\mathcal{B}_{2}\bigr)=\bigl\{\widehat{\mathrm{F}}_{\mathrm{C}}\bigl|k,l\bigr\rangle\bigr\}_{k,l\in\left\{ 0,1\right\} }$
is a vector basis of $\mathscr{E}_{2}$. Using the contrapositive
statement, if $\widehat{\mathrm{F}}_{\mathrm{C}}\bigl(\mathcal{B}_{2}\bigr)$
is not a vector basis of $\mathscr{E}_{2}$, then it follows that
the controlled-$\widehat{\mathrm{F}}$ gate is irreversible.
\item \emph{Matrix factorizations}. Any matrix factorization of $F$ can
be directly applied to $F_{\mathrm{C}}$:
\begin{equation}
F_{\mathrm{C}}=\left(\begin{array}{c|c}
I & 0\\
\hline 0 & F
\end{array}\right)=\left(\begin{array}{c|c}
I & 0\\
\hline 0 & \prod_{k}F_{k}
\end{array}\right)=\prod_{k}\left(\begin{array}{c|c}
I & 0\\
\hline 0 & F_{k}
\end{array}\right).\label{eq:2.3.9}
\end{equation}
In other words, if the $\widehat{\mathrm{F}}$-gate is equivalent
to a set of $\widehat{\mathrm{F}}_{k}$-gates connected in series,
then the controlled-$\widehat{\mathrm{F}}$ gate is equivalent to
a set of controlled-$\widehat{\mathrm{F}}_{k}$ gates connected in
series.
\item \emph{Algebraic structure}. The matrix $F_{\mathrm{C}}$ inherits
the algebraic structure of $F$: if $F\in\mathrm{U}\left(2\right)$,
then $F_{\mathrm{C}}\in\mathrm{U}\left(4\right)$; if $F\in\mathrm{GL}\left(2,\mathbb{C}\right)$,
then $F_{\mathrm{C}}\in\mathrm{GL}\left(4,\mathbb{C}\right)$; if
$F\in\mathfrak{gl}\left(2,\mathbb{C}\right)$, then $F_{\mathrm{C}}\in\mathfrak{gl}\left(4,\mathbb{C}\right)$
(the definition of the Lie groups $\mathrm{U}\left(n\right)$ and
$\mathrm{GL}\left(n,\mathbb{C}\right)$, and the Lie algebra $\mathfrak{gl}\left(n,\mathbb{C}\right)$
is detailed in ref.\,\cite{key-S21}). Note that in QC, $F_{\mathrm{C}}$
can only belong to $\mathrm{U}\left(4\right)$.
\item \emph{Universality}. A universal matrix of a controlled-$\widehat{\mathrm{F}}$
gate must be able to describe all the possible $4\times4$ matrix
transformations of the form given by Supplementary Equation \ref{eq:2.3.4}.
Therefore, $F_{\mathrm{C}}$ is a universal matrix of a controlled-$\widehat{\mathrm{F}}$
gate if and only if $F$ is a universal matrix of a U-, G-, or M-gate.
\item \emph{PIP implementation}. In QC, the theoretical strategies to design
controlled gates are focused on factorizing $F_{\mathrm{C}}$ as a
function of single-qubit gates and the controlled-NOT (CNOT) gate
\cite{key-S22}. Despite the fact that these schemes could be extrapolated
to APC, the implementation of a controlled anbit gate using PIP technology
does not require these intricate design strategies. Remarkably, \emph{APC
has its own design strategies} \emph{and architectures}, as demonstrated
in Figs.\,4b, 4c and 4d of the paper.
\end{enumerate}

\subsubsection*{2.3.2 Multiple control anbits}

\addcontentsline{toc}{subsubsection}{2.3.2  Multiple control anbits}

\paragraph*{Definition and formalism.}

Consider a single-anbit $\widehat{\mathrm{F}}$-gate (a U-, G-, or
M-gate) operating on a target anbit $\bigl|t\bigr\rangle$. The controlled-$\widehat{\mathrm{F}}$
gate with $n$ control anbits $\bigl|c_{1,2,\ldots,n}\bigr\rangle\in\left\{ \bigl|0\bigr\rangle,\bigl|1\bigr\rangle\right\} $
is a linear operator $\widehat{\mathrm{F}}_{\mathrm{C}}$ defined
as:
\begin{equation}
\widehat{\mathrm{F}}_{\mathrm{C}}:=\underset{n-1\textrm{ times}}{\underbrace{\widehat{\mathrm{1}}\otimes\ldots\otimes\widehat{\mathrm{1}}}}\otimes\widehat{\mathrm{F}}^{c_{1}c_{2}\ldots c_{n}}=\widehat{\mathrm{1}}^{\otimes\left(n-1\right)}\otimes\widehat{\mathrm{F}}^{c_{1}c_{2}\ldots c_{n}},\label{eq:2.3.10}
\end{equation}
which transforms the input $\bigl|c_{1},c_{2},\ldots,c_{n}\bigr\rangle\otimes\bigl|t\bigr\rangle$
into the output:
\begin{equation}
\widehat{\mathrm{F}}_{\mathrm{C}}\bigl(\bigl|c_{1},c_{2},\ldots,c_{n}\bigr\rangle\otimes\bigl|t\bigr\rangle\bigr)=\bigl|c_{1},c_{2},\ldots,c_{n}\bigr\rangle\otimes\widehat{\mathrm{F}}^{c_{1}c_{2}\ldots c_{n}}\bigl|t\bigr\rangle.\label{eq:2.3.11}
\end{equation}
The transformation $\widehat{\mathrm{F}}$ is applied to $\bigl|t\bigr\rangle$
when $\bigl|c_{1},c_{2},\ldots,c_{n}\bigr\rangle=\bigl|1,1,\ldots,1\bigr\rangle$
or, otherwise, $\bigl|t\bigr\rangle$ remains invariant at the output.
Supplementary Figure 8 depicts the symbolic representation of $\widehat{\mathrm{F}}_{\mathrm{C}}$,
the same as that of a multi-controlled quantum gate for simplicity.
Nevertheless, in contrast to QC, note that $\widehat{\mathrm{F}}_{\mathrm{C}}$
will be a non-unitary operation when $\widehat{\mathrm{F}}\widehat{\mathrm{F}}^{\dagger}\neq\widehat{1}$,
that is, when $\widehat{\mathrm{F}}$ is a G- or M-gate.
\noindent \begin{center}
\includegraphics[width=9cm,height=4.5cm,keepaspectratio]{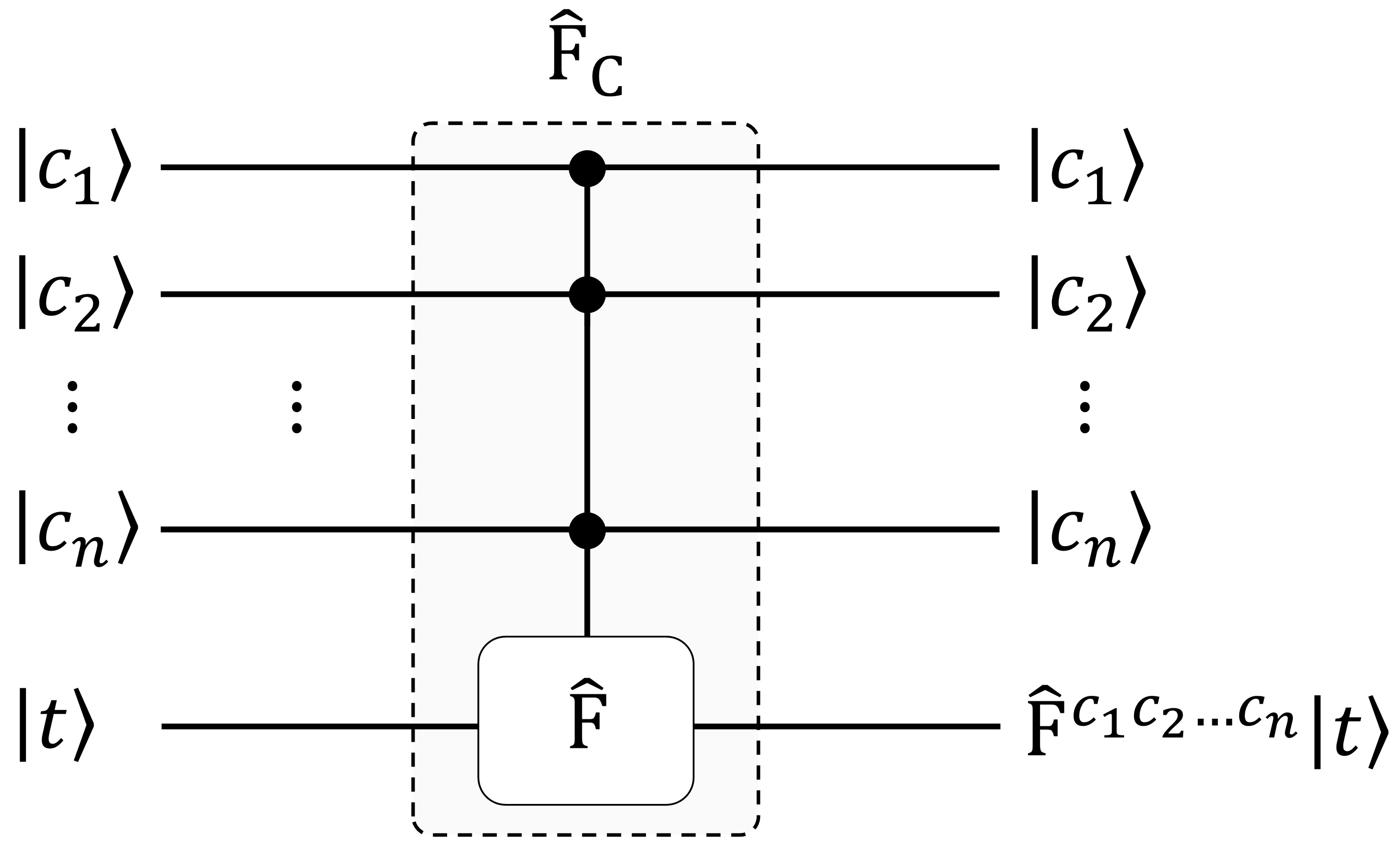}
\par\end{center}

\noindent \textbf{\small{}Supplementary Figure 8.}{\small{} Functional
scheme and symbolic representation of a controlled-$\widehat{\mathrm{F}}$
gate with $n$ control anbits $\bigl|c_{1,2,\ldots,n}\bigr\rangle\in\left\{ \bigl|0\bigr\rangle,\bigl|1\bigr\rangle\right\} $
and a single target anbit $\bigl|t\bigr\rangle$. Inspired in a multi-controlled
quantum gate, the single-anbit operation $\widehat{\mathrm{F}}$ (a
U-, G-, or M-gate) is applied to $\bigl|t\bigr\rangle$ when $\bigl|c_{1},c_{2},\ldots,c_{n}\bigr\rangle=\bigl|1,1,\ldots,1\bigr\rangle$
or, otherwise, $\bigl|t\bigr\rangle$ remains invariant at the output.\\}{\small \par}

Now, the Hilbert space is $\mathscr{E}_{n+1}=\mathscr{E}_{1}^{\otimes\left(n\right)}$
and the canonical (orthonormal) vector basis is found to be:
\begin{equation}
\mathcal{B}_{n+1}=\left\{ \bigl|k_{1}\bigr\rangle\otimes\bigl|k_{2}\bigr\rangle\otimes\ldots\otimes\bigl|k_{n+1}\bigr\rangle\right\} _{k_{1,\ldots,n+1}\in\left\{ 0,1\right\} }.\label{eq:2.3.12}
\end{equation}
Hence, the matrix representation of $\widehat{\mathrm{F}}_{\mathrm{C}}$
associated to $\mathcal{B}_{n+1}$ is:
\begin{equation}
F_{\mathrm{C}}=M_{\mathcal{B}_{n+1}}^{\mathcal{B}_{n+1}}\bigl(\widehat{\mathrm{F}}_{\mathrm{C}}\bigr)=\left(\begin{array}{c|c}
I & 0\\
\hline 0 & F
\end{array}\right),\label{eq:2.3.13}
\end{equation}
where $F_{\mathrm{C}}\in M_{2^{n+1}}\left(\mathbb{C}\right)$, $I$
is the identity matrix of size $\bigl(2^{n+1}-2\bigr)\times\bigl(2^{n+1}-2\bigr)$
and $F\in M_{2}\left(\mathbb{C}\right)$ is the matrix representation
of $\widehat{\mathrm{F}}$ associated to the canonical basis $\mathcal{B}_{1}=\left\{ \left|0\right\rangle ,\left|1\right\rangle \right\} $
of $\mathscr{E}_{1}$ (see Supplementary Equation \ref{eq:2.1.3}).
Moreover, the ket-bra representation of $\widehat{\mathrm{F}}_{\mathrm{C}}$
can be calculated from the entries of $F_{\mathrm{C}}$ in the same
way as in a controlled gate with a single control anbit (see p.\,\pageref{eq:2.3.5}).

\paragraph*{Properties.}

A multi-controlled anbit gate has the same properties as a controlled
gate with a single control anbit, see p.\,\pageref{par:Properties.}.
Nonetheless, now we work with endomorphisms $\widehat{\mathrm{F}}_{\mathrm{C}}$
of the Hilbert space $\mathscr{E}_{n+1}=\mathscr{E}_{1}^{\otimes\left(n\right)}$
and matrices $F_{\mathrm{C}}$ belonging to $\mathrm{U}\bigl(2^{n+1}\bigr)$,
$\mathrm{GL}\bigl(2^{n+1},\mathbb{C}\bigr)$ or $\mathfrak{gl}\bigl(2^{n+1},\mathbb{C}\bigr)$
when $F$ is a U-, G-, or M-gate, respectively.

In addition, it should be noted that the PIP implementation of a multi-controlled
gate can be designed as in a single-controlled gate: using an electro-optic
or an all-optical architecture. In the former case, the amplitudes
of the control anbits $\bigl|c_{k}\bigr\rangle=c_{k,0}\bigl|0\bigr\rangle+c_{k,1}\bigl|1\bigr\rangle$
($k\in\left\{ 1,\ldots,n\right\} $) are encoded by the electrical
control signals of the PIP platform and the amplitudes of the target
anbit $\bigl|t\bigr\rangle=t_{0}\bigl|0\bigr\rangle+t_{1}\bigl|1\bigr\rangle$
are encoded by a 2D optical wave (Supplementary Figure\,9a). In the
latter case, the amplitudes of $\bigl|c_{1},c_{2},\ldots,c_{n}\bigr\rangle\otimes\bigl|t\bigr\rangle$
associated to the vector basis $\mathcal{B}_{n+1}$ are encoded by
optical waves (Supplementary Figure 9b). Here, the reduced forward
transfer matrix of the system must be $F_{\mathrm{C}}$. Therefore,
we require to use $2^{n+1}-2$ waveguides to implement the submatrix
$I$ of $F_{\mathrm{C}}$ in combination with the MCA of the submatrix
$F$ (given by Fig.\,3b if $F$ is a U-gate or Fig.\,3c if $F$
is a G- or M-gate). 
\noindent \begin{center}
\includegraphics[width=15.4cm,height=8cm,keepaspectratio]{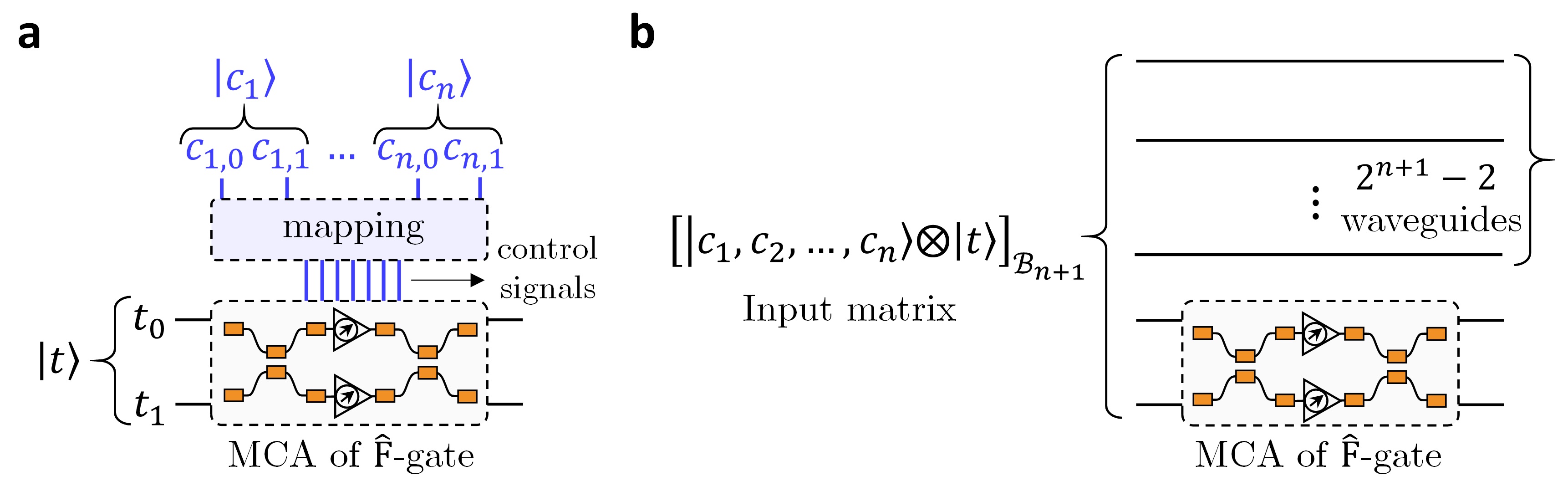}
\par\end{center}

\noindent \textbf{\small{}Supplementary Figure 9.}{\small{} PIP implementation
of a multi-controlled gate with $n$ control anbits $\bigl|c_{1}\bigr\rangle,\ldots,\bigl|c_{n}\bigr\rangle$
and a single target anbit $\bigl|t\bigr\rangle$, whose functionality
is shown in Supplementary Figure\,8. }\textbf{\small{}a}{\small{}
Electro-optic design. The PIP circuit is the MCA of the single-anbit
operation $\widehat{\mathrm{F}}$ associated to the target anbit (here
we depict the MCA of an M-gate to cover the general case). The basic
PIP devices are controlled by electrical signals (blue lines) mapped
with the amplitudes of the control anbits $\bigl|c_{k}\bigr\rangle=c_{k,0}\bigl|0\bigr\rangle+c_{k,1}\bigl|1\bigr\rangle$
($k\in\left\{ 1,\ldots,n\right\} $), e.g., via software \cite{key-S23}.
The optical inputs encode the amplitudes of $\bigl|t\bigr\rangle=t_{0}\bigl|0\bigr\rangle+t_{1}\bigl|1\bigr\rangle$
(black lines). }\textbf{\small{}b}{\small{} All-optical design. The
optical inputs encode the amplitudes of the tensor product $\bigl|c_{1},c_{2},\ldots,c_{n}\bigr\rangle\otimes\bigl|t\bigr\rangle$,
given by the matrix $\bigl[\bigl|c_{1},c_{2},\ldots,c_{n}\bigr\rangle\otimes\bigl|t\bigr\rangle\bigr]_{\mathcal{B}_{n+1}}$,
where $\left[\cdot\right]_{\mathcal{B}_{n+1}}$ is the component isomorphism
(see p.\,}\pageref{subsec:1.4}{\small{}). The whole structure implements
the matrix $F_{\mathrm{C}}$ given by Supplementary Equation }\ref{eq:2.3.13}{\small{}.\\}{\small \par}

As seen, the all-optical architecture entails a higher footprint than
that of the electro-optic design, which only requires to establish
a mapping between the control anbits and the electrical control signals
of the PIP circuit (e.g. via software \cite{key-S23}). In this fashion,
the same MCAs as those of the U-, G-, and M-gates (depicted in Fig.\,3
of the paper) may be employed to perform multi-controlled operations
of each kind of gate via an electro-optic design. 

\paragraph*{A basic example.}

So far, we have developed the theory of the multi-controlled gates
in abstract terms. To clarify these concepts, let us include an illustrative
example: the Toffoli (or CCNOT) gate, a multi-controlled operation
with 2 control anbits and 1 target anbit. Specifically, the matrix
of this gate is found to be:

\newpage{}

\begin{equation}
U_{\mathrm{CCN}}:=\left(\begin{array}{c|c}
I_{6\times6} & 0\\
\hline 0 & \sigma_{x}
\end{array}\right)=\left(\begin{array}{cccccccc}
1 & 0 & 0 & 0 & 0 & 0 & 0 & 0\\
0 & 1 & 0 & 0 & 0 & 0 & 0 & 0\\
0 & 0 & 1 & 0 & 0 & 0 & 0 & 0\\
0 & 0 & 0 & 1 & 0 & 0 & 0 & 0\\
0 & 0 & 0 & 0 & 1 & 0 & 0 & 0\\
0 & 0 & 0 & 0 & 0 & 1 & 0 & 0\\
0 & 0 & 0 & 0 & 0 & 0 & 0 & 1\\
0 & 0 & 0 & 0 & 0 & 0 & 1 & 0
\end{array}\right),\label{eq:2.3.14}
\end{equation}
where:
\begin{equation}
\sigma_{x}:=\left(\begin{array}{cc}
0 & 1\\
1 & 0
\end{array}\right),\label{eq:2.3.15}
\end{equation}
is a Pauli matrix, a single-anbit U-gate termed as the NOT gate within
the literature of QC \cite{key-S22}. Therefore, the matrix $F$ of
Supplementary Equation \ref{eq:2.3.13} is the Pauli matrix $\sigma_{x}$
in this example. Furthermore, the ket-bra representation can be calculated
from $U_{\mathrm{CCN}}$ as:
\begin{align}
\widehat{\mathrm{U}}_{\mathrm{CCN}} & =\bigl|0,0,0\bigr\rangle\bigl\langle0,0,0\bigr|+\bigl|0,0,1\bigr\rangle\bigl\langle0,0,1\bigr|+\bigl|0,1,0\bigr\rangle\bigl\langle0,1,0\bigr|+\bigl|0,1,1\bigr\rangle\bigl\langle0,1,1\bigr|\nonumber \\
 & +\bigl|1,0,0\bigr\rangle\bigl\langle1,0,0\bigr|+\bigl|1,0,1\bigr\rangle\bigl\langle1,0,1\bigr|+\bigl|1,1,0\bigr\rangle\bigl\langle1,1,1\bigr|+\bigl|1,1,1\bigr\rangle\bigl\langle1,1,0\bigr|\nonumber \\
 & =\bigl|0\bigr\rangle\bigl\langle0\bigr|\otimes\widehat{1}\otimes\widehat{1}+\bigl|1\bigr\rangle\bigl\langle1\bigr|\otimes\bigl|0\bigr\rangle\bigl\langle0\bigr|\otimes\widehat{1}+\bigl|1\bigr\rangle\bigl\langle1\bigr|\otimes\bigl|1\bigr\rangle\bigl\langle1\bigr|\otimes\widehat{\mathrm{\sigma}}_{x},\label{eq:2.3.16}
\end{align}
with $\widehat{\mathrm{\sigma}}_{x}=\bigl|0\bigr\rangle\bigl\langle1\bigr|+\bigl|1\bigr\rangle\bigl\langle0\bigr|$.
\noindent \begin{center}
\includegraphics[width=15.4cm,height=8cm,keepaspectratio]{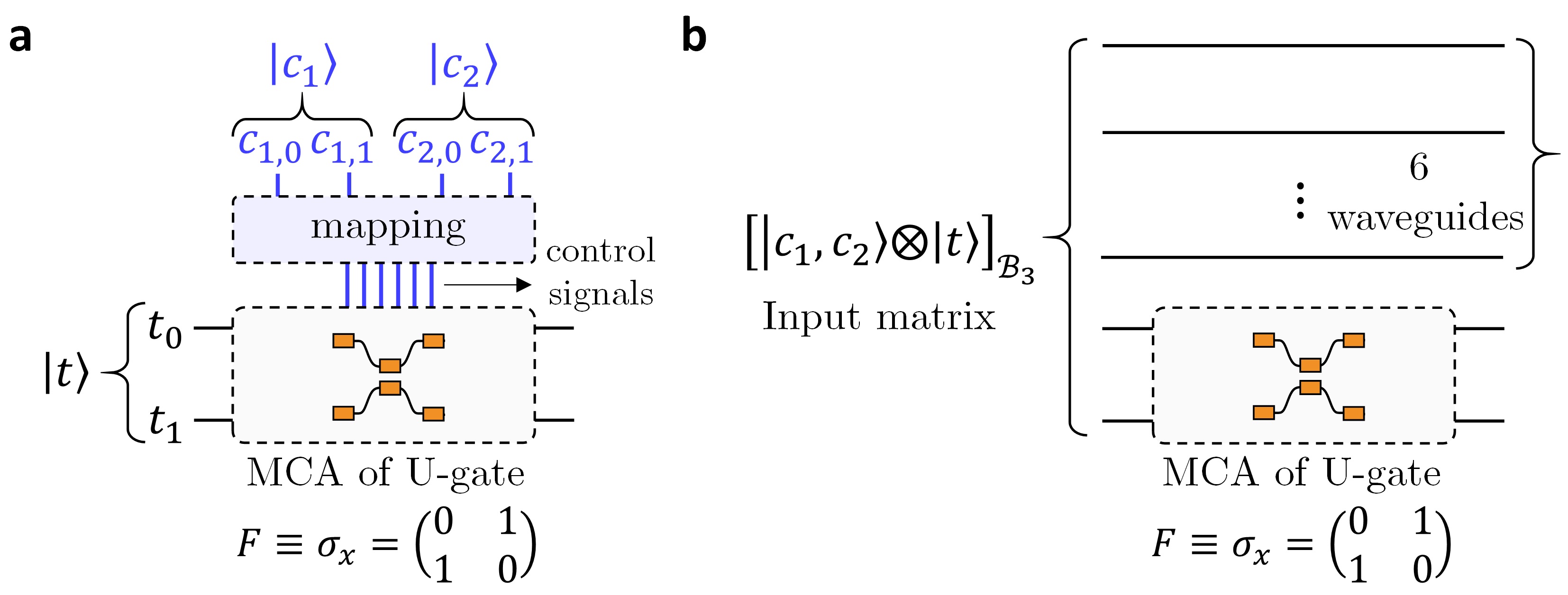}
\par\end{center}

\noindent \textbf{\small{}\label{Supplementary-Figure-10.}Supplementary
Figure 10.}{\small{} PIP implementation of the Toffoli (or CCNOT)
gate, a multi-controlled anbit gate with 2 control anbits $\bigl|c_{1}\bigr\rangle$,
$\bigl|c_{2}\bigr\rangle$ and 1 target anbit $\bigl|t\bigr\rangle$.}\textbf{\small{}
a}{\small{} Electro-optic design. The PIP circuit is the MCA of a
U-gate since $F\equiv\sigma_{x}$ is a $2\times2$ unitary matrix.
}\textbf{\small{}b}{\small{} All-optical design implementing the matrix
$U_{\mathrm{CCN}}$ given by Supplementary Equation }\ref{eq:2.3.14}{\small{}.\\}{\small \par}

The PIP implementation of the CCNOT gate can be carried out via an
electro-optic or an all-optical design, see Supplementary Figure 10.
Both architectures are the same as those of Supplementary Figure 9,
but restricted to the case of 2 control anbits and using the MCA of
a U-gate to induce the $\sigma_{x}$-matrix transformation on the
amplitudes of $\bigl|t\bigr\rangle$. It is worth mentioning that
Supplementary Figure 10a is the same circuit as that of Fig. 4d of
the paper, but including an additional control anbit $\bigl|c_{2}\bigr\rangle$
in the electrical domain. This result demonstrates the possibility
of \emph{scaling} the controlled gates to the case of multiple control
anbits without requiring extra PIP devices in the optical circuits,
a fundamental feature of APC that cannot be found in optical QC.

Finally, in Supplementary Note 5, we report how to use the Toffoli
gate to implement the Boolean logic of digital computation in APC,
see p.\,\pageref{sec:5}.

\newpage{}

\section*{Supplementary Note 3: sequential design\label{sec:3}}

\addcontentsline{toc}{section}{Supplementary Note 3: sequential design}

\noindent Here, we include additional information about the fundamental
pieces required to construct sequential APC architectures: the fan-in
(FI) and fan-out (FO) gates. Furthermore, the two sequential systems
depicted in Fig.\,5b and 5d of the main text are discussed in more
detail. Both examples will allow us to establish the theoretical strategies
to analyse and design sequential architectures in APC.

\subsection*{3.1 Fan-in and fan-out gates}

\addcontentsline{toc}{subsection}{3.1 Fan-in and fan-out gates}

\noindent In this subsection, we first study the definition and mathematical
properties of the FI and FO gates and, subsequently, we propose a
common PIP implementation for both multi-anbit operations.

\subsubsection*{3.1.1 Fan-in: anbit addition }

\addcontentsline{toc}{subsubsection}{3.1.1  Fan-in: anbit addition}

\paragraph*{Definition and formalism.}

Consider the single-anbit Hilbert space $\mathscr{E}_{1}$. Let us
define the FI operation (anbit addition) as the linear mapping $\widehat{\mathrm{G}}_{\mathrm{FI}}$
of $\mathscr{E}_{2}:=\mathscr{E}_{1}\times\mathscr{E}_{1}$ which
transforms the input $\bigl(\bigl|\psi\bigr\rangle,\bigl|\varphi\bigr\rangle\bigr)$
into the output:
\begin{equation}
\widehat{\mathrm{G}}_{\mathrm{FI}}\left(\bigl|\psi\bigr\rangle,\bigl|\varphi\bigr\rangle;n,m\right):=\bigl(n\bigl(\bigl|\psi\bigr\rangle+\bigl|\varphi\bigr\rangle\bigr),m\bigl(\bigl|\psi\bigr\rangle-\bigl|\varphi\bigr\rangle\bigr)\bigr),\label{eq:3.1.1}
\end{equation}
where $n,m\in\mathbb{C}-\left\{ 0\right\} $ are parameters of the
operation ($n=m=1$ will be assumed as the \emph{default values} and,
in such a case, the parameters can be omitted on the left-hand side
of the above equation).\textcolor{red}{{} }Supplementary Figure 11 shows
the functional scheme and symbolic representation of the FI gate.
In most sequential APC architectures, the second output anbit $m\bigl(\bigl|\psi\bigr\rangle-\bigl|\varphi\bigr\rangle\bigr)$
will be considered as a \emph{garbage} anbit and only the first output
anbit $n\bigl(\bigl|\psi\bigr\rangle+\bigl|\varphi\bigr\rangle\bigr)$
will be utilised to build a feedback loop. Nevertheless, the aim of
defining two output anbits is that, in this way, we will be able to
describe the FI gate with a square matrix, see below.
\noindent \begin{center}
\includegraphics[width=8cm,height=2.8cm,keepaspectratio]{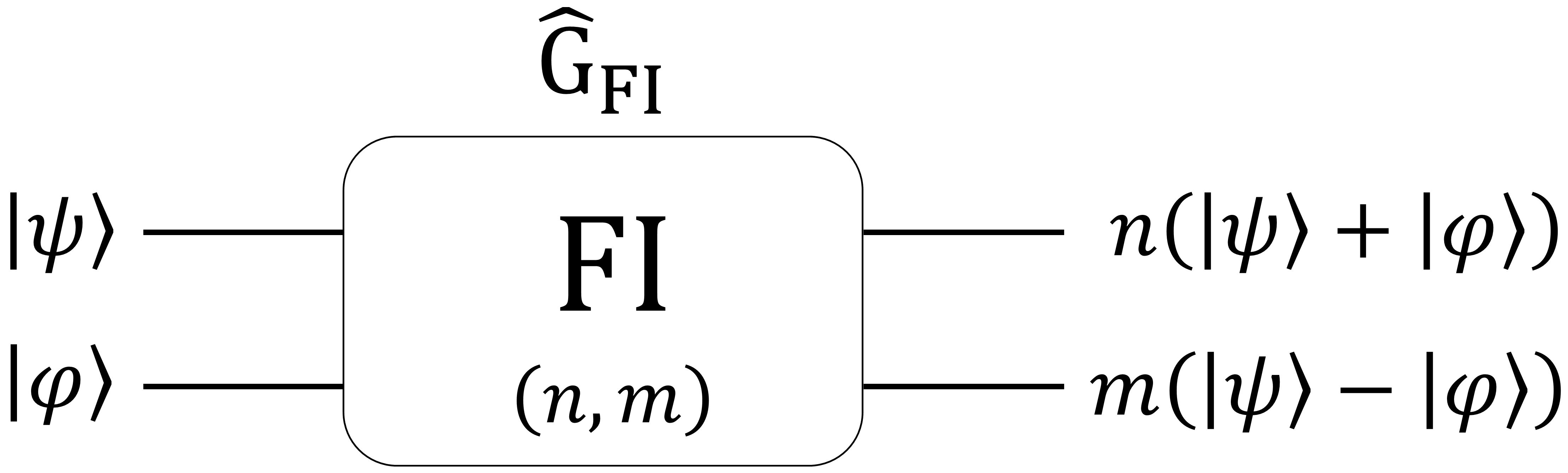}
\par\end{center}

\noindent \textbf{\small{}Supplementary Figure 11.}{\small{} Functional
scheme and symbolic representation of the fan-in (FI) gate. The values
of the (complex) parameters $n,m$ should be indicate, except for
the case $n=m=1$, which are defined as the default values.\\}{\small \par}

Taking into account that the FI gate is an operation of two input
and output anbits, the underlying Hilbert space is $\mathscr{E}_{2}$
and the canonical (orthonormal) basis is given by Supplementary Equation
\ref{eq:1.5.12}, reproduced here for clarity:
\begin{equation}
\mathcal{B}_{2}=\left\{ \left(\bigl|0\bigr\rangle,\bigl|\mathbf{0}\bigr\rangle\right),\left(\bigl|1\bigr\rangle,\bigl|\mathbf{0}\bigr\rangle\right),\left(\bigl|\mathbf{0}\bigr\rangle,\bigl|0\bigr\rangle\right),\left(\bigl|\mathbf{0}\bigr\rangle,\bigl|1\bigr\rangle\right)\right\} .\label{eq:3.1.2}
\end{equation}
The matrix representation of $\widehat{\mathrm{G}}_{\mathrm{FI}}$
associated to $\mathcal{B}_{2}$ is:
\begin{align}
G_{\mathrm{FI}} & =M_{\mathcal{B}_{2}}^{\mathcal{B}_{2}}\bigl(\widehat{\mathrm{G}}_{\mathrm{FI}}\bigr)\nonumber \\
 & =\left(\begin{array}{cccc}
\bigl[\widehat{\mathrm{G}}_{\mathrm{FI}}\left(\bigl|0\bigr\rangle,\bigl|\mathbf{0}\bigr\rangle\right)\bigr]_{\mathcal{B}_{2}} & \bigl[\widehat{\mathrm{G}}_{\mathrm{FI}}\left(\bigl|1\bigr\rangle,\bigl|\mathbf{0}\bigr\rangle\right)\bigr]_{\mathcal{B}_{2}} & \bigl[\widehat{\mathrm{G}}_{\mathrm{FI}}\left(\bigl|\mathbf{0}\bigr\rangle,\bigl|0\bigr\rangle\right)\bigr]_{\mathcal{B}_{2}} & \bigl[\widehat{\mathrm{G}}_{\mathrm{FI}}\left(\bigl|\mathbf{0}\bigr\rangle,\bigl|1\bigr\rangle\right)\bigr]_{\mathcal{B}_{2}}\end{array}\right)\nonumber \\
 & =\left(\begin{array}{cccc}
\bigl[\left(n\bigl|0\bigr\rangle,m\bigl|0\bigr\rangle\right)\bigr]_{\mathcal{B}_{2}} & \bigl[\left(n\bigl|1\bigr\rangle,m\bigl|1\bigr\rangle\right)\bigr]_{\mathcal{B}_{2}} & \bigl[\left(n\bigl|0\bigr\rangle,-m\bigl|0\bigr\rangle\right)\bigr]_{\mathcal{B}_{2}} & \bigl[\left(n\bigl|1\bigr\rangle,-m\bigl|1\bigr\rangle\right)\bigr]_{\mathcal{B}_{2}}\end{array}\right)\nonumber \\
 & =\left(\begin{array}{c|c}
nI & nI\\
\hline mI & -mI
\end{array}\right).\label{eq:3.1.3}
\end{align}
Thus, using the component isomorphism $\left[\cdot\right]_{\mathcal{B}_{2}}$,
Supplementary Equation \ref{eq:3.1.1} can be recast as:
\begin{equation}
\left[\bigl(n\bigl(\bigl|\psi\bigr\rangle+\bigl|\varphi\bigr\rangle\bigr),m\bigl(\bigl|\psi\bigr\rangle-\bigl|\varphi\bigr\rangle\bigr)\bigr)\right]_{\mathcal{B}_{2}}=G_{\mathrm{FI}}\left[\bigl(\bigl|\psi\bigr\rangle,\bigl|\varphi\bigr\rangle\bigr)\right]_{\mathcal{B}_{2}},\label{eq:3.1.4}
\end{equation}
leading to the following input-output matrix relation:
\begin{equation}
\left(\begin{array}{c}
n\left(\psi_{0}+\varphi_{0}\right)\\
n\left(\psi_{1}+\varphi_{1}\right)\\
m\left(\psi_{0}-\varphi_{0}\right)\\
m\left(\psi_{1}-\varphi_{1}\right)
\end{array}\right)=\left(\begin{array}{c|c}
\begin{array}{cc}
n & 0\\
0 & n
\end{array} & \begin{array}{cc}
n & 0\\
0 & n
\end{array}\\
\hline \begin{array}{cc}
m & 0\\
0 & m
\end{array} & \begin{array}{cc}
-m & 0\\
0 & -m
\end{array}
\end{array}\right)\left(\begin{array}{c}
\psi_{0}\\
\psi_{1}\\
\varphi_{0}\\
\varphi_{1}
\end{array}\right).\label{eq:3.1.5}
\end{equation}
Setting $n=m=1$, the above expression describes the computational
system depicted in Fig.\,5a of the paper. 

\label{Cartesian vs tensor product in FI}As seen, defining the FI
operation via the \emph{Cartesian product}, we are able to: (\emph{i})
work with (square) matrices providing a linear nature to this gate,
(\emph{ii}) independently transform the anbit amplitudes $\psi_{0}$,
$\psi_{1}$, $\varphi_{0}$ and $\varphi_{1}$. Alternatively, we
could define the FI gate via the \emph{tensor product}. Nonetheless,
in such a scenario, it is worthy to highlight that this version of
the FI gate would have a nonlinear nature because we could not describe
this multi-anbit operation via a matrix. The proof of this statement
can be done in three steps. Firstly, note that the definition of the
FI gate via the tensor product would be of the form: 
\begin{equation}
\widehat{\mathrm{G}}_{\mathrm{FI}}\bigl|\psi,\varphi\bigr\rangle:=\bigl|\psi+\varphi\bigr\rangle\otimes\bigl|\psi-\varphi\bigr\rangle=\sum_{k,l}\bigl(\psi_{k}+\varphi_{k}\bigr)\bigl(\psi_{l}-\varphi_{l}\bigr)\bigl|k,l\bigr\rangle.\label{eq:3.1.6}
\end{equation}
Secondly, we should derive the ``apparent'' matrix representation
of $\widehat{\mathrm{G}}_{\mathrm{FI}}$ associated to the canonical
basis $\mathcal{B}_{2}$ of $\mathscr{E}_{1}\otimes\mathscr{E}_{1}$
(given by Supplementary Equation \ref{eq:1.5.4}):
\begin{equation}
G_{\mathrm{FI}}=M_{\mathcal{B}_{2}}^{\mathcal{B}_{2}}\bigl(\widehat{\mathrm{G}}_{\mathrm{FI}}\bigr)=\bigl(\begin{array}{ccc}
\cdots & \bigl[\widehat{\mathrm{G}}_{\mathrm{FI}}\bigl|k,l\bigr\rangle\bigr]_{\mathcal{B}_{2}} & \cdots\end{array}\bigr)=\left(\begin{array}{cccc}
0 & 0 & 0 & 0\\
0 & -1 & 0 & 0\\
0 & 0 & -1 & 0\\
0 & 0 & 0 & 0
\end{array}\right).\label{eq:3.1.7}
\end{equation}
Thirdly, we should compare the output generated by $\widehat{\mathrm{G}}_{\mathrm{FI}}$
with the output generated by $G_{\mathrm{FI}}$. The operator $\widehat{\mathrm{G}}_{\mathrm{FI}}$
leads to the output described by Supplementary Equation \ref{eq:3.1.6},
whose components associated to $\mathcal{B}_{2}$ are:
\begin{equation}
\bigl[\widehat{\mathrm{G}}_{\mathrm{FI}}\bigl|\psi,\varphi\bigr\rangle\bigr]_{\mathcal{B}_{2}}=\left(\begin{array}{c}
\bigl(\psi_{0}+\varphi_{0}\bigr)\bigl(\psi_{0}-\varphi_{0}\bigr)\\
\bigl(\psi_{0}+\varphi_{0}\bigr)\bigl(\psi_{1}-\varphi_{1}\bigr)\\
\bigl(\psi_{1}+\varphi_{1}\bigr)\bigl(\psi_{0}-\varphi_{0}\bigr)\\
\bigl(\psi_{1}+\varphi_{1}\bigr)\bigl(\psi_{1}-\varphi_{1}\bigr)
\end{array}\right).\label{eq:3.1.8}
\end{equation}
However, the matrix $G_{\mathrm{FI}}$ gives rise to a different output:
\begin{equation}
G_{\mathrm{FI}}\left[\bigl|\psi,\varphi\bigr\rangle\right]_{\mathcal{B}_{2}}=\left(\begin{array}{cccc}
0 & 0 & 0 & 0\\
0 & -1 & 0 & 0\\
0 & 0 & -1 & 0\\
0 & 0 & 0 & 0
\end{array}\right)\left(\begin{array}{c}
\psi_{0}\varphi_{0}\\
\psi_{0}\varphi_{1}\\
\psi_{1}\varphi_{0}\\
\psi_{1}\varphi_{1}
\end{array}\right)=\left(\begin{array}{c}
0\\
-\psi_{0}\varphi_{1}\\
-\psi_{1}\varphi_{0}\\
0
\end{array}\right).\label{eq:3.1.9}
\end{equation}
Consequently, $\widehat{\mathrm{G}}_{\mathrm{FI}}$ has no matrix
representation. This implies that the global output (Supplementary
Equation \ref{eq:3.1.8}) is different from the linear superposition
of the outputs generated by each vector of the basis $\mathcal{B}_{2}$
(Supplementary Equation \ref{eq:3.1.9}). In other words, $\widehat{\mathrm{G}}_{\mathrm{FI}}$
is a nonlinear mapping of $\mathscr{E}_{1}\otimes\mathscr{E}_{1}$.
For this reason, the FI gate is defined by using the Cartesian product
as indicated by Supplementary Equation \ref{eq:3.1.1}.

\paragraph*{Properties.}

The FI gate illustrated in Supplementary Figure 11 exhibits the following
properties:
\begin{enumerate}
\item \emph{Automorphism}. The linear operator $\widehat{\mathrm{G}}_{\mathrm{FI}}$
is an automorphism of the Hilbert space $\mathscr{E}_{2}=\mathscr{E}_{1}\times\mathscr{E}_{1}$,
i.e., a bijective endomorphism given that the FI gate is reversible,
as demonstrated in property 2.
\item \emph{Reversibility}. From Supplementary Equation \ref{eq:3.1.3},
it follows that:
\begin{equation}
\det\bigl(G_{\mathrm{FI}}\bigr)=4n^{2}m^{2}\neq0,\label{eq:3.1.10}
\end{equation}
that is, the FI gate is reversible.
\item \emph{Commutativity}. The operation $\widehat{\mathrm{G}}_{\mathrm{FI}}$
holds the commutative property:
\begin{equation}
\widehat{\mathrm{G}}_{\mathrm{FI}}\left(\bigl|\psi\bigr\rangle,\bigl|\varphi\bigr\rangle;n,m\right)=\widehat{\mathrm{G}}_{\mathrm{FI}}\left(\bigl|\varphi\bigr\rangle,\bigl|\psi\bigr\rangle;n,-m\right).\label{eq:3.1.11}
\end{equation}
The order of the input anbits can be commuted preserving the state
of the output anbits, provided that the sign of the second parameter
is inverted.
\item \emph{FI-FO connection}. The FO operation (defined below) of the anbit
$\bigl|\psi\bigr\rangle$ can be performed by the FI gate setting
$\bigl|\varphi\bigr\rangle=\bigl|\mathbf{0}\bigr\rangle$ and $n,m\in\mathbb{R}^{+}$
in Supplementary Equation \ref{eq:3.1.1}. Moreover, a perfect cloning
of $\bigl|\psi\bigr\rangle$ will be carried out taking $n=m=1$.
\item \emph{Algebraic structure}. If $\left|n\right|^{2}=\left|m\right|^{2}=1/2$,
then $G_{\mathrm{FI}}\in\mathrm{U}\left(4\right)$ and FI is a 2-anbit
U-gate. Otherwise, $G_{\mathrm{FI}}\in\mathrm{GL}\left(4,\mathbb{C}\right)$
and FI is a 2-anbit G-gate.
\item \emph{\label{enu:Universality-and-PIP}Universality and PIP implementation}.
A universal matrix of the FI gate is of the form given by Supplementary
Equation \ref{eq:3.1.3}. A PIP implementation of such a matrix preserves
the universality if the values of the parameters $n$ and $m$ may
be optically tuned. Along this line, note that when $n=m$ then $G_{\mathrm{FI}}=G_{\mathrm{FI}}^{T}$
and, therefore, if the circuit preserves the Lorentz reciprocity,
we may conclude that the circuit will also preserve the FB symmetry
(as inferred from comment 6 on page \pageref{eq:2.1.10}). Remarkably,
the PIP implementation of the FI gate depicted in Fig.\,5a and Supplementary
Figure 13 satisfies this property.
\end{enumerate}

\subsubsection*{3.1.2 Fan-out: anbit cloning }

\addcontentsline{toc}{subsubsection}{3.1.2  Fan-out: anbit cloning}

\paragraph*{Definition and formalism.}

Consider the single-anbit Hilbert space $\mathscr{E}_{1}$. Let us
define the FO operation (anbit cloning) as the linear mapping $\widehat{\mathrm{M}}_{\mathrm{FO}}$
of $\mathscr{E}_{2}:=\mathscr{E}_{1}\times\mathscr{E}_{1}$ which
transforms the input $\bigl(\bigl|\psi\bigr\rangle,\bigl|\varphi\bigr\rangle\bigr)$
into the output:
\begin{equation}
\widehat{\mathrm{M}}_{\mathrm{FO}}\left(\bigl|\psi\bigr\rangle,\bigl|\varphi\bigr\rangle;n,m\right):=\bigl(n\bigl|\psi\bigr\rangle,m\bigl|\psi\bigr\rangle\bigr),\label{eq:3.1.12}
\end{equation}
where $n,m\in\mathbb{R}^{+}$ are parameters of the operation ($n=m=1$
will be assumed as the \emph{default values} and, in such a case,
the parameters can be omitted on the left-hand side of the above equation).
Here, note that $\bigl|\varphi\bigr\rangle$ is an \emph{ancilla}
anbit since we are only interested in cloning a single input anbit
($\bigl|\psi\bigr\rangle$). Nonetheless, the aim of defining two
input anbits is that, in this vein, we will be able to describe the
FO operation via a square matrix, see below. Supplementary Figure
12 shows the functional scheme and symbolic representation of the
FO gate. 
\noindent \begin{center}
\includegraphics[width=7.5cm,height=2.6cm,keepaspectratio]{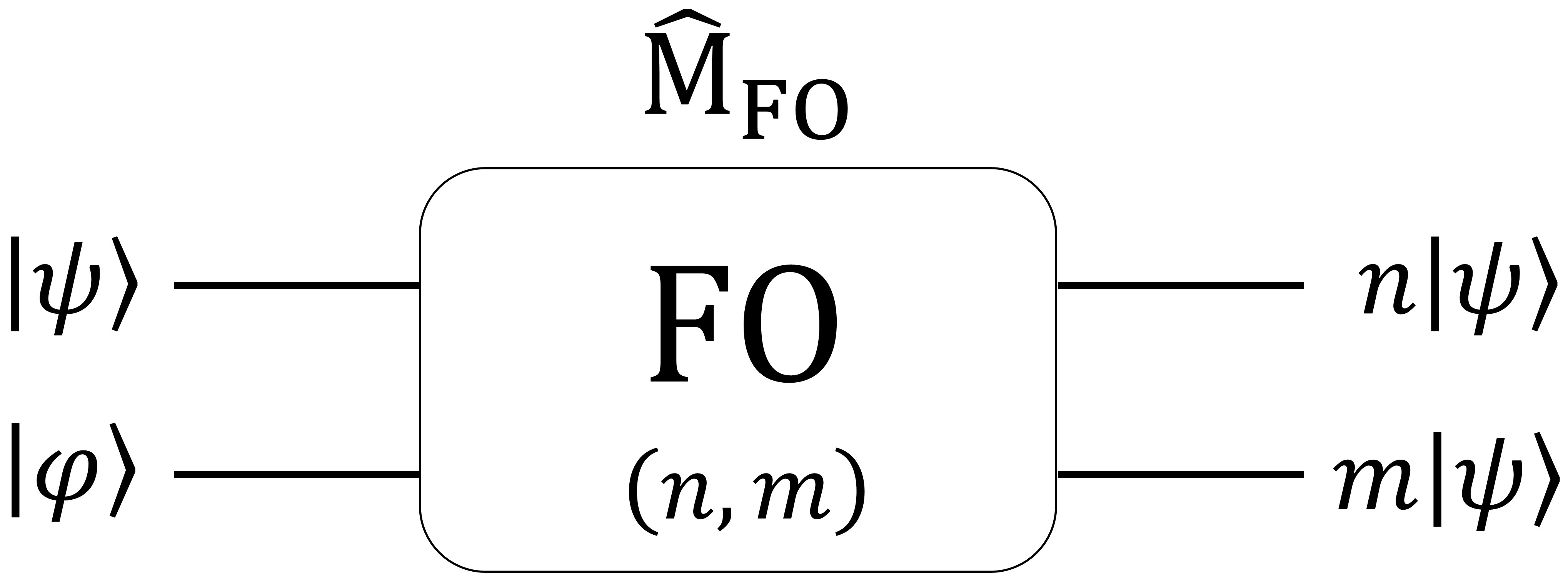}
\par\end{center}

\noindent \textbf{\small{}Supplementary Figure 12.}{\small{} Functional
scheme and symbolic representation of the fan-out (FO) gate. The values
of the (positive real) parameters $n,m$ should be indicate, except
for the case $n=m=1$, which are defined as the default values and
give rise to a perfect cloning of the input anbit $\bigl|\psi\bigr\rangle$.\\}{\small \par}

The goal of the FO gate is to introduce an operation in APC that allows
us to duplicate the state of an anbit with the same norm (perfect
cloning, $n=m=1$) or a different norm (imperfect cloning, $n\neq1$
or $m\neq1$), but without introducing an additional phase. To this
end, the parameters $n$ and $m$ must be assumed positive real constants.

Bearing in mind that the FO gate is an operation of two input and
output anbits, the underlying Hilbert space is $\mathscr{E}_{2}$
and the canonical (orthonormal) basis is $\mathcal{B}_{2}$, given
by Supplementary Equation \ref{eq:3.1.2}. In this scenario, the matrix
representation of $\widehat{\mathrm{M}}_{\mathrm{FO}}$ associated
to $\mathcal{B}_{2}$ is:
\begin{align}
M_{\mathrm{FO}} & =M_{\mathcal{B}_{2}}^{\mathcal{B}_{2}}\bigl(\widehat{\mathrm{M}}_{\mathrm{FO}}\bigr)\nonumber \\
 & =\left(\begin{array}{cccc}
\bigl[\widehat{\mathrm{M}}_{\mathrm{FO}}\left(\bigl|0\bigr\rangle,\bigl|\mathbf{0}\bigr\rangle\right)\bigr]_{\mathcal{B}_{2}} & \bigl[\widehat{\mathrm{M}}_{\mathrm{FO}}\left(\bigl|1\bigr\rangle,\bigl|\mathbf{0}\bigr\rangle\right)\bigr]_{\mathcal{B}_{2}} & \bigl[\widehat{\mathrm{M}}_{\mathrm{FO}}\left(\bigl|\mathbf{0}\bigr\rangle,\bigl|0\bigr\rangle\right)\bigr]_{\mathcal{B}_{2}} & \bigl[\widehat{\mathrm{M}}_{\mathrm{FO}}\left(\bigl|\mathbf{0}\bigr\rangle,\bigl|1\bigr\rangle\right)\bigr]_{\mathcal{B}_{2}}\end{array}\right)\nonumber \\
 & =\left(\begin{array}{cccc}
\bigl[\left(n\bigl|0\bigr\rangle,m\bigl|0\bigr\rangle\right)\bigr]_{\mathcal{B}_{2}} & \bigl[\left(n\bigl|1\bigr\rangle,m\bigl|1\bigr\rangle\right)\bigr]_{\mathcal{B}_{2}} & \bigl[\left(\bigl|\mathbf{0}\bigr\rangle,\bigl|\mathbf{0}\bigr\rangle\right)\bigr]_{\mathcal{B}_{2}} & \bigl[\left(\bigl|\mathbf{0}\bigr\rangle,\bigl|\mathbf{0}\bigr\rangle\right)\bigr]_{\mathcal{B}_{2}}\end{array}\right)\nonumber \\
 & =\left(\begin{array}{c|c}
nI & 0\\
\hline mI & 0
\end{array}\right).\label{eq:3.1.13}
\end{align}
Hence, applying the component isomorphism $\left[\cdot\right]_{\mathcal{B}_{2}}$
to Supplementary Equation \ref{eq:3.1.12}, the input-output relation
of the FO gate can be expressed by the matrix relation:
\begin{equation}
\left(\begin{array}{c}
n\psi_{0}\\
n\psi_{1}\\
m\psi_{0}\\
m\psi_{1}
\end{array}\right)=\left(\begin{array}{c|c}
\begin{array}{cc}
n & 0\\
0 & n
\end{array} & \begin{array}{cc}
0 & 0\\
0 & 0
\end{array}\\
\hline \begin{array}{cc}
m & 0\\
0 & m
\end{array} & \begin{array}{cc}
0 & 0\\
0 & 0
\end{array}
\end{array}\right)\left(\begin{array}{c}
\psi_{0}\\
\psi_{1}\\
\varphi_{0}\\
\varphi_{1}
\end{array}\right).\label{eq:3.1.14}
\end{equation}
It should be noted that the third and fourth columns of $M_{\mathrm{FO}}$
are found to be null as a direct consequence that the FO gate must
clone $\bigl|\psi\bigr\rangle$ at the output for any ancilla anbit
$\bigl|\varphi\bigr\rangle$ at the input. The main drawback of a
matrix with two null columns is that its optical implementation will
be restricted to a few number of PIP architectures. 

In order to circumvent this technological limitation, let us take
$\bigl|\varphi\bigr\rangle=\bigl|\mathbf{0}\bigr\rangle$. As a result,
the entries of the third and fourth columns of $M_{\mathrm{FO}}$
can be regarded as degrees of freedom ($\alpha_{kl}\in\mathbb{C}$
with $k\in\left\{ 1,\ldots,4\right\} $ and $l\in\left\{ 3,4\right\} $):
\begin{equation}
\left(\begin{array}{c}
n\psi_{0}\\
n\psi_{1}\\
m\psi_{0}\\
m\psi_{1}
\end{array}\right)=\left(\begin{array}{c|c}
\begin{array}{cc}
n & 0\\
0 & n
\end{array} & \begin{array}{cc}
\alpha_{13} & \alpha_{14}\\
\alpha_{23} & \alpha_{24}
\end{array}\\
\hline \begin{array}{cc}
m & 0\\
0 & m
\end{array} & \begin{array}{cc}
\alpha_{33} & \alpha_{34}\\
\alpha_{43} & \alpha_{44}
\end{array}
\end{array}\right)\left(\begin{array}{c}
\psi_{0}\\
\psi_{1}\\
0\\
0
\end{array}\right).\label{eq:3.1.15}
\end{equation}
Remarkably, this new version of $M_{\mathrm{FO}}$ can be implemented
by a larger gamut of PIP circuits than in Supplementary Equation \ref{eq:3.1.14}.
As a by-product, the same optical implementation of the FI gate could
also be employed to implement $M_{\mathrm{FO}}$, leading to a common
PIP architecture for both operations (see below p.\,\pageref{subsec:3.3.3-PIP-implementation}).

Finally, we should discuss the importance of defining the FO gate
by using the \emph{Cartesian product}. In the same way as in the FI
operation, the Cartesian product provides a linear nature to the FO
gate. Alternatively, we could define the FO gate via the \emph{tensor
product}. Nevertheless, in such a case, the FO gate would have a nonlinear
nature. The proof of this statement may be performed in two steps.
Firstly, note that the construction of the FO gate via the tensor
product would be a mapping of $\mathscr{E}_{1}\otimes\mathscr{E}_{1}$
defined as $\widehat{\mathrm{M}}_{\mathrm{FO}}\bigl|\psi,\varphi\bigr\rangle:=\bigl|\psi,\psi\bigr\rangle$.
Secondly, we can directly verify that this mapping has a nonlinear
nature using \emph{reductio ad absurdum} (i.e. we assume a linear
behaviour to find an inconsistent conclusion):
\begin{align}
\widehat{\mathrm{M}}_{\mathrm{FO}}\bigl|\psi,\varphi\bigr\rangle & =\widehat{\mathrm{M}}_{\mathrm{FO}}\bigl(\psi_{0}\bigl|0,\varphi\bigr\rangle+\psi_{1}\bigl|1,\varphi\bigr\rangle\bigr)\nonumber \\
 & =\psi_{0}\widehat{\mathrm{M}}_{\mathrm{FO}}\bigl|0,\varphi\bigr\rangle+\psi_{1}\widehat{\mathrm{M}}_{\mathrm{FO}}\bigl|1,\varphi\bigr\rangle\nonumber \\
 & =\psi_{0}\bigl|0,0\bigr\rangle+\psi_{1}\bigl|1,1\bigr\rangle\neq\bigl|\psi,\psi\bigr\rangle.\label{eq:3.1.16}
\end{align}
The nonlinear behaviour of this version of the FO operation (defined
via the tensor product) has extensively been discussed in QC within
the context of the no-cloning theorem \cite{key-S24}.

\paragraph*{Properties.}

As commented above, any $4\times4$ matrix of the form:
\begin{equation}
M_{\mathrm{FO}}=\left(\begin{array}{c|c}
nI & M_{12}\\
\hline mI & M_{22}
\end{array}\right),\label{eq:3.1.17}
\end{equation}
with $M_{12}$ and $M_{22}$ being $2\times2$ arbitrary submatrices,
is able to carry out an FO operation of the input anbit $\bigl|\psi\bigr\rangle$
when the ancilla anbit is taken to be null $\bigl|\varphi\bigr\rangle=\bigl|\mathbf{0}\bigr\rangle$.
Since this matrix is more general than Supplementary Equation \ref{eq:3.1.13},
let us study its properties:
\begin{enumerate}
\item \emph{Endomorphism} \emph{family and FI-FO connection}. Specifically,
$M_{\mathrm{FO}}$ describes a family of endomorphisms of $\mathscr{E}_{2}=\mathscr{E}_{1}\times\mathscr{E}_{1}$.
In fact, note that the FI gate (see Supplementary Equation \ref{eq:3.1.3})
is also described by $M_{\mathrm{FO}}$ taking $M_{12}\equiv nI$
and $M_{22}\equiv-mI$. 
\item \emph{Reversibility}. Given that $\det\bigl(M_{\mathrm{FO}}\bigr)$
depends on the value of the submatrices $M_{12}$ and $M_{22}$, we
conclude that the FO gate may be reversible ($\det\bigl(M_{\mathrm{FO}}\bigr)\neq0$)
or irreversible ($\det\bigl(M_{\mathrm{FO}}\bigr)=0$).
\item \emph{Non-commutativity}. In contrast to the FI operation, the order
of the input anbits cannot be commuted in the FO gate.
\item \emph{Algebraic structure}. From the expressions $M_{\mathrm{FO}}^{\dagger}M_{\mathrm{FO}}=I$
and $M_{\mathrm{FO}}M_{\mathrm{FO}}^{\dagger}=I$, we infer that $M_{\mathrm{FO}}\in\mathrm{U}\left(4\right)$
(a 2-anbit U-gate) if and only if the next necessary and sufficient
conditions are fulfilled: (1)\,$n^{2}+m^{2}=1$, (2)\,$nM_{12}+mM_{22}=0$,
(3)\,$M_{12}^{\dagger}M_{12}+M_{22}^{\dagger}M_{22}=I$,\linebreak{}
(4) $M_{12}M_{12}^{\dagger}=\bigl(1-n^{2}\bigr)I$, (5) $M_{22}M_{12}^{\dagger}=-nmI$,
(6) $M_{22}M_{22}^{\dagger}=\bigl(1-m^{2}\bigr)I$. For instance,
$n=m=1/\sqrt{2}$ and $M_{12}=-M_{22}=\bigl(1/\sqrt{2}\bigr)I$ satisfy
the above conditions. On the other hand, $M_{\mathrm{FO}}\in\mathrm{GL}\left(4,\mathbb{C}\right)$
(a 2-anbit G-gate) by taking parameters $n,m$ and submatrices $M_{12}$
and $M_{22}$ guaranteeing that $\det\bigl(M_{\mathrm{FO}}\bigr)\neq0$.
As an example, $n=m=1$ and $M_{12}=-M_{22}=I$ satisfy these conditions.
Finally, the cases where $\det\bigl(M_{\mathrm{FO}}\bigr)=0$ correspond
to an irreversible 2-anbit M-gate.
\end{enumerate}

\subsubsection*{3.1.3 PIP implementation: a common architecture\label{subsec:3.3.3-PIP-implementation}}

\addcontentsline{toc}{subsubsection}{3.1.3  PIP implementation: a common architecture}

\noindent From the study of the FI and FO gates, we know that both
operations are connected via the equation:
\begin{equation}
\widehat{\mathrm{M}}_{\mathrm{FO}}\left(\bigl|\psi\bigr\rangle,\bigl|\varphi\bigr\rangle;n,m\right)=\widehat{\mathrm{G}}_{\mathrm{FI}}\left(\bigl|\psi\bigr\rangle,\bigl|\mathbf{0}\bigr\rangle;n,m\right).\label{eq:3.1.18}
\end{equation}
Thus, the PIP circuit of the FI gate will also be able to perform
the FO operation by taking $\bigl|\varphi\bigr\rangle=\bigl|\mathbf{0}\bigr\rangle$.
Supplementary Figure 13 shows the optical circuit of the FI gate,
whose (reduced) forward transfer matrix is $G_{\mathrm{FI}}$ (Supplementary
Equation \ref{eq:3.1.3}). 
\noindent \begin{center}
\includegraphics[width=11cm,height=6cm,keepaspectratio]{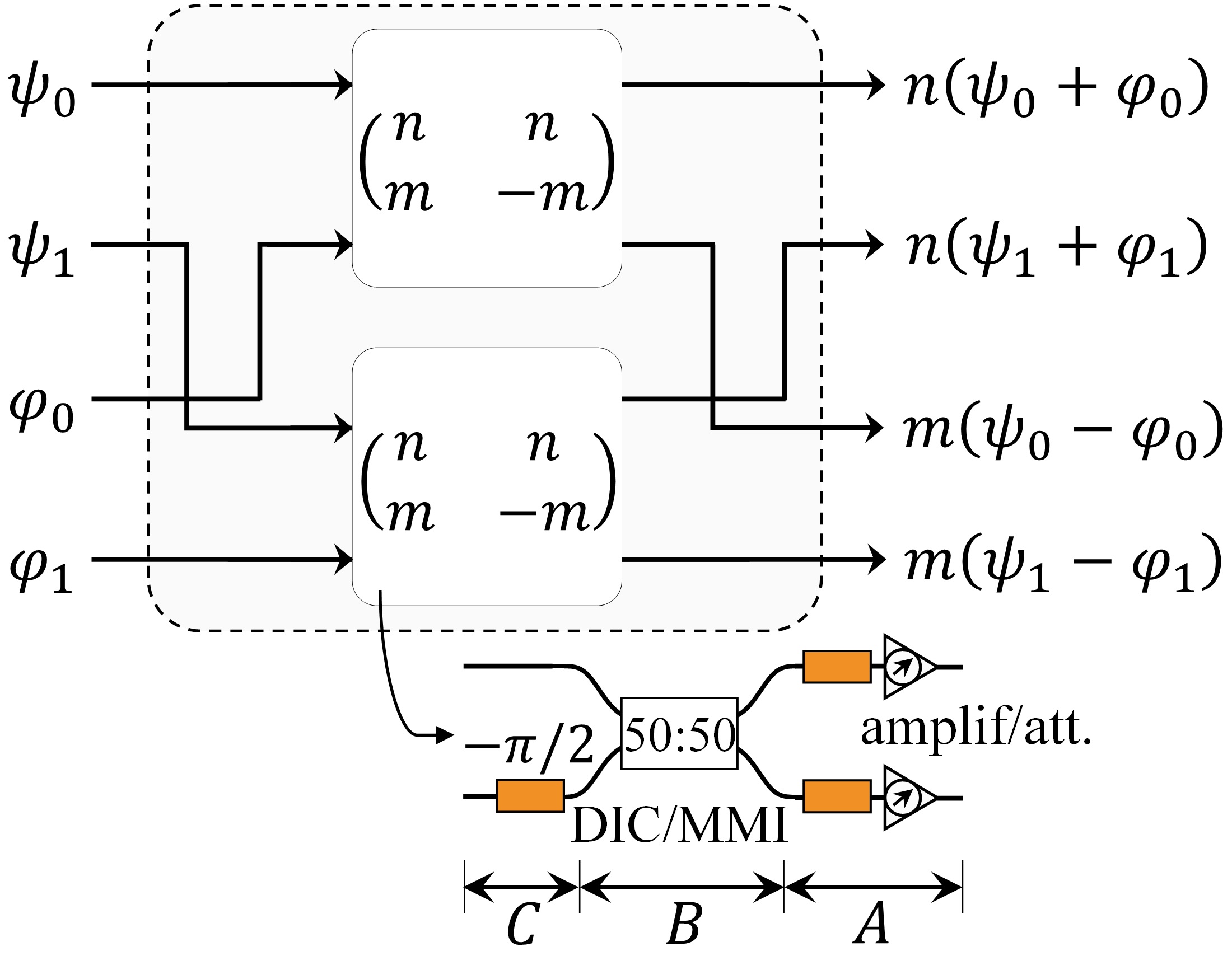}
\par\end{center}

\noindent \textbf{\small{}Supplementary Figure 13.}{\small{} PIP implementation
of both fan-in (FI) and fan-out (FO) gates using an optical circuit
whose (reduced) forward transfer matrix is given by $G_{\mathrm{FI}}$
(Supplementary Equation \ref{eq:3.1.3}). The circuit carries out
the FI operation by mapping the anbit amplitudes of $\bigl|\psi\bigr\rangle\times\bigl|\varphi\bigr\rangle$
into the anbit amplitudes of $n\bigl(\bigl|\psi\bigr\rangle+\bigl|\varphi\bigr\rangle\bigr)\times m\bigl(\bigl|\psi\bigr\rangle-\bigl|\varphi\bigr\rangle\bigr)$,
with $n,m\in\mathbb{C}-\left\{ 0\right\} $. In addition, the FO operation
of $\bigl|\psi\bigr\rangle$ can be performed by setting $\bigl|\varphi\bigr\rangle=\bigl|\mathbf{0}\bigr\rangle$
and using positive real parameters $n,m$.\\}{\small \par}

This architecture is constructed from two identical $2\times2$ subsystems
implementing the reduced forward transfer matrix depicted within the
boxes, which may be factorised as:
\begin{equation}
\left(\begin{array}{cc}
n & n\\
m & -m
\end{array}\right)\equiv\underset{A}{\underbrace{\left(\begin{array}{cc}
\sqrt{2}n & 0\\
0 & -i\sqrt{2}m
\end{array}\right)}}\underset{B}{\underbrace{\frac{1}{\sqrt{2}}\left(\begin{array}{cc}
1 & i\\
i & 1
\end{array}\right)}}\underset{C}{\underbrace{\left(\begin{array}{cc}
1 & 0\\
0 & -i
\end{array}\right)}}.\label{eq:3.1.19}
\end{equation}
The matrices $A,B,C$ can be implemented by PSs, tunable optical attenuators/amplifiers
and a 50:50 beam splitter (the PIP implementation of each matrix is
sketched in the figure for clarity). Along this line, it should be
noted that Fig.\,5a of the paper emerges from Supplementary Figure
13 by restricting the parameters to the default values $n=m=1$.

Finally, let us conclude with a brief discussion about the reciprocity
and FB symmetry of such a structure. Assuming reciprocal and FB-symmetric
attenuators/amplifiers, the global system will preserve the Lorentz
reciprocity by exhibiting a symmetric $S$ matrix:
\begin{equation}
S=\left(\begin{array}{c|c}
0 & G_{\mathrm{FI}}^{T}\\
\hline G_{\mathrm{FI}} & 0
\end{array}\right),\label{eq:3.1.20}
\end{equation}
and will preserve (break) the FB symmetry when $n=m$ ($n\neq m$).
This can be easily verified by comparing the reduced forward and backward
transfer matrices of the $2\times2$ subsystems integrating the whole
structure, which are found to be:
\begin{equation}
\widetilde{T}_{\mathrm{f}}=\left(\begin{array}{cc}
n & n\\
m & -m
\end{array}\right),\ \ \ \widetilde{T}_{\mathrm{b}}=\left(\begin{array}{cc}
n & m\\
n & -m
\end{array}\right).\label{eq:3.1.21}
\end{equation}
As seen, $\widetilde{T}_{\mathrm{f}}=\widetilde{T}_{\mathrm{b}}$
when $n=m$. These conclusions are in line with property 6 of the
FI gate (see p.\,\pageref{enu:Universality-and-PIP}).

\subsection*{3.2 First sequential system of the paper}

\addcontentsline{toc}{subsection}{3.2 First sequential system of the paper}

\noindent In this subsection, we will analyse in detail the \emph{first}
sequential architecture reported in the main text (Fig.\,5b). This
scheme is the simplest sequential system that can be built in APC,
composed by two FI/FO gates along with two M-gates that complete the
feedback loop. While in the paper we use FI/FO gates with parameters
fixed at the default value 1, let us consider here FI/FO operations
with arbitrary parameters to carry out a complete and rigorous analysis
of this first sequential scheme, see Supplementary Figure 14.
\noindent \begin{center}
\includegraphics[width=11cm,height=7cm,keepaspectratio]{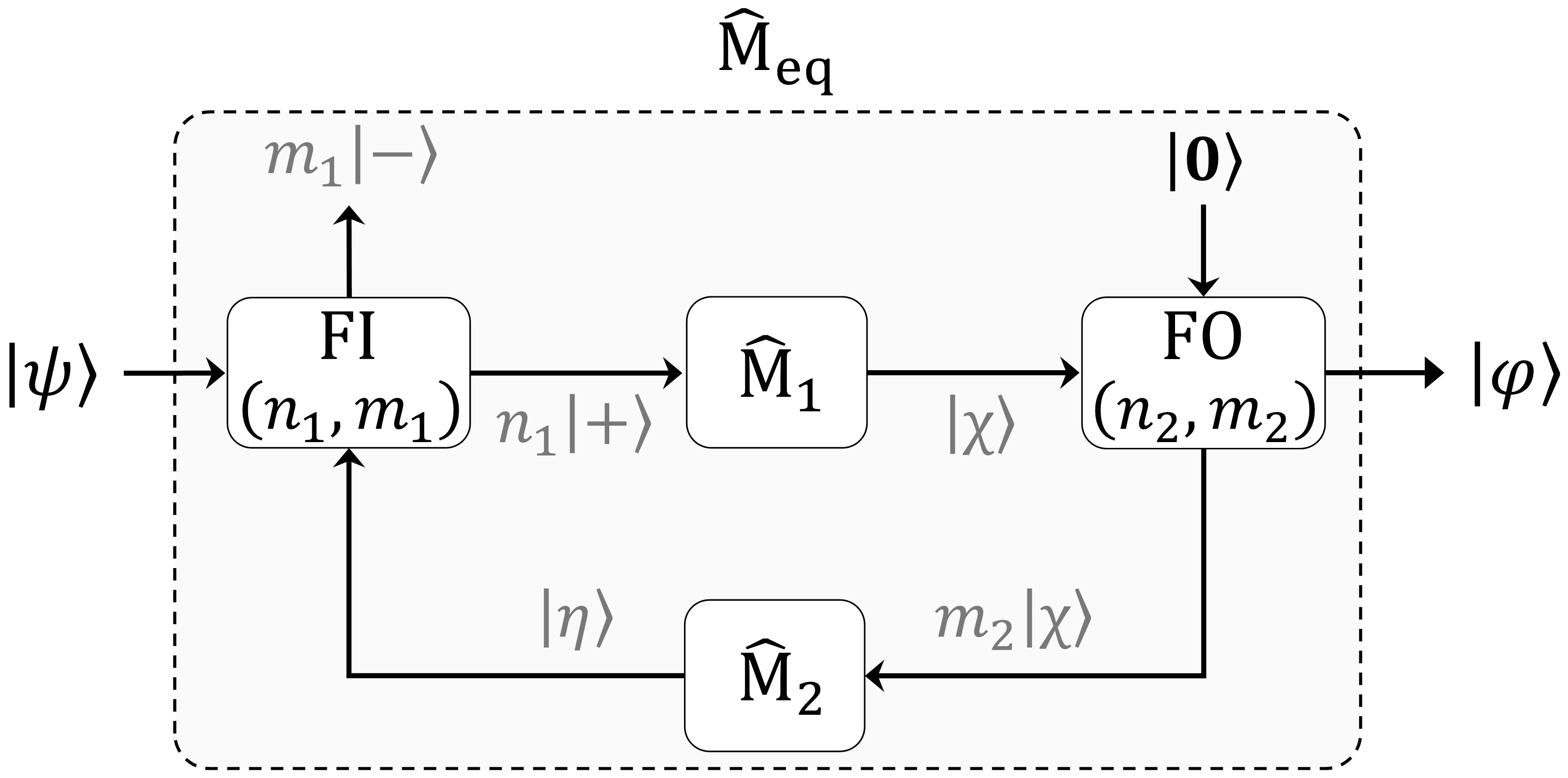}
\par\end{center}

\noindent \textbf{\small{}Supplementary Figure 14.}{\small{} Sequential
computational system of a single anbit, composed by 2 single-anbit
M-gates ($\widehat{\mathrm{M}}_{1}$ and $\widehat{\mathrm{M}}_{2}$),
1 FI gate and 1 FO gate. For the sake of completeness, we use here
FI/FO operations with arbitrary parameters $n_{1},m_{1}\in\mathbb{C}-\left\{ 0\right\} $
and $n_{2},m_{2}\in\mathbb{R}^{+}$ (however, in the paper, these
parameters are set to the default value 1 for simplicity).}{\small \par}

\newpage{}

\paragraph*{Analysis.}

The strategy to analyse any sequential architecture in APC is the
same as that of a combinational scheme: we must calculate the input-output
relation, in this example $\bigl|\varphi\bigr\rangle=\widehat{\mathrm{M}}_{\mathrm{eq}}\bigl|\psi\bigr\rangle$.
To this end, we should find the relation of $\widehat{\mathrm{M}}_{\mathrm{eq}}$
with the M-gates $\widehat{\mathrm{M}}_{1}$ and $\widehat{\mathrm{M}}_{2}$
following the anbit transformations that appear in the feedback loop.
The simplest way is to start from the output anbit:
\begin{equation}
\bigl|\varphi\bigr\rangle=n_{2}\bigl|\chi\bigr\rangle=n_{1}n_{2}\widehat{\mathrm{M}}_{1}\bigl|+\bigr\rangle=n_{1}n_{2}\widehat{\mathrm{M}}_{1}\bigl(\bigl|\psi\bigr\rangle+\bigl|\eta\bigr\rangle\bigr).\label{eq:3.2.1}
\end{equation}
Next, taking into account that $\bigl|\eta\bigr\rangle=m_{2}\widehat{\mathrm{M}}_{2}\bigl|\chi\bigr\rangle=\bigl(m_{2}/n_{2}\bigr)\widehat{\mathrm{M}}_{2}\bigl|\varphi\bigr\rangle$,
the above equation becomes:
\begin{equation}
\bigl|\varphi\bigr\rangle=n_{1}n_{2}\widehat{\mathrm{M}}_{1}\bigl|\psi\bigr\rangle+n_{1}m_{2}\widehat{\mathrm{M}}_{1}\widehat{\mathrm{M}}_{2}\bigl|\varphi\bigr\rangle.\label{eq:3.2.2}
\end{equation}
Therefore, the input-output relation of the system is found to be:
\begin{equation}
\bigl|\varphi\bigr\rangle=n_{1}n_{2}\bigl(\thinspace\widehat{1}-n_{1}m_{2}\widehat{\mathrm{M}}_{1}\widehat{\mathrm{M}}_{2}\bigr)^{-1}\widehat{\mathrm{M}}_{1}\bigl|\psi\bigr\rangle.\label{eq:3.2.3}
\end{equation}
This implies that the sequential architecture is described by an \emph{equivalent}
linear operator of the form:
\begin{align}
\widehat{\mathrm{M}}_{\mathrm{eq}} & \equiv n_{1}n_{2}\bigl(\thinspace\widehat{1}-n_{1}m_{2}\widehat{\mathrm{M}}_{1}\widehat{\mathrm{M}}_{2}\bigr)^{-1}\widehat{\mathrm{M}}_{1},\label{eq:3.2.4}
\end{align}
with $n_{1}\in\mathbb{C}-\left\{ 0\right\} $ and $n_{2},m_{2}\in\mathbb{R}^{+}$. 

Given that the system is an operation of a single anbit, the underlying
Hilbert space is $\mathscr{E}_{1}$ (with canonical basis $\mathcal{B}_{1}=\left\{ \bigl|0\bigr\rangle,\bigl|1\bigr\rangle\right\} $).
Hence, the matrix representation of $\widehat{\mathrm{M}}_{\mathrm{eq}}$
associated to $\mathcal{B}_{1}$ is:
\begin{equation}
M_{\mathrm{eq}}=M_{\mathcal{B}_{1}}^{\mathcal{B}_{1}}\bigl(\widehat{\mathrm{M}}_{\mathrm{eq}}\bigr)=n_{1}n_{2}\left(I-n_{1}m_{2}M_{1}M_{2}\right)^{-1}M_{1},\label{eq:3.2.5}
\end{equation}
where $M_{1}$ and $M_{2}$ are the matrix representations of $\widehat{\mathrm{M}}_{1}$
and $\widehat{\mathrm{M}}_{2}$, respectively. Setting $n_{1}=n_{2}=m_{2}\equiv1$,
Supplementary Equation \ref{eq:3.2.5} is reduced to the expression
of $M_{\mathrm{eq}}$ reported in the main text.

\paragraph*{Properties.}

The single-anbit sequential system of Supplementary Figure 14 has
the next properties:
\begin{enumerate}
\item \emph{Endomorphism}. The linear operator $\widehat{\mathrm{M}}_{\mathrm{eq}}$
is an endomorphism of the Hilbert space $\mathscr{E}_{1}$.
\item \emph{Existence}. Let us rewrite Supplementary Equation \ref{eq:3.2.5}
as $M_{\mathrm{eq}}=n_{1}n_{2}G^{-1}M_{\mathrm{1}}$, with $G:=I-n_{1}m_{2}M_{1}M_{2}$.
As can be seen, the existence of this sequential system (described
by $M_{\mathrm{eq}}$) directly depends on the existence of $G^{-1}$,
that is:
\begin{equation}
\exists M_{\mathrm{eq}}\Leftrightarrow\exists G^{-1}\Leftrightarrow\det\left(G\right)\neq0\Leftrightarrow\det\bigl(I-n_{1}m_{2}M_{1}M_{2}\bigr)\neq0.\label{eq:3.2.6}
\end{equation}
Contrariwise, the sequential system could not be built because $G$
would be singular.
\item \emph{Reversibility}. The gate $\widehat{\mathrm{M}}_{\mathrm{eq}}$
is reversible (automorphism) if and only if $\det\bigl(G^{-1}\bigr)\neq0$
and $\det\bigl(M_{1}\bigr)\neq0$.
\item \emph{Equivalent system}. Since $\widehat{\mathrm{M}}_{\mathrm{eq}}$
may be reversible or irreversible, then the anbit transformation can
be regarded as a single-anbit combinational M-gate. In addition, it
is interesting to highlight that the system is equivalent to two different
single-anbit gates $M_{\mathrm{1}}$ and $n_{1}n_{2}G^{-1}$ connected
in series (Supplementary Figure 15). In this scenario, note that $n_{1}n_{2}G^{-1}$
actually describes a sequential system similar to that of Supplementary
Figure 14, but with the M-gates being $I$ and $M_{\mathrm{1}}M_{\mathrm{2}}$.
In particular, this system equivalence could be employed to simplify
computational architectures that combine both combinational and sequential
gates.
\noindent \begin{center}
\includegraphics[width=10cm,height=6cm,keepaspectratio]{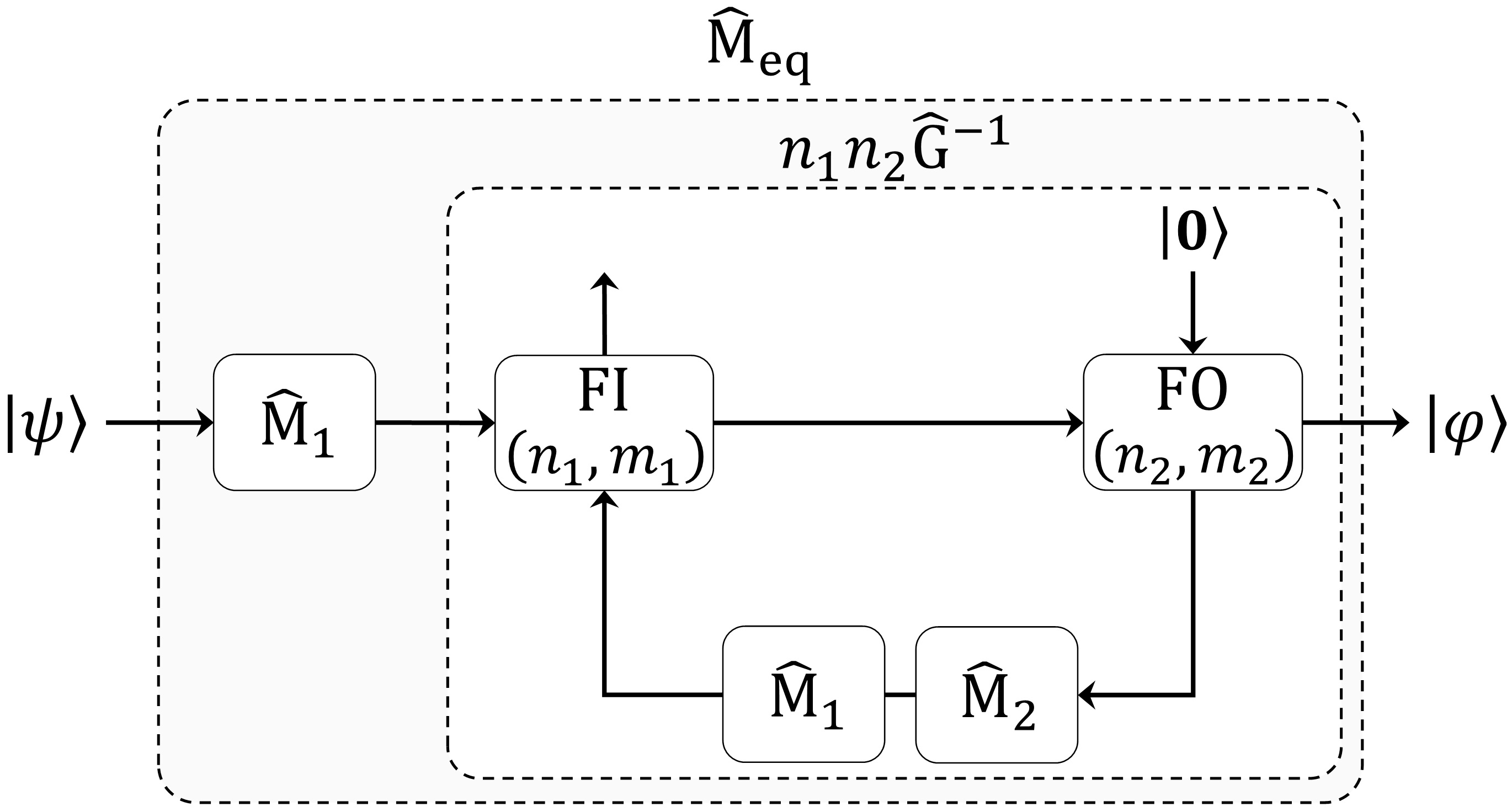}
\par\end{center}

\noindent \textbf{\small{}Supplementary Figure 15.}{\small{} Equivalent
sequential system. The gate $\widehat{\mathrm{M}}_{1}$ can be extracted
from the upper branch of the feedback loop following this scheme.
The operator $\widehat{\mathrm{M}}_{\mathrm{eq}}$ is the same as
that of the original system shown in Supplementary Figure 14.}{\small \par}
\item \emph{Algebraic structure}. The matrix $M_{\mathrm{eq}}$ belongs
to $\mathrm{U}\left(2\right)$ (i.e. the system is equivalent to a
U-gate) if $M_{1}$ and $n_{1}n_{2}G^{-1}$ are $2\times2$ unitary
matrices. Here, $n_{1}n_{2}G^{-1}\in\mathrm{U}\left(2\right)$ if
the following necessary and sufficient condition is satisfied:
\begin{equation}
A^{\dagger}A-\frac{1}{n_{1}^{\ast}n_{2}}A-\frac{1}{n_{1}n_{2}}A^{\dagger}=\left(1-\frac{1}{\bigl|n_{1}\bigr|^{2}n_{2}^{2}}\right)I,\label{eq:3.2.7}
\end{equation}
where $A:=\bigl(m_{2}/n_{2}\bigr)M_{1}M_{2}$ and $\ast$ denotes
the complex conjugate. On the other hand, $M_{\mathrm{eq}}$ belongs
to $\mathrm{GL}\left(2,\mathbb{C}\right)$ (i.e. the system is equivalent
to a G-gate) if $M_{1}$ is a general linear matrix (here, we implicitly
assume that $G\in\mathrm{GL}\left(2,\mathbb{C}\right)$ since it is
a necessary and sufficient condition for the existence of $M_{\mathrm{eq}}$,
as discussed above). Finally, $M_{\mathrm{eq}}$ is an M-gate if $M_{1}$
is a singular matrix (irreversible operation).
\item \emph{PIP implementation}. Using the circuits depicted in Figs.\,3c
and 5a of the paper to implement the gates of the feedback loop illustrated
in Supplementary Figure 14, we will obtain an optical structure that
breaks the Lorentz reciprocity and the FB symmetry.
\end{enumerate}

\subsection*{3.3 Second sequential system of the paper}

\addcontentsline{toc}{subsection}{3.3 Second sequential system of the paper}

\noindent Finally, we will analyse the \emph{second} sequential architecture
reported in the main text (Fig.\,5d). For clarity, we reproduce this
scheme in Supplementary Figure 16 including the anbits employed in
the mathematical discussions and considering FI/FO operations with
arbitrary parameters to carry out a complete and rigorous analysis.
\noindent \begin{center}
\includegraphics[width=11cm,height=7cm,keepaspectratio]{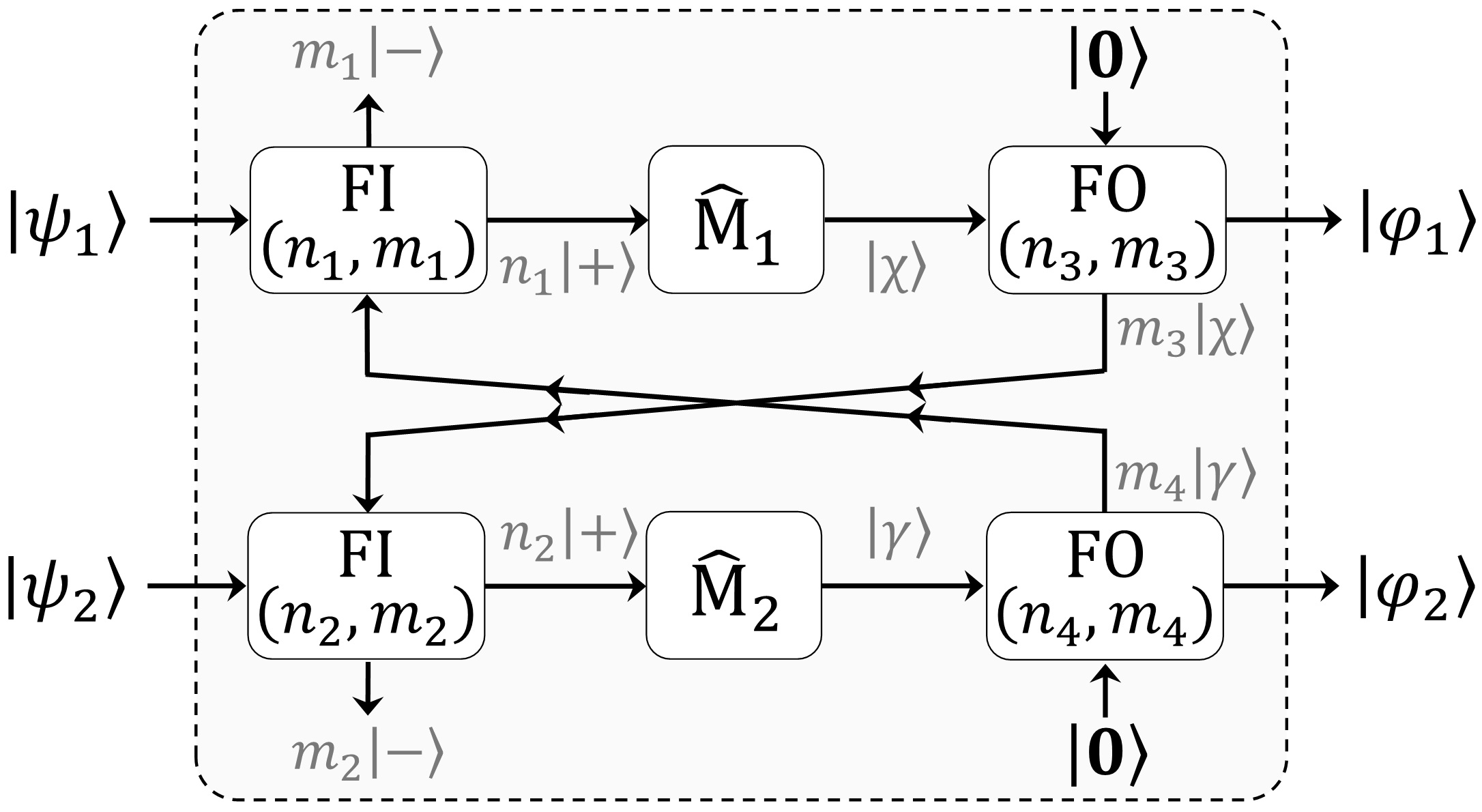}
\par\end{center}

\noindent \textbf{\small{}Supplementary Figure 16.}{\small{} Sequential
computational system of two anbits, composed by 2 single-anbit M-gates
($\widehat{\mathrm{M}}_{1}$ and $\widehat{\mathrm{M}}_{2}$), 2 FI
gates and 2 FO gates. For completeness, we use FI/FO operations with
arbitrary parameters $n_{1,2},m_{1,2}\in\mathbb{C}-\left\{ 0\right\} $
and $n_{3,4},m_{3,4}\in\mathbb{R}^{+}$.\\}{\small \par}

The goal is to find the input-output relation of the structure, as
well as to demonstrate that such a mapping is equivalent to the input-output
relation of the multi-anbit combinational architecture shown in Fig.\,5c
of the paper. With this approach in mind, we should start by calculating
the output anbits $\bigl|\varphi_{1}\bigr\rangle$ and $\bigl|\varphi_{2}\bigr\rangle$
independently. The output anbit $\bigl|\varphi_{1}\bigr\rangle$ can
be expressed as:
\begin{align}
\bigl|\varphi_{1}\bigr\rangle & =n_{3}\bigl|\chi\bigr\rangle\nonumber \\
 & =n_{1}n_{3}\widehat{\mathrm{M}}_{1}\bigl|+\bigr\rangle\nonumber \\
 & =n_{1}n_{3}\widehat{\mathrm{M}}_{1}\bigl(\bigl|\psi_{1}\bigr\rangle+m_{4}\bigl|\gamma\bigr\rangle\bigr)\nonumber \\
 & =n_{1}n_{3}\widehat{\mathrm{M}}_{1}\bigl|\psi_{1}\bigr\rangle+n_{1}n_{3}m_{4}\widehat{\mathrm{M}}_{1}\bigl|\gamma\bigr\rangle\nonumber \\
 & =n_{1}n_{3}\widehat{\mathrm{M}}_{1}\bigl|\psi_{1}\bigr\rangle+n_{1}n_{2}n_{3}m_{4}\widehat{\mathrm{M}}_{1}\widehat{\mathrm{M}}_{2}\bigl|+\bigr\rangle\nonumber \\
 & =n_{1}n_{3}\widehat{\mathrm{M}}_{1}\bigl|\psi_{1}\bigr\rangle+n_{1}n_{2}n_{3}m_{4}\widehat{\mathrm{M}}_{1}\widehat{\mathrm{M}}_{2}\bigl(\bigl|\psi_{2}\bigr\rangle+m_{3}\bigl|\chi\bigr\rangle\bigr)\nonumber \\
 & =n_{1}n_{3}\widehat{\mathrm{M}}_{1}\bigl|\psi_{1}\bigr\rangle+n_{1}n_{2}n_{3}m_{4}\widehat{\mathrm{M}}_{1}\widehat{\mathrm{M}}_{2}\bigl(\bigl|\psi_{2}\bigr\rangle+\frac{m_{3}}{n_{3}}\bigl|\varphi_{1}\bigr\rangle\bigr)\nonumber \\
 & =n_{1}n_{3}\widehat{\mathrm{M}}_{1}\bigl|\psi_{1}\bigr\rangle+n_{1}n_{2}n_{3}m_{4}\widehat{\mathrm{M}}_{1}\widehat{\mathrm{M}}_{2}\bigl|\psi_{2}\bigr\rangle+n_{1}n_{2}m_{3}m_{4}\widehat{\mathrm{M}}_{1}\widehat{\mathrm{M}}_{2}\bigl|\varphi_{1}\bigr\rangle.\label{eq:3.3.1}
\end{align}
Hence, $\bigl|\varphi_{1}\bigr\rangle$ is found to be:
\begin{align}
\bigl|\varphi_{1}\bigr\rangle & =n_{1}n_{3}\bigl(\thinspace\widehat{1}-n_{1}n_{2}m_{3}m_{4}\widehat{\mathrm{M}}_{1}\widehat{\mathrm{M}}_{2}\bigr)^{-1}\widehat{\mathrm{M}}_{1}\bigl|\psi_{1}\bigr\rangle\nonumber \\
 & +n_{1}n_{2}n_{3}m_{4}\bigl(\thinspace\widehat{1}-n_{1}n_{2}m_{3}m_{4}\widehat{\mathrm{M}}_{1}\widehat{\mathrm{M}}_{2}\bigr)^{-1}\widehat{\mathrm{M}}_{1}\widehat{\mathrm{M}}_{2}\bigl|\psi_{2}\bigr\rangle.\label{eq:3.3.2}
\end{align}
Repeating the same procedure to calculate $\bigl|\varphi_{2}\bigr\rangle$,
we obtain:
\begin{align}
\bigl|\varphi_{2}\bigr\rangle & =n_{1}n_{2}m_{3}n_{4}\bigl(\thinspace\widehat{1}-n_{1}n_{2}m_{3}m_{4}\widehat{\mathrm{M}}_{2}\widehat{\mathrm{M}}_{1}\bigr)^{-1}\widehat{\mathrm{M}}_{2}\widehat{\mathrm{M}}_{1}\bigl|\psi_{1}\bigr\rangle\nonumber \\
 & +n_{2}n_{4}\bigl(\thinspace\widehat{1}-n_{1}n_{2}m_{3}m_{4}\widehat{\mathrm{M}}_{2}\widehat{\mathrm{M}}_{1}\bigr)^{-1}\widehat{\mathrm{M}}_{2}\bigl|\psi_{2}\bigr\rangle.\label{eq:3.3.3}
\end{align}
For simplicity, let us compact the above expressions as:
\begin{align}
\bigl|\varphi_{1}\bigr\rangle & =\widehat{\mathrm{A}}_{1}\bigl|\psi_{1}\bigr\rangle+\widehat{\mathrm{A}}_{2}\bigl|\psi_{2}\bigr\rangle,\label{eq:3.3.4}\\
\bigl|\varphi_{2}\bigr\rangle & =\widehat{\mathrm{B}}_{1}\bigl|\psi_{1}\bigr\rangle+\widehat{\mathrm{B}}_{2}\bigl|\psi_{2}\bigr\rangle,\label{eq:3.3.5}
\end{align}
where the operators $\widehat{\mathrm{A}}_{1,2}$ and $\widehat{\mathrm{B}}_{1,2}$
can be identified by comparing the corresponding equations. Using
the \emph{Cartesian} \emph{product}, the input-output relation of
the whole sequential structure is:
\begin{align}
\bigl(\bigl|\varphi_{1}\bigr\rangle,\bigl|\varphi_{2}\bigr\rangle\bigr) & =\bigl(\widehat{\mathrm{A}}_{1}\bigl|\psi_{1}\bigr\rangle+\widehat{\mathrm{A}}_{2}\bigl|\psi_{2}\bigr\rangle,\widehat{\mathrm{B}}_{1}\bigl|\psi_{1}\bigr\rangle+\widehat{\mathrm{B}}_{2}\bigl|\psi_{2}\bigr\rangle\bigr)\nonumber \\
 & =\bigl(\widehat{\mathrm{A}}_{1}\bigl|\psi_{1}\bigr\rangle,\widehat{\mathrm{B}}_{1}\bigl|\psi_{1}\bigr\rangle\bigr)+\bigl(\widehat{\mathrm{A}}_{2}\bigl|\psi_{2}\bigr\rangle,\widehat{\mathrm{B}}_{2}\bigl|\psi_{2}\bigr\rangle\bigr)\nonumber \\
 & =\bigl(\widehat{\mathrm{A}}_{1}\times\widehat{\mathrm{B}}_{1}\bigr)\thinspace\bigl(\bigl|\psi_{1}\bigr\rangle,\bigl|\psi_{1}\bigr\rangle\bigr)+\bigl(\widehat{\mathrm{A}}_{2}\times\widehat{\mathrm{B}}_{2}\bigr)\thinspace\bigl(\bigl|\psi_{2}\bigr\rangle,\bigl|\psi_{2}\bigr\rangle\bigr).\label{eq:3.3.6}
\end{align}
Furthermore, the input-output relation of the multi-anbit combinational
architecture of Fig.\,5c of the paper is:
\begin{align}
\bigl(\bigl|\varphi_{1}\bigr\rangle,\bigl|\varphi_{2}\bigr\rangle\bigr) & =\bigl(\widehat{\mathrm{M}}_{3}\bigl|\psi_{1}\bigr\rangle+\widehat{\mathrm{M}}_{5}\bigl|\psi_{2}\bigr\rangle,\widehat{\mathrm{M}}_{4}\bigl|\psi_{1}\bigr\rangle+\widehat{\mathrm{M}}_{6}\bigl|\psi_{2}\bigr\rangle\bigr)\nonumber \\
 & =\bigl(\widehat{\mathrm{M}}_{3}\bigl|\psi_{1}\bigr\rangle,\widehat{\mathrm{M}}_{4}\bigl|\psi_{1}\bigr\rangle\bigr)+\bigl(\widehat{\mathrm{M}}_{5}\bigl|\psi_{2}\bigr\rangle,\widehat{\mathrm{M}}_{6}\bigl|\psi_{2}\bigr\rangle\bigr)\nonumber \\
 & =\bigl(\widehat{\mathrm{M}}_{3}\times\widehat{\mathrm{M}}_{4}\bigr)\thinspace\bigl(\bigl|\psi_{1}\bigr\rangle,\bigl|\psi_{1}\bigr\rangle\bigr)+\bigl(\widehat{\mathrm{M}}_{5}\times\widehat{\mathrm{M}}_{6}\bigr)\thinspace\bigl(\bigl|\psi_{2}\bigr\rangle,\bigl|\psi_{2}\bigr\rangle\bigr).\label{eq:3.3.7}
\end{align}
As seen, Supplementary Equations \ref{eq:3.3.6} and \ref{eq:3.3.7}
are equivalent mathematical expressions.

\newpage{}

\section*{Supplementary Note 4: different versions of APC\label{sec:4}}

\addcontentsline{toc}{section}{Supplementary Note 4: different versions of APC}

\noindent The anbit is conceived as a vector function belonging to
a two-dimensional (2D) Hilbert space. However, as commented in the
main text, we have the possibility of defining the unit of information
of APC in a Hilbert space with dimension $d\neq2$, leading to different
versions of APC, termed as $d$-APC. In this section, we discuss how
to construct the computation theory (unit of information and basic
gates) for the cases $d=1$ and $d>2$.

\subsection*{4.1 One-dimensional APC (1-APC)}

\addcontentsline{toc}{subsection}{4.1 One-dimensional APC (1-APC)}

\subsubsection*{4.1.1 Unit of information: the analog dit}

\addcontentsline{toc}{subsubsection}{4.1.1 Unit of information: the analog dit}

\noindent Now, the unit of information is a 1D vector function $\bigl|\psi\left(t\right)\bigr\rangle=\psi_{0}\left(t\right)\bigl|0\bigr\rangle$
referred to as the \emph{analog dit} (or \emph{andit} for short),
where $\psi_{0}$ is a scalar complex function termed as the andit
amplitude and $\bigl|0\bigr\rangle$ is a constant unit vector. The
user information is encoded in the module and phase of $\psi_{0}$.

In the same vein as an anbit, a 1D andit may be implementable in PIP
via different modulation formats (see p.\,\pageref{sec:1}). For
instance, using a SEM approach, $\psi_{0}$ can be encoded by an optical
wave packet propagated by the fundamental mode $\bigl|0\bigr\rangle$
of a single-mode waveguide (Supplementary Figure 17). For coherence
with the paper, the temporal shape of $\psi_{0}$ is assumed rectangular,
although alternative physical implementations of a 1D andit can be
proposed by exploring diverse temporal shapes of $\psi_{0}$. In addition,
the following features of a 1D andit should be highlighted:
\begin{itemize}
\item \emph{Hilbert space}. The single-andit vector space $\mathscr{E}_{1}=\mathrm{span}\left\{ \bigl|0\bigr\rangle\right\} $
along with the standard complex inner product $\bigl\langle\cdot|\cdot\bigr\rangle$
lead to a 1D Hilbert space with canonical (normal) basis $\mathcal{B}_{1}=\left\{ \bigl|0\bigr\rangle\right\} $
and a finite norm $\left\Vert \cdot\right\Vert $. Specifically, $\left\Vert \psi\right\Vert ^{2}$
describes the optical power ($\mathcal{P}$) propagated by the waveguide
of Supplementary Figure 17: $\mathcal{P}=\left\Vert \psi\right\Vert ^{2}=\bigl\langle\psi|\psi\bigr\rangle=\left|\psi_{0}\right|^{2}$. 
\item \emph{Andit period}. The andit period $T_{\mathrm{ANDIT}}$ is the
time interval where $\psi_{0}\left(t\right)$ is defined.
\item \emph{Measurement and EDFs}. The number of EDFs that can be used to
encode the user information depends solely on the kind of andit measurement
employed at the receiver. A coherent measurement retrieves the module
and phase of $\psi_{0}$ (2 EDFs), while a differential measurement
only recovers $\left|\psi_{0}\right|^{2}$ (1 EDF).
\item \emph{Geometric representation}. A 1D andit can be geometrically represented
by using a polar diagram illustrating the module and phase of $\psi_{0}$
in the complex plane (Supplementary Figure 17).
\item \emph{Multiple andits}. A computational system with $n$ andits will
require to operate in a Hilbert space $\mathscr{E}_{n}$ that can
be constructed from $\mathscr{E}_{1}$ using the tensor or the Cartesian
product. In the former case, we will obtain the Hilbert space $\mathscr{E}_{n}=\mathscr{E}_{1}^{\otimes\left(n-1\right)}$,
with $\dim\bigl(\mathscr{E}_{n}\bigr)=\dim\bigl(\mathscr{E}_{1}\bigr)^{n}=1$
and canonical basis $\mathcal{B}_{n}=\left\{ \left|0,0,\ldots,0\right\rangle \right\} $.
In the latter case, we will obtain the Hilbert space $\mathscr{E}_{n}=\mathscr{E}_{1}^{\times\left(n-1\right)}$,
with $\dim\bigl(\mathscr{E}_{n}\bigr)=n\dim\bigl(\mathscr{E}_{1}\bigr)=n$
and canonical basis $\mathcal{B}_{n}=\bigl\{\bigl(\left|0\right\rangle ,\left|\mathbf{0}\right\rangle ,\ldots,\left|\mathbf{0}\right\rangle \bigr),\bigl(\left|\mathbf{0}\right\rangle ,\left|0\right\rangle ,\ldots,\left|\mathbf{0}\right\rangle \bigr),\ldots,\bigl(\left|\mathbf{0}\right\rangle ,\left|\mathbf{0}\right\rangle ,\ldots,\left|0\right\rangle \bigr)\bigr\}$.
As we will see below, in 1-APC, it is more useful to work with the
Cartesian product to define multi-andit operations.
\end{itemize}
\newpage{}
\noindent \begin{center}
\includegraphics[width=13cm,height=6cm,keepaspectratio]{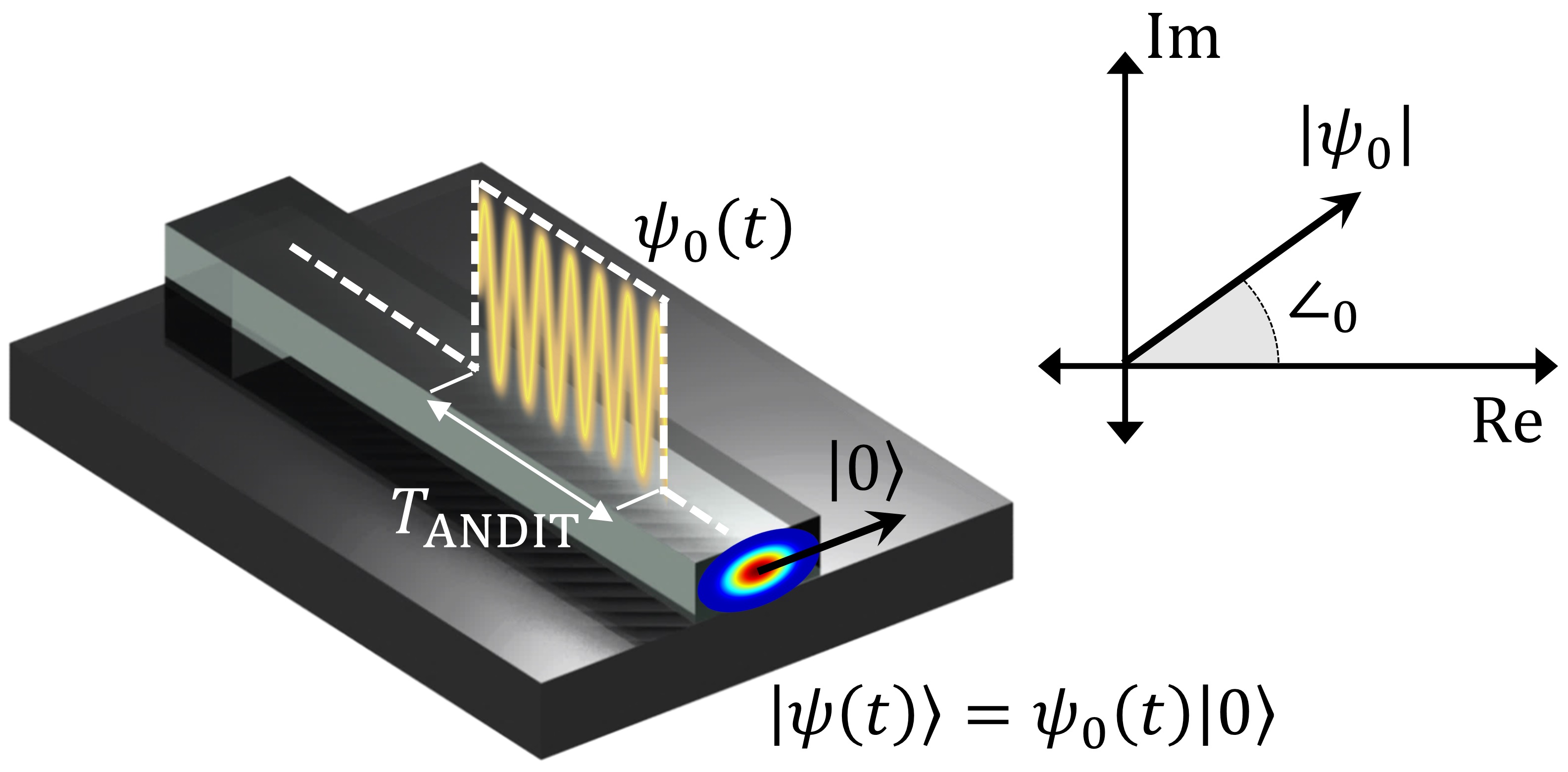}
\par\end{center}

\noindent \textbf{\small{}Supplementary Figure 17.}{\small{} Physical
implementation of a one-dimensional andit $\bigl|\psi\left(t\right)\bigr\rangle=\psi_{0}\left(t\right)\bigl|0\bigr\rangle$
using PIP technology and a space-encoding modulation (SEM). The andit
amplitude $\psi_{0}=\left|\psi_{0}\right|e^{i\angle_{0}}$ is encoded
by an optical wave packet (or complex envelope) propagated by the
fundamental mode $\bigl|0\bigr\rangle$ of a single-mode waveguide.
The andit is geometrically represented by using a polar diagram illustrating
the module $\left|\psi_{0}\right|$ and phase $\angle_{0}$ in the
complex plane.}{\small \par}

\subsubsection*{4.1.2 Single-andit linear gates}

\addcontentsline{toc}{subsubsection}{4.1.2 Single-andit linear gates}

\paragraph*{Definition.}

In the same way as in 2-APC, a single-andit linear gate will be defined
as a linear mapping $\widehat{\mathrm{F}}$ (or endomorphism) of $\mathscr{E}_{1}$.
The operator $\widehat{\mathrm{F}}$ describes a transformation between
an input andit $\bigl|\psi\bigr\rangle=\psi_{0}\bigl|0\bigr\rangle$
and an output andit $\bigl|\varphi\bigr\rangle=\varphi_{0}\bigl|0\bigr\rangle$.

\paragraph*{General properties.}

The properties of a single-andit linear gate in 1-APC are the same
as those of a single-anbit linear gate in 2-APC, but the following
details should be highlighted:
\begin{itemize}
\item \emph{Scalar representation}. Instead of using a matrix representation,
the input-output relation $\bigl|\varphi\bigr\rangle=\widehat{\mathrm{F}}\bigl|\psi\bigr\rangle$
can equivalently be expressed via a scalar expression of the form
$\varphi_{0}=F\psi_{0}$, where $F=\left\langle 0\right|\widehat{\mathrm{F}}\left|0\right\rangle $.
Utilising the component isomorphism $\left[\cdot\right]_{\mathcal{B}_{1}}$,
it is direct to demonstrate the above equivalence:
\begin{equation}
\varphi_{0}=\bigl[\bigl|\varphi\bigr\rangle\bigr]_{\mathcal{B}_{1}}=\bigl[\widehat{\mathrm{F}}\bigl|\psi\bigr\rangle\bigr]_{\mathcal{B}_{1}}=\psi_{0}\bigl[\widehat{\mathrm{F}}\bigl|0\bigr\rangle\bigr]_{\mathcal{B}_{1}}=\psi_{0}\left\langle 0\right|\widehat{\mathrm{F}}\left|0\right\rangle \equiv F\psi_{0}.\label{eq:4.1.1}
\end{equation}
\item \emph{Reversibility}. By definition, the gate is reversible if and
only if $\left|F\right|\neq0$. The inverse gate is described by the
inverse operator $\widehat{\mathrm{F}}^{-1}$, whose scalar representation
is $F^{-1}=1/F$.
\item \emph{Geometric representation}. A single-andit linear gate may be
geometrically interpreted as a trajectory between two different points
in the complex plane, from $\psi_{0}$ to $\varphi_{0}$.
\item \emph{Classes of linear gates}. We use the same classification as
in 2-APC by considering U-, G-, and M-gates. 
\end{itemize}

\paragraph*{U-gates.}

Since a U-gate must preserve the norm of the input andit at the output
andit ($\left|\varphi_{0}\right|=\left|\psi_{0}\right|$), this class
of gates is associated to a unitary operator $\widehat{\mathrm{F}}=e^{i\delta}\bigl|0\bigr\rangle\bigl\langle0\bigr|$
or scalar phase function $F=e^{i\delta}$. Hence, a U-gate only induces
a global phase shifting and, consequently, the MCA is a PS.

\paragraph*{G- and M-gates.}

The most general scalar complex function that can be defined is of
the form $F=\left|F\right|e^{i\delta}$, with arbitrary module $\left|F\right|$
and arbitrary phase $\delta$. Such a linear mapping will describe
a G-gate when $\left|F\right|\neq0$ (reversible operation) and will
describe an M-gate when $\left|F\right|=0$ (irreversible operation).
The MCA is a PS connected in series with a tunable optical amplifier/attenuator
accounting for the transformations $e^{i\delta}$ and $\left|F\right|$,
respectively.

\subsubsection*{4.1.3 Dual-andit linear gates}

\addcontentsline{toc}{subsubsection}{4.1.3 Dual-andit linear gates}

\noindent Using the Cartesian product, we may define a Hilbert space
$\mathscr{E}_{2}=\mathscr{E}_{1}\times\mathscr{E}_{1}$ with canonical
basis $\mathcal{B}_{2}=\bigl\{\bigl(\left|0\right\rangle ,\left|\mathbf{0}\right\rangle \bigr),\bigl(\left|\mathbf{0}\right\rangle ,\left|0\right\rangle \bigr)\bigr\}$
which is isomorphic to the Hilbert space of a single anbit in 2-APC
($\mathscr{E}_{1}^{(\textrm{2-APC})}=\mathrm{span}\left\{ \left|0\right\rangle ,\left|1\right\rangle \right\} $).
The isomorphism between both spaces is established by performing the
identifications $\bigl(\left|0\right\rangle ,\left|\mathbf{0}\right\rangle \bigr)\equiv\left|0\right\rangle $
and $\bigl(\left|\mathbf{0}\right\rangle ,\left|0\right\rangle \bigr)\equiv\left|1\right\rangle $.
In this way, any vector belonging to $\mathscr{E}_{2}$ in 1-APC of
the form: 
\begin{equation}
\bigl(\left|\alpha\right\rangle ,\left|\beta\right\rangle \bigr)=\bigl(\alpha_{0}\left|0\right\rangle ,\beta_{0}\left|0\right\rangle \bigr)=\alpha_{0}\bigl(\left|0\right\rangle ,\left|\mathbf{0}\right\rangle \bigr)+\beta_{0}\bigl(\left|\mathbf{0}\right\rangle ,\left|0\right\rangle \bigr),\label{eq:4.1.2}
\end{equation}
has associated an anbit belonging to $\mathscr{E}_{1}^{(\textrm{2-APC})}$
of the form $\left|\psi\right\rangle =\alpha_{0}\left|0\right\rangle +\beta_{0}\left|1\right\rangle $.
Using this isomorphism, we can directly extrapolate the theory and
circuits of the single-anbit linear gates in 2-APC (see Fig.\,3)
to design and implement dual-andit linear gates in 1-APC. As seen,
the Cartesian product paves the way to extrapolate the theory from
2-APC to 1-APC, a feature that cannot be found by using the tensor
product.

On the other hand, the study of multi-andit linear gates in 1-APC
will be investigated simultaneously along with the development of
multi-anbit linear gates in 2-APC in future works. Here, the optical
structures proposed by Reck and Clements in refs.\,\cite{key-S25,key-S26}
could be of paramount importance to implement multi-andit U-gates
in 1-APC.

\subsubsection*{4.1.4 Controlled gates}

\addcontentsline{toc}{subsubsection}{4.1.4 Controlled gates}

\noindent In 1-APC, a controlled gate should be defined in a different
way from a controlled gate in\linebreak{}
2-APC given that a control andit cannot commute between two different
states. In this case, we may establish that a single-andit operation
$\widehat{\mathrm{F}}$ is applied to a target andit $\left|t\right\rangle =t_{0}\left|0\right\rangle $
when the amplitude of the control andit $\left|c\right\rangle =c_{0}\left|0\right\rangle $
takes a specific complex value. In such a scenario, we can also propose
an electro-optic design to implement a controlled gate, where the
target and control andits are respectively encoded by 1D optical and
electrical signals.

\subsubsection*{4.1.5 Sequential design}

\addcontentsline{toc}{subsubsection}{4.1.5 Sequential design}

\noindent The fundamentals of sequential architectures in 1-APC can
be directly extrapolated from 2-APC by restricting the analysis to
1D vectors. The FI and FO gates (the basic pieces to construct sequential
schemes) can be implemented using similar circuits to those of proposed
in Fig.\,5a and Supplementary Figure 13. Concretely, in Fig.\,5a,
we should only work with the inputs $\psi_{0}$ and $\varphi_{0}$
and the outputs $\psi_{0}+\varphi_{0}$ and $\psi_{0}-\varphi_{0}$
to carry out FI/FO operations of 1D andits. Furthermore, note that
the analysis and design of the sequential architectures shown in Figs.\,5b
and 5d are the same in 1-APC and 2-APC.

\subsubsection*{4.1.6 Nonlinear andit gates}

\addcontentsline{toc}{subsubsection}{4.1.6 Nonlinear andit gates}

\noindent The most general definition of a single-andit gate (including
both linear and nonlinear contributions) is given by the expression
$\widehat{\mathrm{F}}\bigl|\psi\bigr\rangle:=f_{0}\bigl(\psi_{0}\bigr)\bigl|0\bigr\rangle$,
with $f_{0}\in\mathcal{F}\left(\mathbb{C},\mathbb{C}\right)$. Thus,
$\widehat{\mathrm{F}}$ will induce a nonlinear transformation on
the input andit $\bigl|\psi\bigr\rangle$ when the scalar function
$f_{0}$ has a nonlinear behaviour. The theory of nonlinear gates
in 1-APC is similar to that of nonlinear gates in 2-APC. Here, we
should only restrict the mathematical formalism of equations 5 and
6 of the paper to 1D. 

This kind of gates can be implemented in PIP, e.g., by exploiting
the Pockels and Kerr effects in a \emph{single-mode} waveguide to
carry out second- and third-order nonlinear andit operations, respectively.
As an example, stimulating the self-phase modulation effect in a single-mode
waveguide (similar to that of depicted in Supplementary Figure 17),
a nonlinear andit transformation of the form $\widehat{\mathrm{F}}\bigl|\psi\bigr\rangle=\psi_{0}\exp\bigl(-i\gamma\bigl|\psi_{0}\bigr|^{2}L_{\mathrm{eff}}\bigr)\bigl|0\bigr\rangle$
may be obtained ($\gamma$ and $L_{\mathrm{eff}}$ are nonlinear parameters
of the waveguide \cite{key-S27}).

\subsection*{4.2 Multi-dimensional APC (\emph{d}-APC)}

\addcontentsline{toc}{subsection}{4.2 Multi-dimensional APC (\textit{d}-APC)}

\noindent In $d$-APC with $d>2$, the unit of information is a $d$-dimensional
vector function:
\begin{equation}
\bigl|\psi\left(t\right)\bigr\rangle=\sum_{n=0}^{d-1}\psi_{n}\left(t\right)\bigl|n\bigr\rangle,\label{eq:4.2.1}
\end{equation}
also termed as andit, where $\psi_{0},\psi_{1},\ldots,\psi_{d-1}$
are scalar complex functions termed as the andit amplitudes and $\bigl|0\bigr\rangle,\bigl|1\bigr\rangle,\ldots,\bigl|d-1\bigr\rangle$
are constant orthonormal vectors. The user information is encoded
in the moduli and phases of the andit amplitudes.\\\\
\noindent \begin{center}
\includegraphics[width=14cm,height=7cm,keepaspectratio]{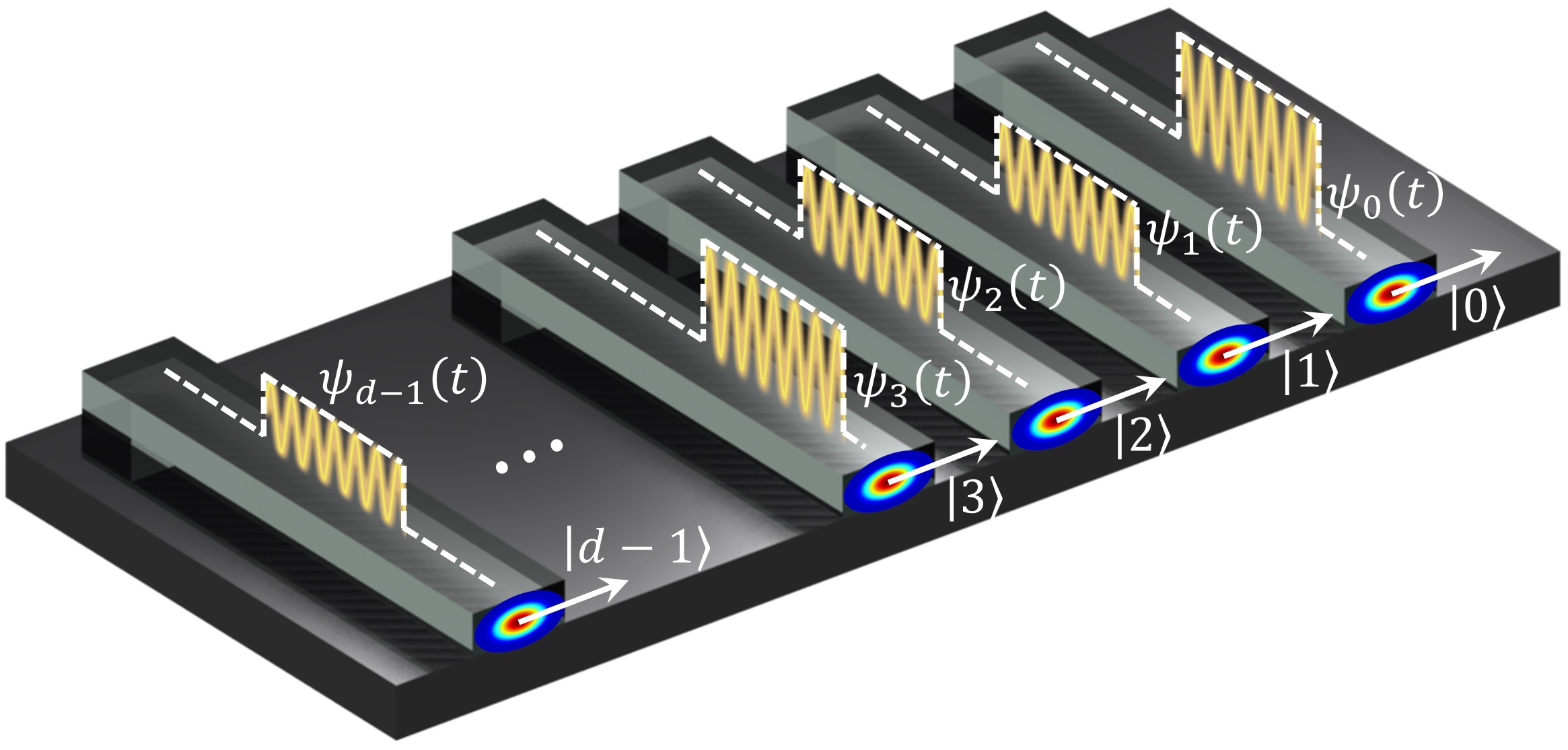}
\par\end{center}

\noindent \textbf{\small{}Supplementary Figure 18.}{\small{} Physical
implementation of a $d$-dimensional andit using PIP technology and
a space-encoding modulation (SEM). The andit amplitudes $\psi_{0},\psi_{1},\psi_{2},\psi_{3},\ldots,\psi_{d-1}$
are encoded by $d$ different optical wave packets (or complex envelopes)
propagated by the fundamental modes $\bigl|0\bigr\rangle,\bigl|1\bigr\rangle,\bigl|2\bigr\rangle,\bigl|3\bigr\rangle,\ldots,\bigl|d-1\bigr\rangle$
of $d$ different single-mode waveguides.}{\small \par}

\newpage{}

A $d$-dimensional andit can be implemented in PIP via different modulation
formats that make use of the space, mode, frequency and time domain
of light. These modulation formats are similar to those proposed for
the anbit in Supplementary Note 1 (p.\,\pageref{sec:1}), but generalised
for a $d$-dimensional vector. As an example, in a SEM technique,
we can associate $\bigl|0\bigr\rangle,\bigl|1\bigr\rangle,\ldots,\bigl|d-1\bigr\rangle$
to the fundamental modes of $d$ different single-mode waveguides
and the andit amplitudes can be encoded by $d$ different optical
wave packets propagated by these fundamental modes, see Supplementary
Figure 18. Moreover, although the temporal shape of the andit amplitudes
is assumed rectangular (for coherence with the main text), diverse
options may be explored in forthcoming works by employing optical
wave packets with different shapes. 

The properties of a $d$-dimensional andit are similar to those of
an anbit. In particular, now, we will have $2d$ or $2d-1$ EDFs when
using coherent or differential measurement, respectively. In the latter
case, the global phase of the andit amplitudes cannot be recovered.

Once we have defined the unit of information of $d$-APC, the subsequent
steps to construct this version of the computation theory are to study
single-andit linear gates, controlled gates, FI/FO gates, sequential
architectures and nonlinear andit gates. Despite the fact that the
theoretical details of these points are out of the scope of this work,
the following considerations are in order:
\begin{itemize}
\item From a theoretical perspective, the fundamentals of combinational
and sequential computing systems in $d$-APC are the same as those
of $2$-APC. The main difference appears in the mathematical formalism
of nonlinear gates, that will require the use of multi-dimensional
multi-variable Taylor series in $d$-APC.
\item From a technological perspective, the PIP implementation of single-andit
gates would require basic devices with $d$ inputs and $d$ outputs.
Since the mainstream PIP devices are $1\times1$ and $2\times2$ optical
systems \cite{key-S4}, a possible difficulty might arise to implement
these gates. Nonetheless, a suitable PIP implementation of $d$-APC
could be proposed using the approaches reported in refs.\,\cite{key-S25,key-S26,key-S28},
based on reconfigurable PSs, beam splitters (combiners) and micro-ring
resonators that allow to integrate the desired number of inputs and
outputs within the same optical system. As an example, the results
of ref.\,\cite{key-S28} can be directly applied to implement single-andit
U-gates in $3$-APC.
\item In QC, the optical implementation of universal quantum gates may be
simplified by combining the use of qubits and qudits \cite{key-S29}.
In the same vein, $d$-APC might be envisioned as a theoretical tool
that could allow us to simplify the PIP implementation of the analog
logic of $2$-APC by utilising anbits and andits within the same circuit.
\end{itemize}
\newpage{}

\section*{Supplementary Note 5: implementing other computing theories\label{sec:5}}

\addcontentsline{toc}{section}{Supplementary Note 5: implementing other computing theories}

\noindent In this section, we will discuss how to describe the unit
of information and basic operations of digital computation (DC), neuromorphic
computation (NC) and quantum computation (QC) using the mathematical
framework of APC.

\subsection*{5.1 Digital computation }

\addcontentsline{toc}{subsection}{5.1 Digital computation}

\noindent The implementation of DC with APC architectures requires
to describe the digital bit and the Boolean logic using the anbit
and its basic gates. Remarkably, an anbit $\bigl|\psi\bigr\rangle=\psi_{0}\bigl|0\bigr\rangle+\psi_{1}\bigl|1\bigr\rangle$
is able to describe the digital bit by taking $\psi_{0}\in\left\{ 0,1\right\} $
and $\psi_{1}=1-\psi_{0}$. Here, the state of the anbit is $\bigl|0\bigr\rangle$
or $\bigl|1\bigr\rangle$, but the possibility of having a linear
combination of both states is discarded.

Furthermore, the description of the Boolean logic can be carried out
in APC keeping in mind that any Boolean function can be decomposed
in terms of NAND gates \cite{key-S30}. Thus, implementing the NAND
operation of DC utilising basic anbit gates, we will be able to perform
any Boolean function in APC.
\noindent \begin{center}
\includegraphics[width=15.5cm,height=8cm,keepaspectratio]{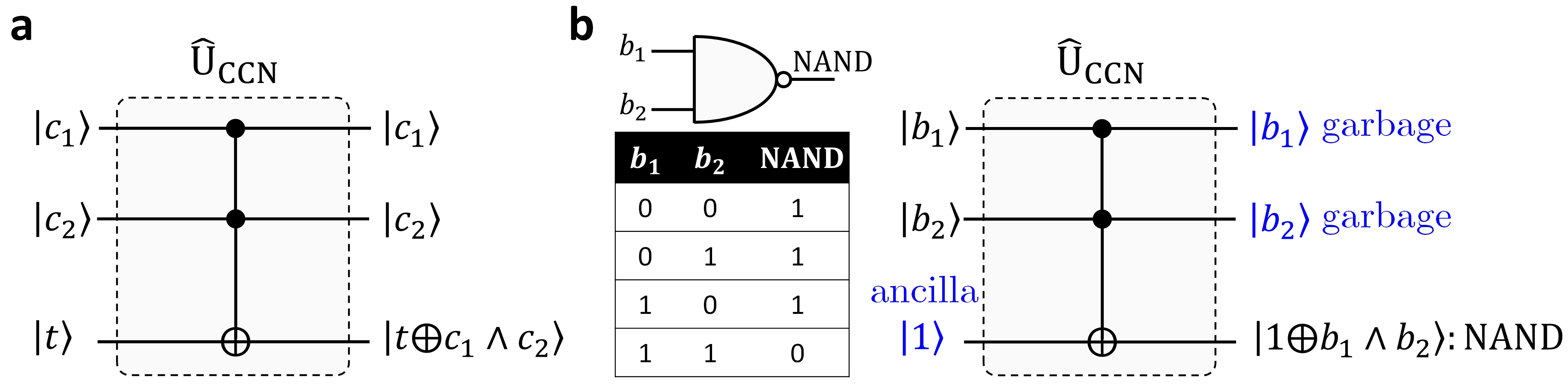}
\par\end{center}

\noindent \textbf{\small{}Supplementary Figure 19.}{\small{} Implementation
of Boolean logic in APC. }\textbf{\small{}a}{\small{} Functional scheme
of the Toffoli (or CCNOT) gate, described by Supplementary Equation
}\ref{eq:5.1.1}{\small{}, a multi-anbit controlled gate with 2 control
anbits $\bigl|c_{1}\bigr\rangle$, $\bigl|c_{2}\bigr\rangle$ and
1 target anbit $\bigl|t\bigr\rangle$. }\textbf{\small{}b}{\small{}
NAND operation of DC (defined via its truth table) implemented by
the Toffoli gate in APC.\\}{\small \par}

In this scenario, note that the linear unitary mapping of the Toffoli
(or CCNOT) gate given by Supplementary Equation \ref{eq:2.3.16} can
be recast of the form \cite{key-S22,key-S31}:
\begin{align}
\bigl|c_{1},c_{2},t\bigr\rangle\overset{_{\textrm{CCNOT}}}{\longrightarrow} & \widehat{\mathrm{U}}_{\mathrm{CCN}}\bigl|c_{1},c_{2},t\bigr\rangle=\bigl|c_{1},c_{2},t\oplus c_{1}\wedge c_{2}\bigr\rangle,\label{eq:5.1.1}
\end{align}
where $\oplus$ is the modulo-2 addition and $\wedge$ is the AND
operation (Supplementary Figure 19a). This expression allows us to
infer that the Toffoli gate may be employed to implement the NAND
gate:
\begin{equation}
\mathrm{NAND}\bigl(b_{1},b_{2}\bigr):=\lnot\bigl(b_{1}\wedge b_{2}\bigr)=1\oplus b_{1}\wedge b_{2},\label{eq:5.1.2}
\end{equation}
by performing the following identification:
\begin{equation}
\bigl|b_{1},b_{2},\mathrm{NAND}\bigl(b_{1},b_{2}\bigr)\bigr\rangle=\bigl|b_{1},b_{2},1\oplus b_{1}\wedge b_{2}\bigr\rangle\equiv\widehat{\mathrm{U}}_{\mathrm{CCN}}\bigl|b_{1},b_{2},1\bigr\rangle,\label{eq:5.1.3}
\end{equation}
where $b_{1}$ and $b_{2}$ are digital bits (i.e., $b_{1,2}\in\left\{ 0,1\right\} $)
playing the role of the control anbits in the Toffoli gate. Moreover,
the target anbit is set to $\bigl|t\bigr\rangle=\bigl|1\bigr\rangle$
at the input (Supplementary Figure 19b). As seen, the use of the Toffoli
gate in APC to implement the NAND gate of DC requires 1 ancilla anbit
and 2 garbage anbits (the same remark is found in QC to implement
DC via the quantum Toffoli gate \cite{key-S22,key-S31}). Likewise,
let us remember that the PIP implementation of the Toffoli gate is
shown in Supplementary Figure 10 (see p.\,\pageref{Supplementary-Figure-10.}).

On the other hand, a complete implementation of DC in APC requires
the ability to build sequential digital circuits (e.g., digital memories).
To this end, latches and flip-flops (the basic building blocks of
most sequential digital systems \cite{key-S30}) could be mimicked
in APC by employing sequential anbit computational schemes. This point
should be further investigated in forthcoming works.

\subsection*{5.2 Neuromorphic computation}

\addcontentsline{toc}{subsection}{5.2 Neuromorphic computation}

\noindent NC is a computational model inspired by signal processing
in the brain using the so-called artificial neural networks (ANNs)
\cite{key-S32,key-S33}. The functional principle of an ANN is sketched
in Supplementary Figure 20. Specifically, an ANN is arranged in several
layers, which are composed by small processing units known as \emph{neurons}.
Each neuron processes a 1D complex signal (the unit of information
in NC). In this example, we observe 3 layers with 3 neurons per layer.
Mathematically, the communication between two adjacent layers can
be modelled in two steps \cite{key-S33}:
\begin{enumerate}
\item Firstly, it is performed a matrix transformation of a 3D input vector
$\mathbf{x}=\left(x_{1},x_{2},x_{3}\right)$, composed by 3 different
units of information $x_{1,2,3}$. The output is also a 3D vector
$\mathbf{y}=\left(y_{1},y_{2},y_{3}\right)$ (the dimension of $\mathbf{x}$
and $\mathbf{y}$ depends on the number of neurons per layer). Hence,
this first step is a multi-linear mapping described by a matrix equation
of the form $\mathbf{y}=M\mathbf{x}$, where $M$ is a $3\times3$
matrix with complex entries.
\item Secondly, a nonlinear function $f\in\mathcal{F}\left(\mathbb{C},\mathbb{C}\right)$
is applied on each component $y_{k}$ of the vector $\mathbf{y}$.
The output of this second step is a 3D vector $\mathbf{z}=\left(z_{1},z_{2},z_{3}\right)$
with $z_{k}=f\left(y_{k}\right)$, $\forall k\in\left\{ 1,2,3\right\} $.
\end{enumerate}
\noindent \begin{center}
\includegraphics[width=10cm,height=6cm,keepaspectratio]{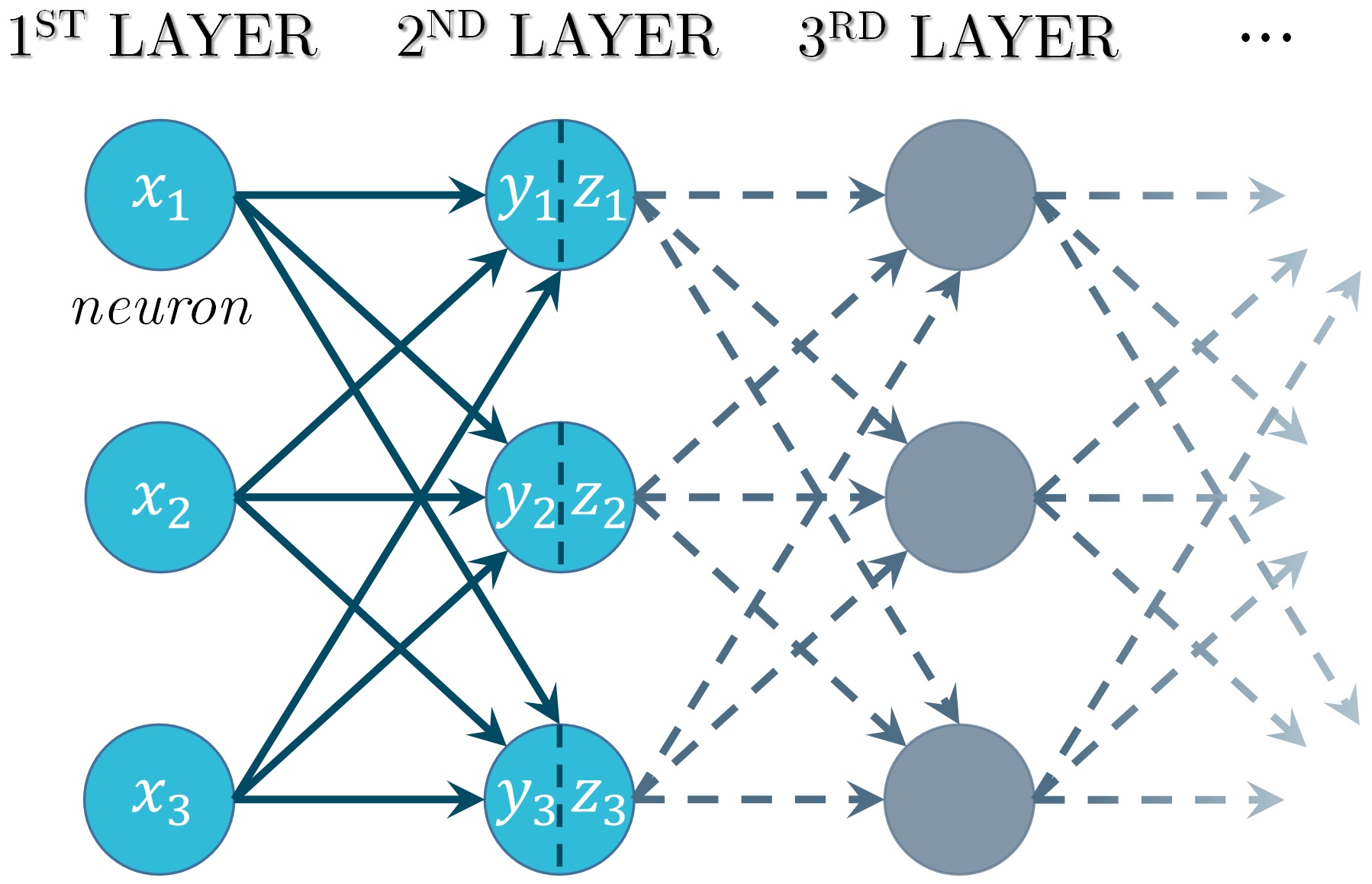}
\par\end{center}

\noindent \begin{center}
\textbf{\small{}Supplementary Figure 20.}{\small{} Artificial neural
network (ANN) with 3 neurons per layer \cite{key-S33}.\\ }
\par\end{center}{\small \par}

Therefore, the description of NC is straightforward by using 1-APC
(see Supplementary Note 4, on p.\,\pageref{sec:4}). In order to
explain the connection between NC and 1-APC as simple as possible,
let us describe the ANN depicted in Supplementary Figure 20 via 1-APC.
In such a scenario, the following considerations are in order:
\begin{itemize}
\item Firstly, we should note that the unit of information of NC may be
described by a 1D andit. For instance, the unit of information $x_{1}$
can be described by a 1D andit of the form $\bigl|x_{1}\bigr\rangle=x_{1}\bigl|0\bigr\rangle$.
\item Secondly, the 3D input vector $\mathbf{x}=\left(x_{1},x_{2},x_{3}\right)$
should be written as a function of 3 different 1D andits $\bigl|x_{1}\bigr\rangle=x_{1}\bigl|0\bigr\rangle$,
$\bigl|x_{2}\bigr\rangle=x_{2}\bigl|0\bigr\rangle$ and $\bigl|x_{3}\bigr\rangle=x_{3}\bigl|0\bigr\rangle$
connected via the Cartesian product. Hence, the vector $\mathbf{x}$
may be described by a ket of the form $\bigl|X\bigr\rangle=\bigl|x_{1}\bigr\rangle\times\bigl|x_{2}\bigr\rangle\times\bigl|x_{3}\bigr\rangle$.
In the same vein, the vectors $\mathbf{y}=\left(y_{1},y_{2},y_{3}\right)$
and $\mathbf{z}=\left(z_{1},z_{2},z_{3}\right)$ can respectively
be described by the kets $\bigl|Y\bigr\rangle=\bigl|y_{1}\bigr\rangle\times\bigl|y_{2}\bigr\rangle\times\bigl|y_{3}\bigr\rangle$
and $\bigl|Z\bigr\rangle=\bigl|z_{1}\bigr\rangle\times\bigl|z_{2}\bigr\rangle\times\bigl|z_{3}\bigr\rangle$,
where $\bigl|y_{k}\bigr\rangle=y_{k}\bigl|0\bigr\rangle$ and $\bigl|z_{k}\bigr\rangle=z_{k}\bigl|0\bigr\rangle$,
$\forall k\in\left\{ 1,2,3\right\} $.
\item Thirdly, the linear and nonlinear mappings of the ANN can respectively
be described in 1-APC by a linear operator $\widehat{\mathrm{M}}$
(whose matrix representation must be the $M$-matrix used to transform
$\mathbf{x}$ into $\mathbf{y}$) and a nonlinear operator $\widehat{\mathrm{F}}$
connecting the kets $\bigl|X\bigr\rangle$, $\bigl|Y\bigr\rangle$,
$\bigl|Z\bigr\rangle$ via the expressions $\bigl|Y\bigr\rangle=\widehat{\mathrm{M}}\bigl|X\bigr\rangle$
and $\bigl|Z\bigr\rangle=\widehat{\mathrm{F}}\bigl|Y\bigr\rangle=f\left(y_{1}\right)\bigl|0\bigr\rangle\times f\left(y_{2}\right)\bigl|0\bigr\rangle\times f\left(y_{3}\right)\bigl|0\bigr\rangle$.\\
\end{itemize}

\subsection*{5.3 Quantum computation}

\addcontentsline{toc}{subsection}{5.3 Quantum computation}

\noindent As expected, APC is also able to describe the mathematical
model of QC, but only partially. It is worth mentioning that there
are some fundamental concepts of the qubit and its basic operations
that cannot be modelled with APC since it is a computing theory whose
unit of information is implemented by classical waves. In order to
clarify these concepts, let us take a closer look at the unit of information
and basic operations of QC.

\paragraph*{Unit of information: the qubit.}

An arbitrary state of a qubit is described by a ket of the form \cite{key-S22}:
\begin{equation}
\bigl|\psi\bigr\rangle=\cos\frac{\theta}{2}\bigl|0\bigr\rangle+e^{i\varphi}\sin\frac{\theta}{2}\bigl|1\bigr\rangle,\label{eq:5.3.1}
\end{equation}
with 2 EDFs ($\theta$ and $\varphi$) and a constant norm $\left\Vert \psi\right\Vert ^{2}=\bigl\langle\psi|\psi\bigr\rangle=1$.
Given that an anbit has more EDFs than a qubit, we can use the former
to describe the latter by setting a constant norm equal to $1$ and
using quantum waves to implement the anbit amplitudes. 

As commented in the paper, the main differences between both units
of information are associated to the measurement, the description
of multiple units of information and the entanglement of units of
information:
\begin{itemize}
\item \emph{Measurement}. In quantum mechanics, the wave function collapse
makes an ideal quantum measurement a \emph{non-reversible} operation
in QC \cite{key-S10,key-S22}. It is impossible to reconstruct the
information encoded by the EDFs of a pre-measurement qubit from the
state of a post-measurement qubit. In contrast, in APC, a measurement
is always a \emph{reversible} operation because the number of EDFs
is preserved, see Supplementary Note 1 on p.\,\pageref{subsec:1.2-Anbit-measurement}.
The information encoded by the EDFs of a pre-measurement anbit may
be retrieved from the state of the post-measurement anbit, regardless
of whether a coherent or differential measurement strategy is employed.
Therefore, an ideal quantum measurement cannot be described by the
mathematical framework of APC.\newpage{}
\item \emph{Multiple units of information}. While the description of multiple
anbits can be carried out using the tensor product or the Cartesian
product in APC, the description of multiple qubits must be exclusively
performed via the tensor product in QC. This composition rule is derived
from the state postulate and the measurement postulate of quantum
mechanics \cite{key-S34}. However, it is direct to infer that this
remark does not impose any limitation on APC to describe multiple
qubits.
\item \emph{Entanglement}. The entanglement of different qubits is possible
thanks to the non-local nature of quantum mechanics \cite{key-S22}.
Unfortunately, this instantaneous non-locality cannot be found in
a deterministic physical theory, e.g., in classical electromagnetism
\cite{key-S35}. Although electromagnetic non-local media are supported
by Maxwell's equations, the time response is causal, that is, there
is a time delay between the incident electric field strength ($\boldsymbol{\mathcal{E}}=\sum_{l}\mathcal{E}_{l}\hat{\mathbf{u}}_{l}$)
and the electric displacement ($\boldsymbol{\mathcal{D}}=\sum_{k}\mathcal{D}_{k}\hat{\mathbf{u}}_{k}$).
Here, it should be noted that the most general non-local linear constitutive
relation between $\boldsymbol{\mathcal{E}}$ and $\boldsymbol{\mathcal{D}}$
is given by the expression (Einstein's summation convention) \cite{key-S36}:
\begin{equation}
\mathcal{D}_{k}\left(\mathbf{r},t\right)=\int_{-\infty}^{\infty}\int_{-\infty}^{t}\varepsilon_{kl}\left(\mathbf{r},\mathbf{r}^{\prime},t,t^{\prime}\right)\mathcal{E}_{l}\left(\mathbf{r}^{\prime},t^{\prime}\right)\mathrm{d}t^{\prime}\mathrm{d}^{3}r^{\prime},\label{eq:5.3.2}
\end{equation}
where $\varepsilon_{kl}$ is the electric permittivity tensor. Specifically,
the time integral precludes\linebreak{}
instantaneous non-local effects in classical electromagnetism. Consequently,
the quantum entanglement cannot be described by APC given that its
technological PIP implementation is based on classical electromagnetic
waves.
\end{itemize}

\paragraph*{Basic qubit operations.}

The basic operations of QC are single-qubit gates and controlled qubit
gates. Interestingly, the mathematical framework of APC is able to
describe all these operations. A single-qubit gate is mathematically
equivalent to a single-anbit U-gate and the theoretical formalism
of a controlled anbit gate is the same as that of a controlled qubit
gate (as commented on p.\,\pageref{eq:2.3.3}). In addition, taking
into account that any qubit logic gate may be composed from single-qubit
gates and controlled qubit gates \cite{key-S22}, we infer that a
multi-qubit operation can also be described by APC. Nonetheless, any
quantum operation exploiting the non-local nature of quantum mechanics
cannot be emulated with APC.

\paragraph*{Conclusion.}

APC is able to \emph{partially} describe the mathematical (and physical)
model of QC given that the ideal quantum measurement and the quantum
entanglement have no classical analogy in our computation theory,
as discussed above. Nevertheless, APC might be regarded as a potential
theoretical and didactic toolbox (implementable with current technology)
that could improve our comprehension about the subtle (but essential)
differences between quantum and classical systems.

\newpage{}

\end{document}